\documentclass[aps,prd,twocolumn,showpacs,superscriptaddress,preprintnumbers, floatfix,nofootinbib]{revtex4-1}
\usepackage{amsmath}
\usepackage{amsfonts}
\usepackage{hyperref, url}
\usepackage{graphicx,subfigure}
\usepackage[normalem]{ulem}
\usepackage{epsfig} 
\usepackage{color,soul}
\usepackage{multirow}
\usepackage{array}
\usepackage[utf8]{inputenc}
\usepackage[T1]{fontenc}
\usepackage[dvipsnames]{xcolor}
\usepackage{comment}
\usepackage{aas_macros}
\usepackage[normalem]{ulem}
\usepackage{ulem}
\hypersetup{colorlinks   = true,
            urlcolor     = blue,
            citecolor    = blue,
            linkcolor    = blue,
            menucolor    = blue,
            anchorcolor  = blue,
            filecolor    = blue}

\enlargethispage{\baselineskip}

\usepackage{graphicx}

\newcommand{\Mpl}{M_{\text{pl}}}

\newcommand{\nn}{\nonumber}

\begin{document}
\preprint{KEK-QUP-2025-0001, KEK-TH-2688, KEK-Cosmo-0371}

\title{Multimessenger Astronomy Beyond the Standard Model: \\New Window from Quantum Sensors}

\author{Jason Arakawa}
\email{arakawaj@udel.edu}
\affiliation{Department of Physics and Astronomy, University of Delaware, Newark, Delaware 19716, USA}
\affiliation{International Center for Quantum-field Measurement Systems for Studies of the Universe and Particles (QUP, WPI),
High Energy Accelerator Research Organization (KEK), Tsukuba, Ibaraki 305-0801, Japan}
\affiliation{Theory Center, Institute of Particle and Nuclear Studies (IPNS), High Energy Accelerator Research Organization (KEK), Tsukuba 305-0801, Japan
}

\author{Muhammad H. Zaheer}
\email{hani@udel.edu}
\affiliation{Department of Physics and Astronomy, University of Delaware, Newark, Delaware 19716, USA}
\affiliation{Department of Electrical \& Computer Engineering, University of Delaware
140 Evans Hall, Newark, DE, USA}

\author{Volodymyr Takhistov}
\email{vtakhist@post.kek.jp}
\affiliation{International Center for Quantum-field Measurement Systems for Studies of the Universe and Particles (QUP, WPI),
High Energy Accelerator Research Organization (KEK), Tsukuba, Ibaraki 305-0801, Japan}
\affiliation{Theory Center, Institute of Particle and Nuclear Studies (IPNS), High Energy Accelerator Research Organization (KEK), Tsukuba 305-0801, Japan
}
\affiliation{Kavli Institute for the Physics and Mathematics of the Universe (WPI), \mbox{The University of Tokyo Institutes for Advanced Study, The University of Tokyo, Kashiwa, Chiba 277-8583, Japan}}
\affiliation{Graduate University for Advanced Studies (SOKENDAI), 
Tsukuba, Ibaraki 305-0801, Japan}

\author{Marianna S. Safronova}
\email{msafrono@udel.edu}
\affiliation{Department of Physics and Astronomy, University of Delaware, Newark, Delaware 19716, USA}
\affiliation{Joint Quantum Institute, National Institute of Standards and Technology and the University of Maryland, College Park, Maryland 20742, USA}

\author{Joshua Eby}
\email{joshaeby@gmail.com}
\affiliation{The Oskar Klein Centre, Department of Physics, Stockholm University, 10691 Stockholm, Sweden}

\author{Charles Cheung}
\email{ccheung@udel.edu}
\affiliation{Department of Physics and Astronomy, University of Delaware, Newark, Delaware 19716, USA}

\date{\today}

\begin{abstract}
Ultralight bosonic (ULB) fields with mass $m_{\phi} \ll 1$~eV often arise in theories beyond the Standard Model (SM). If such fields exist, violent astrophysical events that result in emission of gravitational wave, photon, or neutrino signals could also produce bursts of high-density relativistic ULB fields. Detection of 
such ULB fields in terrestrial or space-based laboratories correlated with other signals from transient astrophysical events opens a novel avenue for multimessenger astronomy. We show that quantum sensors are particularly well-suited to observe emitted scalar and pseudoscalar axion-like ULB fields coupled to SM. We demonstrate that multimessenger astronomy with ULB fields is possible even when accounting for matter screening effects.
\end{abstract}
\maketitle

\section{Introduction}

Multimessenger astronomy provides a powerful framework for investigating the origin and mechanisms of energetic astrophysical events. Gravitational waves (GW), photons ($\gamma$), and neutrinos ($\nu$) offer complementary observational perspectives, unveiling different timescales and processes that contribute to a more comprehensive understanding of complex astrophysical phenomena. While GW and neutrinos propagate over cosmic distances with minimal attenuation, photons can be absorbed or scattered by intervening matter along the line of sight, affecting their detectability. Significant attenuation effects and interactions are especially prominent for propagating charged particles.

A distinct and novel avenue in multimessenger astronomy has been recently proposed, focusing on the detection of feebly-interacting quantum fields beyond the Standard Model (SM) that are linked to transient astrophysical events~\cite{Dailey:2020sxa,Eby:2021ece}. 
The discovery of the Higgs boson confirmed that scalar fields exist and can be of fundamental significance. New scalar and pseudoscalar axion-like particles (ALPs) readily appear in various extensions of the SM, including more fundamental string theory~\cite{Svrcek:2006yi, Arvanitaki:2009fg}, extra-dimensional models~\cite{Choi:2003wr}, and frameworks based on the Peccei-Quinn mechanism~\cite{Kim:1986ax, Ringwald:2012hr}. These fields can naturally be ultralight, with masses $m_{\phi} \ll 1$ eV, such as when they arise as pseudo-Nambu-Goldstone bosons (pNGBs) of spontaneously broken global symmetries. In certain scenarios (e.g.~\cite{Arvanitaki:2009fg}), a large number of such fields with a broad mass spectrum can emerge, significantly impacting astrophysical and cosmological processes.
Considerable experimental efforts are directed towards searching for ultralight bosonic fields (ULBs), particularly in the context of their connection to dark matter (DM), which constitutes the dominant component of matter in the Universe~\cite{Antypas:2022asj}.

Quantum sensors, such as quantum clocks and atom interferometers, are well-equipped to probe the wave-like nature of ULBs, due to their extreme sensitivity to minute variations in the experimental observables \cite{Antypas:2022asj}. In general, quantum technology systems are capable of sensing the collective behavior of the wave-like ULB fields, rather than the energy deposition from individual quanta. For instance, oscillations in atomic transition frequencies in clock experiments can constrain variations in fundamental constants (e.g. the fine structure constant, $\alpha$). These detector systems have been used to search for ULB DM, under the assumption that an appreciable density of ULB DM exists locally near a quantum sensor system. However, in many scenarios ULBs emitted from transient sources remain wave-like \cite{Dailey:2020sxa, Eby:2021ece,Arakawa:2023gyq,Arakawa:2024lqr}, and quantum sensors can operate as telescopes within a multimessenger framework.

\begin{table*}[t]
\centering
\begin{tabular}{c}
\textbf{Glossary of Terms}
\end{tabular}
\\
\begin{tabular}{| c | l |}
\hline
    $\phi$ & ~~~ULB field\\
    $d^{(n)}_{i}$ &  ~~~Dilatonic coupling of order $n$ and type $i$ \\
    $d_{i,\rm crit}^{(2)}$ & ~~~Critical coupling of type $i$ \\
    $g_{\phi ij}$ & ~~~Axion coupling to SM fields $i$ and $j$\\
    $\rho_i$ & ~~~Energy density of field $i$ in medium\\
    $n(\omega)$ & ~~~Index of refraction of $\phi$\\
    $v_g$ & ~~~Group velocity of $\phi$\\
    $t_*$ & ~~~Intrinsic burst duration at source \\
    $\tilde{t}_*$ & ~~~Signal duration at detector\\
    $\delta t$ & ~~~Time delay between $\phi$ and astrophysical signal\\
    $\eta_{\rm DM}$ & ~~~Fractional frequency sensitivity of experiment to DM signal\\
    $\eta_{*}$ & ~~~Fractional frequency sensitivity of experiment to transient signal\\
    $t_{\rm int,DM}$ & ~~~Integration time for DM experiment to reach sensitivity $\eta_{\rm DM}$\\
    $t_{\rm int, *}$ & ~~~Integration time for transient search\\
    $\tau_{\rm DM}$ & ~~~Coherence time of DM\\
    $\tau_{*}$ & ~~~Coherence time of transient ULB signal\\
\hline
\end{tabular}
\end{table*}

A diverse range of transient astrophysical sources can emit bursts of ULB fields (see e.g.~\cite{Eby:2024mhd} for discussion). This includes explosive emission from boson stars (bosenovae)~\cite{Eby:2016cnq,Levkov:2016rkk,Eby:2021ece,Arakawa:2023gyq,Arakawa:2024lqr}, superradiant emission from rotating binary black holes~\cite{Yoshino:2012kn,Baumann:2021fkf,Takahashi:2024fyq} and supernovae~\cite{Raffelt:1999tx, deGiorgi:2024pjb}, among others. These scenarios offer opportunities to detect new quantum fields that may not constitute a significant fraction of DM. The direct detection of ULB fields depends on their local density near the experiment. If the ambient ULB density is significantly lower than that of DM, conventional direct detection experiments will have suppressed signals.
However, astrophysical transient events provide a promising avenue for enhanced detection prospects, as they could lead to a temporary increase in the local ULB density at terrestrial or space-based experimental sites. With numerous scenarios predicting ULB production, further studies and numerical simulations are required to precisely characterize their spectral emission properties and event rates.

Dedicated experimental searches for coincident multimessenger signals of ULBs with GW or $\gamma$ signals have recently commenced~\cite{Sen:2023fcu,Khamis:2024oqa}, considering that ULB fields interact with SM constituents. Depending on the interaction type, these couplings can induce a range of time-dependent laboratory signatures, including variations in fundamental coupling constants. Specifically, Ref.~\cite{Khamis:2024oqa} investigated pseudoscalar interactions in the context of black hole binary mergers, while Ref.~\cite{Sen:2023fcu} explored quadratic dilatonic couplings. Establishing the feasibility and detection prospects of multimessenger astronomy with new quantum messenger fields requires further investigation.

Importantly, ULBs can experience significant screening effects during propagation, severely limiting their detectability in multimessenger astronomy. This is particularly relevant for quadratic ULB couplings, which induce a back-action effect from SM fields, affecting both ULB propagation and detection~\cite{Stadnik:2021qyf}.
Screening effects can also play a prominent role in cosmology and dark energy theories with scalar fields~\cite{Brax:2013ida}, including chameleons~\cite{Khoury:2003rn}, K-mouflage~\cite{Babichev:2009ee, Brax:2014wla}, dilatons~\cite{Brax:2010gi}, and symmetrons~\cite{Hinterbichler:2011ca}. These effects have been the subject of various experimental searches with laboratory experiments~(e.g.~\cite{Jaffe:2016fsh, yin:2025}) and have also been investigated in the context of DM~\cite{Derevianko:2013oaa, Hees:2018fpg}, 
including the behavior of quadratically-coupled DM around compact objects like the Earth~\cite{burrage2024time,bauer2024validity,del2025enhanced,banerjee2025momentum}.

Screening mechanisms are critical in determining the feasibility of multimessenger astronomy, as in high-density environments, back-action effects can lead to severe screening, causing significant delays between astrophysical messengers and ULB signals—potentially that can exceed the live-time of experiments. In extreme cases, exponential suppression due to screening could render ULB detection entirely unfeasible.

In this work, we systematically investigate screening effects during ULB propagation and detection in the context of multimessenger astronomy, addressing whether such signals can be observed in coincidence with standard astrophysical messengers from the same source despite screening. We focus on three key aspects: source properties, field propagation to detectors at or in the vicinity of Earth, and detection strategies. This analysis builds on the relativistic axion burst framework introduced in Ref.~\cite{Eby:2021ece}, while quantum sensor networks provide a natural platform for correlated multimessenger searches for ultralight fields, as discussed in Ref.~\cite{Dailey:2020sxa}. For quadratic couplings that exhibit screening, the same interaction governing detection also drives screening, leading to an interplay where stronger couplings enhance both signal strength and screening effects. We analyze both scalar and pseudoscalar fields in a general framework, incorporating screening effects and expanding the characterization of potential sources. Our findings further motivate the search for correlated transient signals of ULBs alongside astrophysical messengers such as GWs, photons, and neutrinos. By exploring both positive and negative couplings, we demonstrate that ULB signals can remain detectable alongside SM signals even in the presence of screening effects. This has significant implications for the feasibility of multimessenger astronomy beyond the SM.

The paper is structured as follows. In Sec.~\ref{sec:EFT}, we outline how ULB couplings emerge within the framework of effective field theory (EFT). Sec.~\ref{sec:screening} details the screening mechanism. Then, in Sec.~\ref{sec:sources} we characterize the general properties of emission sources. In Sec.~\ref{sec:propagation}, we examine how screening influences the propagation of ULB fields through interstellar and intergalactic media. The detection prospects for ULBs are discussed in Sec.~\ref{sec:detection}. We present our findings in Sec.~\ref{sec:results} and conclude with an outlook on multimessenger astronomy beyond the SM in Sec.~\ref{sec:conclusion}. Throughout the paper, we adopt natural units, setting $c = \hbar = 1$.

\begin{figure*}
    \centering
    \includegraphics[width = \textwidth]{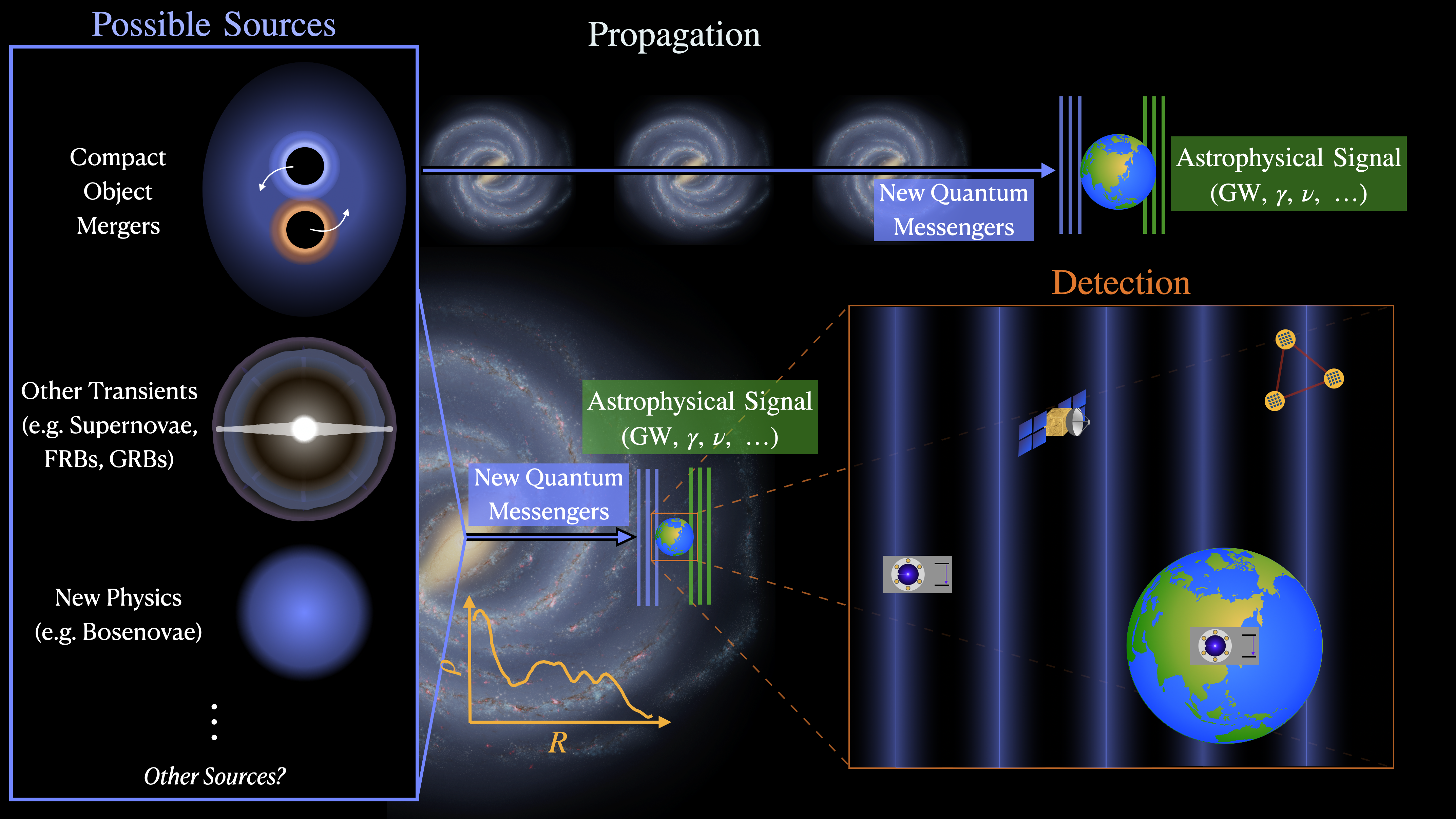}
    \caption{Illustration of multimessenger astronomy potential through synergy of astrophysical electromagnetic (photon, $\gamma$), gravitational wave (GW) and neutrino ($\nu$) messengers together with ULB fields. Here, dots represent potential other astrophysical messengers, such as charged particles.
    Possible sources of standard astrophysical signals accompanied by relativistic bursts of ULB fields
    include transient astrophysical systems such as compact object (black hole, neutron star) mergers or supernovae as well as those originating from new physics beyond SM, such as boson star bosenovae. Two characteristic scenarios of messenger propagation towards detecting experiments are depicted: when sources are extragalactic (top) and fields could traverse ISM of multiple galaxies, as well as when source emission is within the Milky Way Galaxy (bottom) and fields propagate through the Galaxy's ISM whose gas number density varying with galacto-centric distance $R$. Detection of ULB fields is illustrated involving terrestrial Earth-based or space-based quantum sensor experiments such as those based on precision clocks or laser/atom interferometers. ULB field signals from new physics (periwinkle shading), as well as astrophysical messenger signals (green shading) are shown. Milky Way image adopted from NASA~\cite{Hurt_2017}.}
    \label{fig:introdiagram} 
\end{figure*}

\section{Effective Interactions}
\label{sec:EFT}

To analyze in generality the prospects of multi-messenger astronomy 
for scalar and pseudoscalar quantum fields\footnote{We label both scalars and pseudoscalars as $\phi$ and distinguish them as necessary.} $\phi$ with ULB masses $m_{\phi} \ll 1~{\rm eV}$, we consider effective interactions between $\phi$ and the SM fields as characterized by EFT Lagrangian
\begin{align} \label{eq:Lphi}
    \mathcal{L} &= \mathcal{L}_{\rm int} + \frac{1}{2}\partial_{\mu}\phi \,\partial^{\mu} \phi - V(\phi)~, 
\end{align}
where $V(\phi)$ is the potential and $\mathcal{L}_{\rm int}$ accounts for interactions between $\phi$ and the SM constituents. 
Our approach captures an effective description of the relevant physics below certain energies and length scales. This allows for significant model-independence by accounting for a multitude of possible high-energy (ultraviolet, UV) theories that can lead to such interactions.

\subsection{Scalars}

We first consider interactions of a spin-0 parity-even scalar coupled to operators denoted by $i$ consisting of SM fields $\mathcal{O}^i_{\text{SM}}$
\begin{align}
    \mathcal{L}^{(s)}_{\text{int}} = \sum_{n,i} d^{(n)}_i \bigg(\frac{\sqrt{4\pi}\,\phi}{\Mpl}\bigg)^n \mathcal{O}^i_{\text{SM}}\,,
\end{align}
where $d^{(n)}_i$ are dilatonic couplings of $\phi^n$ with $n$ a positive integer and $M_{\rm pl} = 1.2 \times 10^{19}$ GeV is the Planck mass\footnote{The dilatonic couplings can be larger than unity. This is well-defined since the perturbative expansion is in powers of $d_i^{(n)} \phi^n/\Mpl^n$ and remains small for the parameters of interest even for $d_i^{(n)}\gg 1$.}. Considering $n = 1$ and $n = 2$ terms and SM operators $\mathcal{O}^i_{\text{SM}}$ that lead to variation of fundamental constants
\begin{align} 
\label{eq:scalar_Lint}
    \mathcal{L}^{(s)}_{\text{int}} =&~  
    \frac{\sqrt{4\pi}\,\phi}{\Mpl}
    \bigg( -d_{m_e}^{(1)}m_e \overline{e} e 
        + \frac{d_{e}^{(1)}}{4} F_{\mu \nu}F^{\mu \nu}\nonumber \\
        &+ \frac{d_g^{(1)} \beta(g_s)}{2g_s}G_{\mu\nu}^a G^{a\mu\nu}\bigg)\nonumber\\
    & +\frac{4\pi\,\phi^2}{\Mpl^2}
    \bigg(-d_{m_e}^{(2)}m_e \overline{e} e 
        + \frac{d_{e}^{(2)}}{4} F_{\mu \nu}F^{\mu \nu}\nonumber\\
        &+ \frac{d_g^{(2)} \beta(g_s)}{2g_s}G_{\mu\nu}^a G^{a\mu\nu}\bigg)\,,
\end{align}
where $m_e$ is the electron mass, $e$ is the electron field, $F_{\mu \nu}$ is the electromagnetic field strength, $G_{\mu \nu}^a$ is the gluon field strength with color index $a$, $g_s$ is the strong coupling constant and $\beta(g_s)$ is the QCD beta-function encoding behavior of $g_s$ with energy scale. The $d^{(n)}_{m_e}$, $d^{(n)}_{e}$, and $d^{(n)}_{g}$ couplings lead specifically to variations of the electron-to-proton mass ratio $m_e/m_p$, fine structure constant $\alpha$, and the ratio of the quark mass to QCD-scale $m_q/\Lambda_{\rm QCD}$, respectively. 

We consider $d_i^{(n)}$ couplings as free parameters to remain model-independent as possible, although they can be determined for a given UV theory. The couplings of different orders $n$ are not completely independent. For example, each of the linear $n = 1$ operators can induce effective quadratic $n = 2$ operators. This can be accounted for by higher powers in perturbation theory through loop processes. We note that symmetries can forbid different $n$-term contributions\footnote{For instance, a $\mathbb{Z}_2$ symmetry would prevent $n = 1$ interactions, leaving the quadratic $n=2$ interactions as the leading-order EFT contributions.}. However, this depends on the considered theory.

\begin{figure}
    \centering
\includegraphics[width=0.7\linewidth]{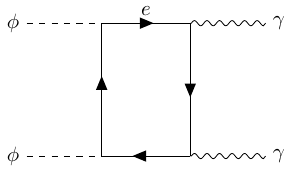}
    \caption{Induced effective  quadratic ULB field coupling to photons arising from loop-level contributions of linear ULB coupling to electrons.}
    \label{fig:quadloop}
\end{figure}

As shown in Fig.~\ref{fig:quadloop}, linear dilatonic couplings can give rise to effective quadratic interactions, even when the Lagrangian parameters satisfy $d_i^{(2)} = 0$. For instance, a linear coupling to electrons can induce a quadratic coupling to photons. It is therefore essential to evaluate the significance of these corrections for the parameter space under consideration. At considered ULB field energies $\omega \ll m_e$, the one-loop corrections to effective quadratic coupling exhibit scaling behavior   
\begin{align} \label{eq:inducscal}
    d_{e,\rm eff}^{(2)}(\omega) \sim d_e^{(2)} + \frac{e^2 (d_{m_e}^{(1)})^2}{4 \pi^2} \frac{\omega^2}{m_e^2}~,
\end{align}
where $\omega$ is the $\phi$ energy and $e$ is the fundamental electric charge, and $\sim$ represents approximate scaling. At higher energies when $\omega > m_e$, the electron field dynamics can no longer be ignored. Then, as the contributions from the electron loop are no longer suppressed and can have an order-one  effect the scaling behavior of the one-loop corrections in Eq.~\eqref{eq:inducscal}. Since we focus on $\omega \lesssim 1$~eV, Eq.~\eqref{eq:inducscal} remains valid for our considerations. 

At the energies relevant for ULB fields that we consider, existing constraints on linear dilatonic couplings from equivalence principle (EP) tests limit $d_{m_e}^{(1)} < 1$ and can even reach $d_{m_e}^{(1)} < 10^{-4}$ \cite{Berge:2017ovy,Hees:2018fpg}. The resulting corrections to quadratic coupling are at most of the order $\mathcal{O}(10^{-20})$ and hence are negligible compared to the bare $d_e^{(2)}$ values that can be explored, as we discuss in Sec.~\ref{sec:results}. Consequently, we can disregard these induced quadratic coupling contributions.

\subsection{ALPs}

ULB pseudoscalar fields are parity-odd and consequently give rise to a distinct set of interaction operators. For ULB pseudoscalars, it is natural to attribute their small masses to $\phi$ being a pNGB arising from a broken continuous symmetry. Although this framework is not completely general, we focus on couplings typically allowed for pNGBs as a representative case. Throughout this discussion, we refer to these fields\footnote{QCD axions arise as pNGBs of $U(1)_{\rm PQ}$ Peccei-Quinn symmetry.} as ALPs.
The relevant linear ALP terms\footnote{In literature the ALP-photon coupling is sometimes defined with an extra $1/4$ factor, and the ALP-fermion coupling with an extra $1/2m_e$ factor where $m_e$ is electron mass. }
read
\begin{align} 
    \label{eq:ALP_Lint}
    \mathcal{L}_{\rm int}^{(\rm ps)} = &~ \phi (g_{\phi \gamma\gamma} F_{\mu\nu} \tilde{F}^{\mu \nu} 
    + g_{\phi gg} G_{\mu\nu}^a \tilde{G}^{a\mu \nu})\nonumber\\ &+ g_{\phi f} (\partial_{\mu}\phi) \overline{f}\gamma^{\mu}\gamma^5 f~,
\end{align}
where $f$ denotes the SM fermion fields and in particular electrons and quarks, $\tilde{F}^{\mu\nu} = \frac{1}{2}\epsilon^{\mu\nu\rho\sigma}F_{\rho\sigma}$ and $\tilde{G}^{a\mu\nu} = \frac{1}{2}\epsilon^{\mu\nu\rho\sigma}G_{\rho\sigma}^a$ are the electromagnetic and gluon dual field strengths. 

The quark and gluon operators can be systematically mapped onto experimentally-relevant nucleon-level operators, such as by using chiral perturbation theory (ChPT)~\cite{Scherer:2002tk}. ChPT is an EFT framework for constructing meson and baryon operators—representing the physically-relevant degrees of freedom after QCD confinement—from the underlying quark and gluon operators.
In particular, the interaction between the ALP field 
$\phi$ and quarks can induce an ALP-axial nuclear current coupling. Additionally, the coupling $\phi$ to gluons can generate an electric dipole moment (EDM) for nucleons, resulting in another ALP-nucleon interaction term involving the electromagnetic field strength tensor $F_{\mu\nu}$. Both effects provide experimentally testable signatures as we discuss in Sec.~\ref{sec:detection}. The interactions can be written as \cite{graham2013new}
\begin{align}
    \label{eq:ALP_nucl}
    \mathcal{L}_{\rm nuc}^{(\rm ps)} =&~  g_{\phi \mathcal{N}\mathcal{N}} (\partial_\mu \phi) \bar{\mathcal{N}} \gamma^\mu \gamma^5 \mathcal{N} \nonumber\\
    &~-  \frac{i}{2} g_{\phi \mathcal{N}\gamma} \phi \bar{\mathcal{N}} \sigma_{\mu\nu} \gamma_5 \mathcal{N} F^{\mu\nu}~.
\end{align}
Here, $\mathcal{N}$ denotes nucleons (the proton field $P$ or the neutron field $N$), and $\gamma^\mu$ are the Dirac matrices\footnote{We also use $\sigma_{\mu\nu}=-\frac{i}{4}[\gamma_\mu,\gamma_\nu]$, and $\gamma^5 = i\gamma^0\gamma^1\gamma^2\gamma^3$.}. Note that $g_{\phi P\gamma} = - g_{\phi N\gamma}$ \cite{pospelov1999theta}.

Analogously to the discussion of scalars and Eq.~\eqref{eq:inducscal}, linear ALP interactions with the SM can lead to additional induced terms that are quadratic in $\phi$~\cite{Beadle:2023flm,Kim:2023pvt}
\begin{align}
\label{eq:ALPquad}
    \mathcal{L}_{\rm quad}^{\rm (ps)} = -c_{\gamma}\frac{\alpha}{16 \pi^2}\frac{\phi^2}{f_a^2} F^{\mu \nu}F_{\mu \nu} -c_e\,m_e\frac{\phi^2}{f_a^2} \bar{e}e~,
\end{align}
where $c_{\gamma}, c_{e} \lesssim 1$ are Wilson coefficients encoding the high energy UV dynamics that generate these couplings. These coefficients arise from loop-level diagrams involving interactions that explicitly break the $U(1)$ shift symmetry. Notably, as we discuss further, such interactions can induce screening effects even when the original Lagrangian includes only linear couplings. For instance, in the case of the QCD axion, a quadratic coupling of this type is generated through interactions with charged pions, yielding  $c_{\gamma} \sim \mathcal{O}(10^{-1})$ and $c_e \sim \mathcal{O}(10^{-6})$~\cite{Beadle:2023flm,Kim:2023pvt}. Here, we focus on the phenomenological implications and do not further discuss specific UV completions.

\section{Screening and Anti-screening}
\label{sec:screening}

\begin{figure*}
    \centering
    \includegraphics[width = 0.8\linewidth]{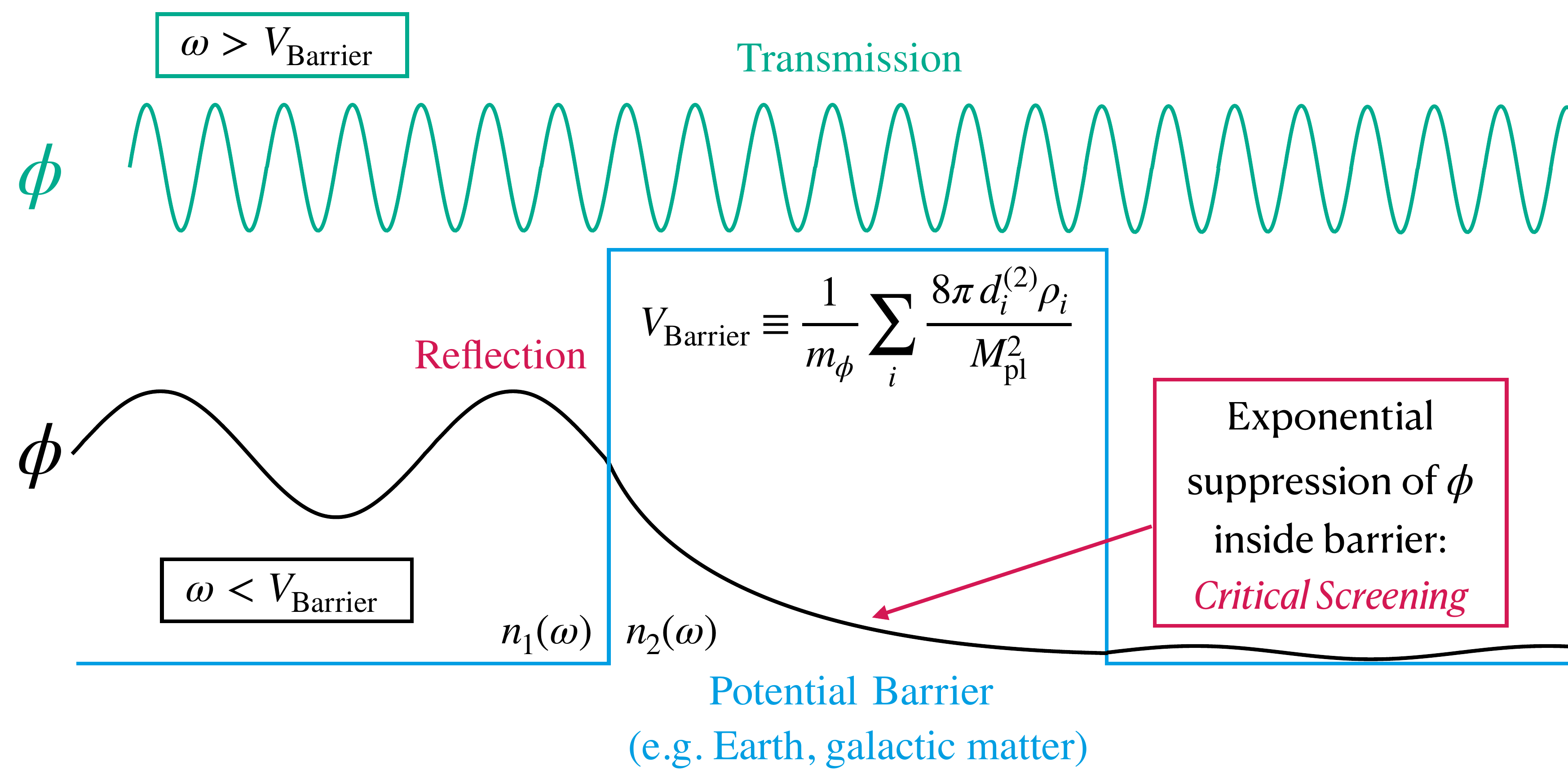}
     \caption{  Schematic representation of medium (e.g. Earth, galactic matter) screening ULB wave, analogous to a finite-width potential barrier in quantum mechanics. Due to the quadratic $\phi-$SM coupling, a dense matter medium like Earth behaves as a potential barrier with a height $V_{\rm barrier} =  \sum_i 8\pi\,d_i^{(2)} \rho_i/(m_{\phi}\Mpl^2)$. If the wave energy $\omega$ is less than $V_{\rm barrier}$, $\phi$ primarily reflects off the barrier, with an exponentially suppressed amplitude inside. The indices of refraction, $n_1(\omega)$ outside and $n_2(\omega)$ inside the barrier, depend on $\phi$'s energy. For $\omega > V_{\rm barrier}$, the wave transmits through the barrier with reduced kinetic energy due to the increased effective mass, resulting in slower propagation.}
    \label{fig:barrier}
\end{figure*}

In the presence of quadratic couplings between either scalar or ALP ULB fields and the SM constituents, the 
$\phi$ fields experience screening effects driven by an environment-dependent effective mass~\cite{Hees:2018fpg}. We consider \emph{screening} to constitute a scenario in which the local SM matter density increases the effective mass of 
$\phi$, and \emph{anti-screening} as the scenario where the matter density reduces it. In the case of screening, we further consider \emph{critical screening} as the regime where the effective mass becomes sufficiently large in a region of space to suppress the field's free propagation, leading to exponential attenuation. 

The screening and anti-screening effects carry significant implications for multimessenger astronomy beyond the SM, as discussed further: (A) the propagation of 
$\phi$, particularly whether the 
$\phi$ signal can arrive with a sufficiently small time delay relative to signals of astrophysical messengers to allow correlation between them; and (B) the prospects for detecting the 
$\phi$ signal itself.
More specifically, the presence of SM matter alters the effective mass of the 
$\phi$ fields, thereby modifying their group velocity $v_g$ and affecting the arrival time of their signals relative to astrophysical messengers. Moreover, in regions with sufficiently high SM matter densities, the effective mass can exceed the total energy of the 
$\phi$ field, effectively preventing its propagation with exponential suppression in those regions. Consequently, it is crucial to appropriately characterize these effects and underlying phenomena.

To examine these effects, we consider a concrete potential $V(\phi)$ corresponding to minimal realization that stabilizes the vacuum in the case of negative dilatonic coupling
\begin{align} \label{eq:phipotscr}
    V(\phi) = \frac{1}{2} m_{\phi}^2 \phi^2 + \frac{\lambda}{4!}\phi^4~,
\end{align}
where $\lambda$ is a coupling constant. Similar effects could also appear in the context of various other potentials. We discuss more general potentials in App.~\ref{app:potential}.

Combining the potential of Eq.~\eqref{eq:phipotscr} with quadratic $\phi$ and SM field interactions described by Eqs.~\eqref{eq:scalar_Lint} and \eqref{eq:ALPquad}, the $\phi$ Lagrangian can be written as
\begin{align} \label{eq:lag1}
    \mathcal{L} &\supset  \frac{1}{2}\partial_{\mu}\phi \partial^{\mu} \phi - \frac{1}{2}(m_{\phi}^2+\beta) \phi^2 - \frac{\lambda}{4!}\phi^4
\end{align}
where we do not include the linear terms from Eq.~\eqref{eq:scalar_Lint} for the discussion of screening. Here,
\begin{align}
    \beta = \begin{cases}
        \pm \sum_i \dfrac{8\pi\,d_i^{(2)} \rho_i}{\Mpl^2}, & ~{\rm scalar}\\
        c_{\gamma} \dfrac{\alpha}{2 \pi^2 f_a^2}\rho_{\gamma} +  c_e \dfrac{1}{f_a^2} \rho_e, & ~{\rm ALP}
    \end{cases} 
\end{align}
translates the $\phi$-SM quadratic couplings into a modification to the $\phi$ mass. This encodes the effect of the environment of the SM fields in the vicinity of $\phi$ via the SM matter densities defined by  
\begin{align} \label{eq:expenv}
    \rho_i &= \langle \rho | \mathcal{O}_{\rm SM}^i|\rho \rangle \nn\\
    &= \begin{cases} 
    \langle \rho| ( m_e \bar{e}e )|\rho\rangle, \\
    \langle \rho | (-\dfrac{1}{4}F_{\mu \nu}F^{\mu \nu} ) |\rho\rangle, \\
    \langle \rho | (-\dfrac{\beta(g_s)}{2g_s}G^a_{\mu \nu}G^{a\mu \nu} )|\rho\rangle,  
\end{cases} 
\end{align} 
where $|\rho\rangle$ characterizes the medium where the expectation values are evaluated. 

The expectation values in Eq.~\eqref{eq:expenv} represent the energy densities of field $i$ within a medium, which constitutes a fraction $f_i < 1$ of the total energy density, given by $\rho_i = f_i \rho$. In astrophysical media, the majority of the total energy density originates from nucleons, with subdominant contributions from electrons and photons. Considering hydrogen, the energy fraction associated with nucleons and electrons can be understood by noting that the proton contributes most of the hydrogen mass, $m_p \simeq m_H$, while the electron mass is significantly smaller, $m_e \simeq 4.4 \times 10^{-4} m_H$. The electromagnetic field can be integrated over an atom, yielding the energy density of the Coulomb field, which is approximately $\rho_{\gamma} \simeq 6.3 \times 10^{-4} \rho_H$. The semi-empirical mass relationships~\cite{krane1991introductory} allow to derive the contributing energy fractions, which generally depend on the chemical composition of the medium.

The $\phi$ vacuum expectation value (VEV) $\langle\phi\rangle$ representing the background value of the field at the minimum of the potential when $\partial V/\partial \phi = 0$ is given by
\begin{equation}
    \langle\phi\rangle = 
    \begin{cases}
         0~, & m_{\phi}^2 + \beta > 0\\
         \sqrt{-\dfrac{6(m^2_{\phi} + \beta)}{\lambda}}~, & m_{\phi}^2 + \beta < 0~.
    \end{cases}
\end{equation}
Hence, matter density contributing to $\beta$ in different regions of space can disrupt spontaneous symmetry breaking and effectively restore the symmetry in regions with low density while enabling symmetry breaking in high-density regions.
Around potential minimum, the total effective mass of $\phi$ is given by  
\begin{align}
\label{eq:meff}
    m_{\rm eff}^2(\beta) =&~ \frac{d^2 V}{d\phi^2}\bigg|_{\phi = \langle \phi \rangle} \notag\\
    =&~ (m_{\phi}^2 + \beta) + \frac{\lambda}{2}\langle\phi\rangle^2 \notag\\
=&~ 
    \begin{cases}
         m_{\phi}^2 + \beta>0~, & m_{\phi}^2 + \beta > 0\\
         -2(m_{\phi}^2 + \beta)> 0~, & m_{\phi}^2 + \beta < 0~.
    \end{cases} 
\end{align}
Hence, environmental dependence of $\beta$ determined by the density of SM fields in the region of space that $\phi$ is propagating through directly affects its effective mass. A non-zero $\langle\phi\rangle$ contributes to the effective mass due to the self-interaction term, with background $\langle\phi\rangle$ acting as a medium with which the propagating $\phi$ wave interacts through the coupling $\lambda$.

From Eq.~\eqref{eq:lag1} the equations of motion are
\begin{align} \label{eq:eom}
\partial_{\mu}\partial^{\mu}\phi+(m_{\phi}^2 +\beta )\phi + \frac{1}{3!}\lambda \phi^3 = 0~.
\end{align}
Then, decomposing $\phi$ into Fourier modes characterized by the wave vector of magnitude $k = |\vec{k}|$ and frequency $\omega$, the associated dispersion relation considering $\phi$ around the minimum of its potential is   
\begin{equation} \label{eq:disp}
k^2 = \omega^2 - m_{\rm eff}^2~.  
\end{equation}

Screening phenomenon can then be intuitively described through the macroscopic refractive index 
$n$, drawing from principles of optics and light propagation in a medium. The index of refraction for $\phi$ propagation can be obtained as 

\begin{equation} \label{eq:indexn}
    n(\omega) = \frac{k}{\omega} = \sqrt{1 - \frac{m_{\rm eff}^2}{\omega^2}}~.
    \end{equation}
This indicates that when 
$\phi$ waves encounter regions with a varying index of refraction, they could undergo reflection or transmission at the interface, similar to a potential barrier in geometric optics or quantum mechanics. By analogy, we can quantify the critical screening, with transmission and reflection coefficients depending on the wave's momentum and the barrier's height and width.

The $\phi$ wave group velocity $v_g$ is given by
\begin{align}
\label{eq:vg}
    v_g &= \dfrac{\partial \omega}{\partial k} = \bigg(n + \omega \frac{dn}{d\omega}\bigg)^{-1}\nonumber\\
    &= \sqrt{1 - \frac{m_{\rm eff}^2}{\omega^2}}~,
\end{align}
which for dispersion relation of Eq.~\eqref{eq:disp} is seen to be equal to $n$ in Eq.~\eqref{eq:indexn}. Since $\omega$ is not a linear function of $k$, the wave envelope becomes distorted during propagation as different $k$ lead to distinct group velocities.

 In Fig.~\ref{fig:barrier} we display Earth's screening of a ULB wave, in analogy to quantum mechanical tunneling through a potential barrier.
For a system with index of refraction $n(\omega)$ and length scale $\mathcal{R}$, such as Earth's radius, the transmission coefficient follows approximately (see e.g.~\cite{Griffiths_Schroeter_2018})
\begin{align}
    \label{eq:transmission}
    T \simeq \dfrac{1}{1 + \dfrac{2m_{\phi}\beta^2}{\omega^4 n(\omega)^2}\sin^2{\bigg(2\,\omega \,n(\omega) \,\mathcal{R}\bigg)}}~.
\end{align} 
A more accurate treatment of wave reflection and transmission can include considerations such as relativistic scalar wave scattering by a dense approximately spherical body in case of stars or planets, like Earth~\cite{Stadnik:2021qyf}.   

Critical screening occurs when the transmission of the field becomes negligibly small. This happens when the refractive index $n(\omega)$ becomes imaginary, leading to an exponential suppression of the field. We consider this to occur when the condition 
\begin{equation} \label{eq:crit}
     2i\omega\, n(\beta_{\rm crit}, \omega)\, \mathcal{R} = 1
\end{equation} 
is satisfied. We define the couplings that induce such $\beta = \beta_{\rm crit}$ as \emph{critical couplings}, denoted by $d^{(2)}_{i,{\rm crit}}$ for scalars and $1/f^2_{a,{\rm crit}}$ for ALPs. In the case of the Earth, when the scalar coupling exceeds the critical value $d_{i}^{(2)}>d_{i,{\rm crit}\oplus}^{(2)}$, the field becomes exponentially suppressed within the Earth, as it is no longer classically allowed to propagate in this region. This behavior closely parallels quantum-mechanical tunneling through a potential barrier.

\begin{figure*}
    \centering
    \includegraphics[width = .7\linewidth]{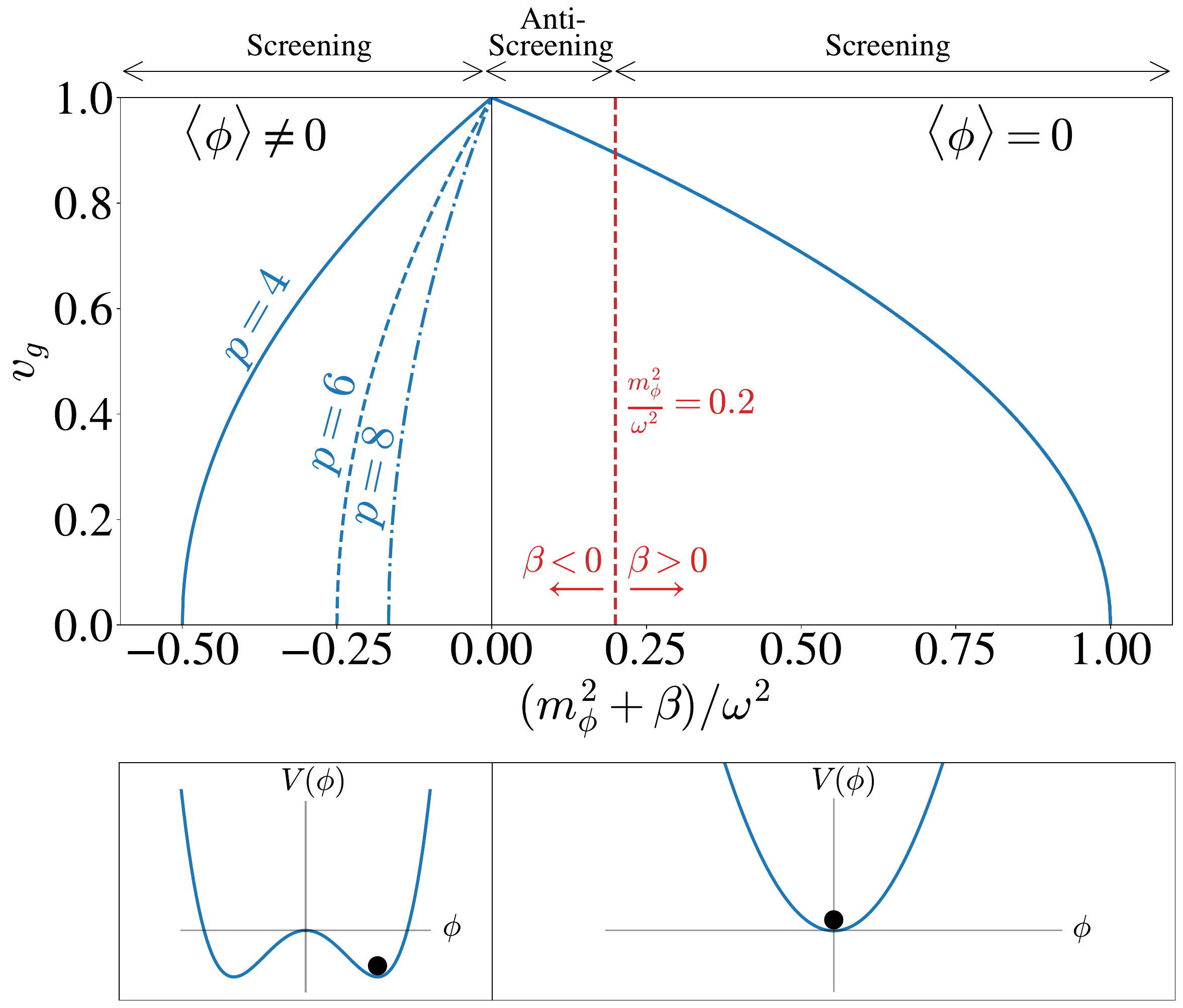}
    \caption{Group velocity of $\phi$ waves as a function of the ratio  $(m_{\phi}^2 +\beta)/\omega^2$ at $V(\phi = \langle \phi \rangle)$, for different choices of potential $p$ values for $\phi^p$ (see App. \ref{app:potential}). Reference bare $\phi$ field mass scenario without matter density contributions and $\beta = 0$ (red dashed line) delineates screening and anti-screening regimes. Schematic potential structure with respect to respective regions in the top figure is displayed on the bottom.}
    \label{fig:vg}
\end{figure*}

\subsection{Case with $\beta > 0$}

When $\beta > 0$, the presence of matter density composed of SM fields increases the effective mass of 
$\phi$, thereby slowing its propagation. This effect is reflected in changes to the index of refraction and a reduction in the group velocity of $\phi$, as described by Eq.~\eqref{eq:vg}, leading to
\begin{align}
    v_g = \sqrt{1 - \frac{m_{\phi}^2 + \beta}{\omega^2}}~,
\end{align}
where $\langle \phi \rangle = 0$ for $\beta>0$.

In the context of multimessenger astronomy, these screening effects can cause significant time delays between the standard astrophysical signals and those of ULB fields, hindering the ability to correlate them. As these signals propagate over vast distances, the group velocity $v_g$ of ULB fields must remain sufficiently close to unity to match the propagation speed in case of GW as well as electromagnetic signals, considering there is no significant medium dispersion. Neutrino signals can also be considered in this context, propagating at ultra-relativistic speeds near to that of light.
Furthermore, if a detector is located in an environment where the coupling constants reach the critical values  $d^{(2)}_{i,{\rm crit}}$ and $1/f^2_{a,{\rm crit}}$, its sensitivity will be exponentially reduced due to suppression of the 
$\phi$ amplitude. 

From Eq.~\eqref{eq:crit}, the critical couplings can be found as
\begin{align}
\label{eq:d_crit}
    d_{i,{\rm crit}}^{(2)} &= \Mpl^2 \dfrac{(2 \mathcal{R})^{-2} +\omega^2 - m_{\phi}^2}{8 \pi \rho_i} \\
    \dfrac{1}{f_{a,{\rm crit}}^2} &= \dfrac{(2\mathcal{R})^{-2} + \omega^2 - m_a^2}{\dfrac{c_{\gamma}\rho_{\gamma}\alpha}{2\pi^2} + c_e\rho_e}~. 
\end{align}
For the Earth,we approximate the density to be a constant $\rho_{E} = M_{E}/V_E \simeq 5.5~{\rm g/cm^3}$. However, we note that densities can be significantly higher in the Earth's core, and our analysis can be further refined by incorporating a full Earth structure model~\cite{Dziewonski:1981xy}.
We then obtain estimates for the critical coupling values considering energies of $\omega \lesssim 1/(2\mathcal{R}_{\oplus})^2$ as 
$d_{e,{\rm crit}}^{(2)\,\oplus} \simeq 3\times 10^{10}$, $d_{m_e,{\rm crit}}^{(2)\,\oplus} \simeq 2\times 10^{11}$, $d_{g,{\rm crit}}^{(2)\,\oplus} \simeq 6\times 10^{7}$ and $1/f_{a,{\rm crit}} \simeq 6 \times 10^{-14}$ GeV$^{-1}$, assuming conservative values of $c_{\gamma}, c_e = 1$.  Notably, this value of $1/f_{a,{\rm crit}}$ is already excluded by constraints from observations of the Sun, white dwarfs, and neutron stars~\cite{balkin2024white,gomez2024constraining,DiLuzio:2021pxd}. 

Detection prospects for quadratic couplings sensitively depend on the detector's environment, with terrestrial-based detectors facing challenges due to screening effects especially if they are located underground. The ULB field amplitude, and consequently the signal, can be exponentially suppressed by the high density environment of Earth. Detectors near or on the surface of Earth are therefore better suited than those underground to detect these signals, avoiding the exponential suppression inside the Earth. However, for even larger dilatonic couplings, Earth's atmosphere could also be the source of screening. This motivates space-based detectors as promising targets for multimessenger ULB signals.

\subsection{Case with $\beta < 0$}

Anti-screening effects can arise when the dilatonic coupling to SM fields is characterized by 
$\beta < 0$. Here, instead of the effective mass of 
$\phi$ strictly increasing in the presence of SM matter, 
$\beta < 0$ can lead to a decrease in the effective mass. This appears to suggest that additional time delays and critical screening might be alleviated. However, this is not generally true. In fact, 
$\beta < 0$ introduces rich phenomenological consequences, including the dependence of the 
$\phi$ vacuum structure on the coupling and the possibility of 
$\beta < 0$ exhibiting both ordinary screening and anti-screening, depending on the environment.

In Fig.~\ref{fig:vg} we illustrate different screening regimes considering the full range of behavior of the $\phi$ wave  group velocity, as a function of $(m_{\phi}^2 + \beta)/\omega^2$ for both positive and negative $\beta$. Unlike the case with $\beta > 0$, the VEV of $\phi$ for $\beta < 0$ depends on the environment. When $\beta < 0$, the overall sign of the $\phi^2$ term in the potential can become negative, leading to a tachyonic instability. This occurs when $m_{\phi}^2 + \beta < 0$ at the origin, making $\phi = 0$ a local maximum. In this scenario, the field $\phi$ rolls down the potential to a new vacuum state, acquiring a non-zero VEV, as illustrated at the bottom of Fig.~\ref{fig:vg}. 

In case only the leading $\mathcal{O}(\phi^2)$ term is considered, the potential lacks a stable vacuum state because it has no minimum. Incorporating $\phi^4$ or higher-order terms alleviates this instability, as they stabilize the potential by introducing a minimum. The specific form of these higher-order terms in $V(\phi)$ depends on the UV dynamics of the theory. We focus on simple scenario for illustration and discuss additional potential forms in App.~\ref{app:potential}.

 From Eq.~\eqref{eq:meff}, the effective mass is seen to always be positive in the minimum energy states.   As $|\beta|$ increases, 
for $m_{\phi}^2 + \beta < 0$ effective mass $m_{\rm eff}$ also increases and exhibiting screening analogously to  $\beta>0$. For $m_{\phi}^2 + \beta<0$ the index of refraction  is 
\begin{align}
     n(\omega) 
     &=  \sqrt{1 + \frac{2(m_{\phi}^2 + \beta)}{\omega^2}}~.
\end{align} 
Hence, when $2(m_{\phi}^2 + \beta)/\omega^2 < -1$, the index of refraction becomes imaginary and the field becomes critically screened, similarly to the case of $\beta>0$. This corresponds to critical couplings of  
\begin{align}
\label{eq:d_crit_negative}
    |d_{i,\rm crit}^{(2)}| &= \Mpl^2\frac{\mathcal{R}^{-2}+\omega^2  + 2m_{\phi}^2}{16\pi\, \rho_i}\\
    \dfrac{1}{f_{a,{\rm crit}}^2} &= \dfrac{\mathcal{R}^{-2} + \omega^2 + 2m_a^2}{2\bigg|\dfrac{c_{\gamma}\rho_{\gamma}\alpha}{2\pi^2} + c_e\rho_e\bigg|}~.
\end{align}
Note that the condition for spontaneous symmetry breaking is
\begin{align}
    d_i^{(2)} &> \Mpl^2 \frac{m_{\phi}^2}{8\pi \rho_i}\\
    \dfrac{1}{f_{a,{\rm crit}}^2} &> \dfrac{m_a^2}{\bigg|\dfrac{c_{\gamma}\rho_{\gamma}\alpha}{2\pi^2} + c_e\rho_e\bigg|}~.
\end{align}

In the ultra-relativistic limit, $\omega^2 \gg m_{\phi}^2$, the critical couplings for positive (Eq.~\ref{eq:d_crit}) and negative (Eq.~\ref{eq:d_crit_negative}) $\beta$ differ only by a factor of two. This is displayed in Fig~\ref{fig:vg}, which also includes a schematic representation of the vacuum structure in different regions of parameter space. Although for $\beta<0$ anti-screening can alleviate the ULB field propagation time delays, this requires $\beta$ to lie within parameter space range of
\begin{equation}
~~~~~~~~~~~~~ -m^2_{\phi} < \beta < 0  ~~~~~ {\rm (anti-screening)}~.
\end{equation}
This regime indicates partial cancellation between $m_{\phi}^2$ and $\beta$ contributions.

From Fig.~\ref{fig:vg} we see that taking the structure of potential $V(\phi)$ into account, $v_g$ is always less than unity unless $(m_{\phi}^2 + \beta)/\omega^2 = 0$.  On the other hand, approximations employed in Ref.~\cite{Derevianko:2021wgw} fail when $m_{\phi}^2 + \beta < 0$, suggesting $v_g>1$. Instead, as we discuss, the tachyonic behavior of the potential induces a $\phi$ VEV, similarly to the Higgs mechanism. Our analysis demonstrates an important point that antiscreening is not a universally expected behavior for negative quadratic dilatonic couplings. Instead, antiscreening applies only within a specific parameter window, beyond which the behavior transitions into screening. This is typical behavior as can be seen from examination of different potentials $\propto \phi^p$ that change the behavior of $v_g$ depicted in Fig.~\ref{fig:vg} considering different $p$ values.

\section{Source Emission}
\label{sec:sources}

We now examine the implications of the previous sections in the context of multimessenger astronomy beyond SM. Proposed multimessenger sources of ULB bursts include black hole mergers~\cite{Dailey:2020sxa}, inspired by the destabilization of superradiant clouds around black holes, which was subsequently explored in greater detail in Refs.~\cite{Aurrekoetxea:2024cqd,Takahashi:2024fyq}. Another prominent mechanism for generating relativistic ULB bursts with potential coincidence signatures is the explosion of boson stars (bosenovae)~\cite{Eby:2021ece,Arakawa:2023gyq,Arakawa:2024lqr}. A wide variety of potential sources and spectra exist, each warranting dedicated investigation.

Rather than focusing on specific potential ULB field emission sources, we aim to determine in generality whether multi-messenger astronomy with bursts of ULB fields is feasible for any source or mechanism. To address this broadly, we adopt a bottom-up approach, characterizing potential sources using a set of phenomenological parameters that capture their essential features. We consider a minimal set of parameters necessary to describe the general behavior of ULB field emission signals. These parameters include the total emitted energy of the source into ULB fields $E_{\rm tot}$, the average energy of the $\phi$ fields $\omega$, the spread in the energy spectrum  $\delta \omega$, the duration of the emission process at the source  $t_*$ as well as the distance from the source to the target detector $R$. Using this framework, any specific source can be mapped onto our results.

The emission characteristics of the source determine the parameters governing the $\phi$ field at the detector location, assumed to be on Earth unless stated otherwise. 
 Considering the harmonic oscillator representation $\phi = \phi_{\rm peak} \cos (\omega t)$, and assuming the energy density is dominated by the kinetic term  $\rho_{\phi} \simeq \frac{1}{2} \dot{\phi}^2$  in the regime $\omega^2 \gg m_{\phi}^2$, where $m_{\phi}$ is the mass of the field and the dot represents a time derivative, the peak scalar field amplitude is approximately 
\begin{equation}
\phi_{\rm peak} \simeq \frac{\sqrt{2 \rho_{*}}}{\omega}~,
\end{equation}
where $\rho_{\phi} = \rho_{*}$ is the energy density at the detector location. 
Thus, the key parameters influencing the detectability of the $\phi$ field are $\omega$ and $\rho_{\ast}$, with the latter given by 
\begin{align}
\label{eq:density}
    \rho_{*} \simeq \frac{E_{\rm tot}}{4\pi R^2 \tilde{x}_*}~,
\end{align}
where $\tilde{x}_*$ represents the thickness of the spherical wave shell of $\phi$ at the detector site. 

The value of $\tilde{x}_*$ depends on the intrinsic burst duration at the source and the extent to which the wave packet spreads during its propagation from the source to Earth. This is given by
\begin{align}
    \tilde{x}_* &\simeq x_* + 
    \delta x_{\rm spread}
\end{align}
where $x_*$ is the width of the spherical wave shell at the source, and $\delta x_{\rm spread}$ represents the additional elongation of the shell thickness due to wave spreading. The total shell thickness, $\tilde{x}_*$, also depends on the signal duration $\tilde{t}_*$, which must be sufficiently long for quantum sensors to achieve required sensitivity.

For the spread $\delta \omega$, we consider restriction from the Heisenberg uncertainty relation for the intrinsic characteristic burst duration $t_*$. Additionally, there is a restriction on the minimal energy $\omega$ that can be emitted. This can be understood in two ways.  First, one can consider the emission of wave packets with $\omega \sim \delta \omega$. Second, for a finite-duration process, the lowest energy signal Fourier mode above the zero-point contributions is given by $\omega = 2 \pi/t_*$. 
Thus, we impose
\begin{align}
\label{Eq:uncert}
\begin{cases}
      t_{*}\delta \omega  \gtrsim 1/2, &  \\
      t_{*} \omega  \gtrsim 2 \pi~. &  \\ 
    \end{cases} 
\end{align}

To determine the feasibility of multi-messenger astronomy under optimal conditions, we consider the best-case scenario. The momentum spread influences wave spreading—causing additional signal suppression proportional to $\sim 1/R$ in addition to the $\sim 1/R^2$ from isotropic signal emission. Hence, for our analysis we assume minimal uncertainty of $\delta \omega = 1/(2 t_*)$.

\section{Signal Propagation}
\label{sec:propagation}

In the context of multimessenger astronomy, the propagation of ULBs from the source to a detector, assumed to be located on Earth unless stated otherwise, is crucial for assessing their detectability in coincidence with standard astrophysical messengers such as GWs, gamma rays, or neutrinos. The (anti)screening phenomenon adds complexity to ULB propagation, involving inputs from microphysical Lagrangian parameters, Standard Model processes, and the properties of the source emission. In this work, we focus on general propagation features, using estimates to illustrate the effects of (anti)screening.

 ULBs can be typically expected to lag behind GW, gamma-ray and neutrino signals, and a long delay prevents signal correlation. However, in cases of anti-screening or when screening remains negligible, ULBs can outpace photons, which themselves may undergo interactions and dispersion. Additionally, wavepacket dispersion during propagation increases signal duration and coherence time. We also derive a relationship between time delay and signal duration.

\subsection{Coincidence Signal Time Delay}

The time delay $\delta t$ between the arrival of standard astrophysical messenger GW, gamma-ray or neutrino signals $t_{\rm astro}$ and $\phi$ signals $t_{\phi}$ can be estimated from the group velocity of the ULBs, $v_g$ and the speed of propagation of astrophysical messengers $v_{\rm astro}$. Note that $v_g$ depends on $\beta$ through effective mass $m_{\rm eff}$ in Eq.~\eqref{eq:vg}, which depends on the density of the SM fields that the $\phi$ field encounters along its path.
The density of SM fields is generally not homogeneous from the source to Earth and hence $\beta$ varies with position. 

The time delay of 
$\phi$ signals can be analyzed relative to standard astrophysical messengers. For GWs, carried by gravitons, the propagation speed is identical to the speed of light, carried by photons, in a vacuum. Multimessenger observations of neutron star mergers, which include both GW and electromagnetic wave counterparts, place stringent limits on any deviation in the propagation speeds of photons and gravitons~\cite{Sakstein:2017xjx}. Although neutrinos have a non-zero mass~\cite{Super-Kamiokande:1998kpq,SNO:2001kpb}, their masses are limited to be sub-eV~\cite{Katrin:2024tvg}. Considering the typical MeV-to-TeV energies of astrophysical neutrinos, we neglect the deviation of their speeds from the speed of light in a vacuum for our analysis.

Photons can slow down when propagating through astrophysical medium due to interactions with charged particles or its overall properties. This results in dispersion caused by interstellar medium. Such interactions cause the phase and group velocity of photons to deviate from the speed of light in a vacuum. In particular, the photon group velocity
\begin{equation}
    v_p = \sqrt{1 - \dfrac{\omega_p^2}{\omega^2}}~,
\end{equation}
where $\omega_p = \sqrt{4 \pi \alpha n_e / m_e} \simeq 1.3 \times 10^{-12}~(n_e/{\rm 1~cm}^3)^{1/2}$~eV is the plasma frequency with $\alpha \simeq 1/137$ being the fine structure constant, $n_e$ is the electron number density of medium and $m_e \simeq 0.511$~MeV is the electron mass. While for energetic photons with keV-MeV or higher 
energies such as those from supernovae or gamma-ray bursts these effects are not typically significant on astrophysical distances, they can become relevant for low-energy photons such as with radio GHz frequencies. In our analysis, we consider that for multimessenger sources of interest these effects for coincidence photons have negligible effect on their speed. 

The arrival time delay between scalar 
$\phi$ signals and signals from standard astrophysical messengers, assumed to travel at the speed of light in a vacuum with $v_{\rm astro}(x) \simeq 1$, can be generally expressed as
\begin{align}
\label{eq:TimeDelay}
    \delta t =&~ t_{\phi}- t_{\rm astro} \notag\\
    =&~ \int_{0}^{R}dx~(v_{\phi}^{-1}(x) - v_{\rm astro}^{-1}(x)) \notag\\
    \simeq&~ \int_{0}^{R}dx~ \bigg(\bigg(1- \dfrac{m_{\rm eff}^2(x)}{\omega^2}\bigg)^{-1/2} - 1 \bigg)\notag\\
    \simeq&~ \int_{0}^{R}dx ~ \bigg(\bigg(1- \dfrac{m^2_{\phi} + \beta(x)}{\omega^2}\bigg)^{-1/2} - 1\bigg),
\end{align} 
where in the last line we consider the $m_{\phi}^2 + \beta > 0$ case as representative, see Fig.~\ref{fig:vg}.  Even in the absence of screening, the $\phi$ signal can lag significantly behind astrophysical messengers depending on its mass $m_{\phi}$ and frequency $\omega$ that determine the $\phi$ group velocity.

For a homogeneous medium described by a constant $\beta$ we can evaluate Eq.~\eqref{eq:TimeDelay} to
\begin{align}
\label{eq:deltat}
    \delta t = R \bigg((1-q^{-2})^{-1/2} - 1\bigg)~.
\end{align}
where we have introduced $q = \omega/m_{\rm eff}$. Then, for a given $\delta t$, the minimum value of $q$ is
\begin{align}
    q = \bigg(1- \frac{1}{(1+\delta t/R)^2}\bigg)^{-1/2}~.
\end{align}
In the ultrarelativistic limit $q \gg 1$ this reduces to
\begin{align} \label{eq:rel}
    q \simeq \sqrt{\frac{R}{2\delta t}}~.
\end{align}
For instance, for ULBs originating from a distance of $R = 10\,{\rm kpc}$,  a coupling parameter $q \gtrsim 10^{3}$ is required to produce a time delay of \(\sim 1\) day. This motivates our assumption of the relativistic limit, $q \gg 1$. 

\subsection{Galactic and Extragalactic Signal Delays}

In the following analysis, we focus on studying a single dilatonic coupling at a time. 
For linear couplings, without screening effects at leading order,  we use the expression of Eq.~\eqref{eq:rel} considering $\beta = 0$ to calculate the time delays. However, for quadratic couplings, the time delay must be evaluated using the integral in Eq.~\eqref{eq:deltat}, with $\beta$ determined by the properties of the media through which $\phi$ propagates.  
We consider two reference benchmarks for the source locations: one within the Galactic Center (GC) of the Milky Way, and the other at an extragalactic (EG) distance of $10~{\rm Mpc}$ from Earth.

For the GC benchmark, the time delay depends on the gas density profile within the Milky Way Galaxy. The parameter $\beta(x)$ now varies with the position of the field, as the Galaxy exhibits an inhomogeneous density distribution. Given a known distribution for $\rho_{\rm SM}(x)$, this expression can be directly integrated for specific values of $m_{\phi}$, $\omega$, and $d_{i}^{(2)}$. We adopt the Galactic matter density distribution of neutral hydrogen H I and molecular hydrogen H$_2$ from Refs.~\cite{Nakanishi_2015,2016PASJ...68...63S}.

For the EG source, the emitted $\phi$ signal can traverse through multiple galaxies. Hence, determining the time delay requires knowledge of the distribution of galaxies as well as the distribution of SM matter within each galaxy. Distribution of gas and dust within interstellar medium (ISM) and intergalactic medium (IGM) is not uniform and with contributions from multiple constituents~\cite{Draine:2011book}.
Here, for simplicity, we focus on Milky Way-like galaxies, considering their number density of $n_{\rm MW} \sim \mathcal{O}(10^6)~{\rm Gpc}^{-3}$ based on the Schechter function~\cite{1976ApJ...203..297S}. We assume that these galaxies have similar homogeneous average ISM  particle densities of $\sim 1$~cm$^{-3}$, with $\rho_{\rm ISM} \simeq 1.67 \times 10^{-24}~{\rm g/cm^3}$ taking hydrogen. Additionally, we consider  IGM to have an average particle density of $\sim 1$~m$^{-3}$, with $\rho_{\rm IGM} \simeq 1.67 \times 10^{-30}~{\rm g/cm^3}$ taking hydrogen. We assume that the electron and nucleon number densities are approximately similar.

In the case of EG source as the ULB signal approaches each galaxy, the orientation of a galaxy becomes relevant, as it alters the propagation length through a galaxy. To account for this effect, we calculate the time delay for two extreme cases: when the ULB fields enter perfectly normal to the galactic planes and when they enter parallel to the galactic planes. The actual time delay lies between these two cases. 

While propagating between galaxies, the $\phi$ fields also traverse the intergalactic medium (IGM). We can thus split in this case the integral of Eq.~\eqref{eq:deltat} piecewise into contributions from the ISM and additionally the IGM. The total distance traveled through galaxies from the source to Earth is given by $N_G D_G$, where $N_G$ is the number of galaxies along the line of sight and $D_G$ is the distance $\phi$ travels through each galaxy. For a characteristic extragalactic scale of $10~{\rm Mpc}$, one approximately has $N_G \simeq 3.5$, including the Milky Way. The remaining distance is traveled through the IGM. Explicitly, this is 

\begin{align}
    \delta t 
     =&~ \int_{0}^{N_G D_G}dx~  \bigg(\bigg(1- \dfrac{m_{\phi}^2 + \dfrac{8\pi\,d_i^{(2)} \rho_{{\rm ISM},i}}{\Mpl^2}}{\omega^2}\bigg)^{-1/2} - 1\bigg) \nonumber \\
    &+\int_{N_G D_G}^{R}dx~ ~ \bigg(\bigg(1- \dfrac{m_{\phi}^2 + \dfrac{8\pi\,d_i^{(2)} \rho_{{\rm IGM},i}}{\Mpl^2}}{\omega^2}\bigg)^{-1/2} - 1\bigg) ~.
\end{align}

Approximating the ISM and IGM densities as constant for simplicity, the integrands are evaluated to be
\begin{align}
    \delta t =&~ N_G D_G \bigg(\bigg(1- \dfrac{m_{\phi}^2 + \dfrac{8\pi\,d_i^{(2)} \rho_{{\rm ISM},i}}{\Mpl^2}}{\omega^2}\bigg)^{-1/2} - 1\bigg)\nonumber\\ 
    &+ (R - N_G D_G)\bigg(\bigg(1- \dfrac{m_{\phi}^2 + \dfrac{8\pi\,d_i^{(2)} \rho_{{\rm IGM},i}}{\Mpl^2}}{\omega^2}\bigg)^{-1/2} \notag\\
    &- 1\bigg)~.
\end{align}
We can therefore determine which dilatonic couplings produce time delays short enough to be correlated with signals from standard astrophysical messengers.

\subsection{Wave Spreading}

Both the signal duration, $\tilde{t}_*$, and the coherence time, $\tau_*$, are initially determined by the burst properties but can be extended during propagation as the $\phi$ field travels to detector. The effective coherence time observed by the detector can be estimated as \cite{Eby:2021ece}
\begin{align}
    \tau_* \simeq \frac{2\pi}{\delta \omega} + \frac{2 \pi R}{q^3 m_{\phi} t_*}~,
\end{align}
and the signal duration is given by
\begin{align}
    \label{eq:tildet_*}
     \tilde{t}_* \simeq t_* + \frac{\delta \omega}{\omega} \int_0^R \frac{dx}{q^2}~.
\end{align}

Here, we further consider the relativistic limit, $q \gg 1$. In this regime, we can establish a direct relationship between the time delay, $\delta t$, and the signal duration, $\tilde{t}_{*}$. For $q \gg 1$, both $\delta t$ of Eq.~\eqref{eq:deltat} and $\tilde{t}_{*}$ of Eq.~\eqref{eq:tildet_*}  can be calculated using an average value of $q$, denoted as $\bar{q}$, which reduces the expressions to 
\begin{equation}
 \delta t  \simeq \frac{R}{2\bar{q}^2}   
\end{equation}
and
\begin{equation}
    \tilde{t}_{*}  \simeq t_* + \frac{\delta \omega}{\omega} \frac{R}{\bar{q}^2}~.
\end{equation}
This implies the relationship,

\begin{figure}[t]
    \centering
    \includegraphics[width=0.99\linewidth]{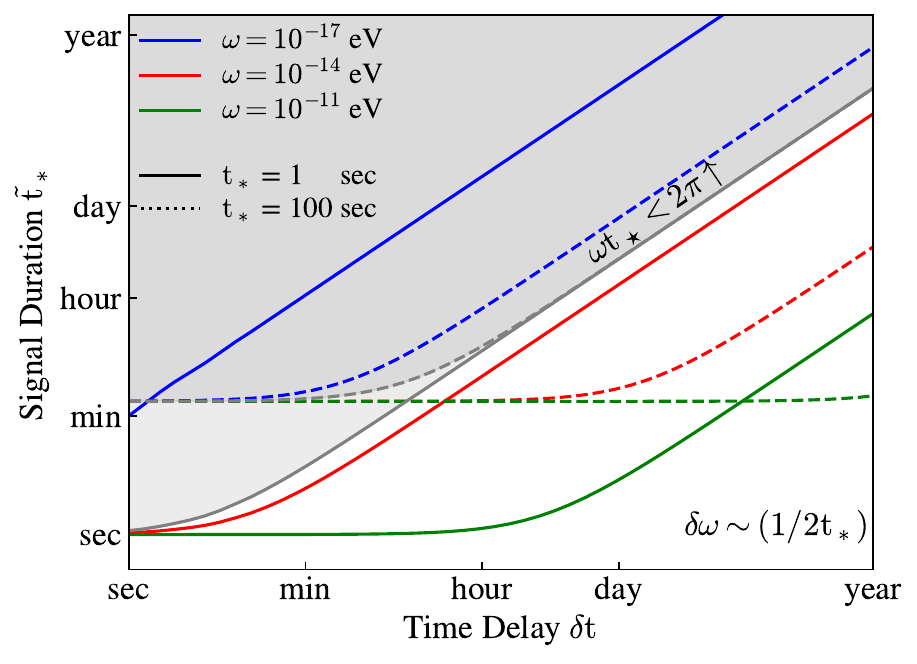}
    \caption{Duration of the signal at the detector, $\tilde{t}_*$, as a function of the time delay, $\delta t$. Different benchmarks for energy and intrinsic burst duration are shown, distinguished by colors and linestyles, respectively. The gray shaded region is excluded by the uncertainty principle relating the intrinsic burst duration and the energy of the emitted ULBs. All curves assume the minimal uncertainty case, $\delta \omega = 1/(2 t_*)$. For larger $\delta \omega$, the nonzero-slope portions of the curves would shift upward linearly.
}
    \label{fig:delay_vs_duration}
\end{figure}

\begin{align}
    \tilde{t}_{*} \simeq t_* + 2\frac{\delta \omega}{\omega} \delta t~.
\end{align}
Note that in the considered regime, the relationship is independent of the distance $R$.

In Fig.~\ref{fig:delay_vs_duration}, we show the signal duration at the detector, $\tilde{t}_{*}$, as a function of the time delay, $\delta t$, for energy and intrinsic burst time benchmarks of $\omega = {10^{-17},10^{-14},10^{-11}}$~eV and $t_* = \{1, 100\}$ s, respectively. 
As before, we assume the minimal uncertainty case, $\delta \omega = 1/(2 t_*)$. As illustrated, a larger $\delta \omega$ would result in increased wave spreading and, consequently, a longer signal duration for a given time delay. As a reference, for time delays that allow correlations with standard astrophysical messenger signals, we consider that the signal duration must satisfy $\tilde{t}_* \lesssim \mathrm{day}$.

We have worked under the assumptions of $q \gg 1$ and $\delta t / R \ll 1$. One might ask how close the source must be to Earth for these limits to no longer hold. For a time delay of $\delta t = 1~\rm{day}$, the source would need to be within approximately a parsec of Earth. Thus, for typical astrophysical distances, both limits are expected to be valid.

\section{Detection Prospects}
\label{sec:detection}
 
\begin{figure*}
    \centering
    \includegraphics[width=0.75\linewidth]{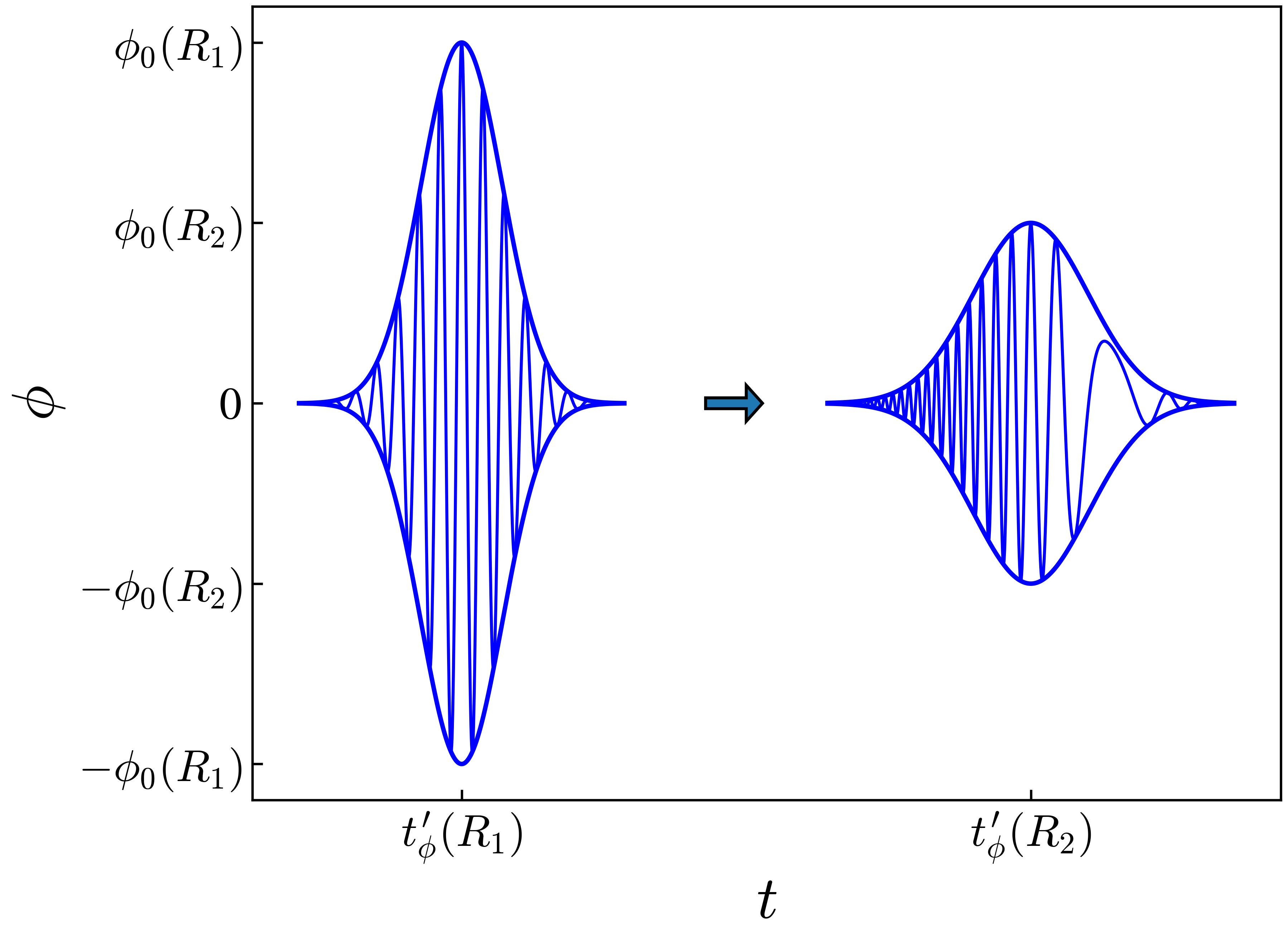} 
    \caption{ULB field Gaussian wavepacket described by Eq.~\eqref{eq:phi} for a burst at two different distances, $R_1$ and $R_2$, corresponding to elapsed times during propagation, $t_\phi^\prime(R_1)$ and $t_\phi^\prime(R_2)$, respectively. The envelope of the wavepacket is attenuated inversely with the distance traveled $R$. 
    At later times $t_\phi^\prime$, the spatial extent of the ULB burst $\tilde{x}_*=\tilde{t}_*$ increases and the oscillation frequency $\omega-(2\delta\omega/\tilde{t}_*)(t-t_\phi^\prime)$ exhibits chirping.
    }
    \label{fig:wavepacket}
\end{figure*}

Once the ULB signal reaches the detector, it can be measured using various quantum technologies, depending on the specific properties of the field and its couplings.
We treat the scalar field $\phi$ as a Gaussian wave packet, as described in \cite{Dailey:2020sxa} 
\begin{align}
\label{eq:phi}
    \phi(t) \simeq &~\frac{\sqrt{2\,\rho_{*}}}{\omega} \exp{\bigg(-\frac{(t-t_\phi)^2}{2 \tilde{t}_*^2}\bigg)}\nonumber\\
    &\times\cos\bigg(\omega (t-t_{\phi}) - \frac{\delta \omega}{\tilde{t}_*}(t-t_{\phi})^2\bigg)~,
\end{align}
assuming minimal uncertainty, $\delta \omega = 1/(2t_*)$.
This wave packet has a spatial extent $\tilde{x}_* = \tilde{t}_*$, an average energy $\omega$, and an energy spread of $\delta \omega$. Consequently, as the signal passes the detector, it exhibits a chirp with a time-dependent frequency.
The detection of a signal depends on the signal duration $\tilde{t}_*$, the scalar field density at the detector $\rho_*$, the detector's sensitivity as a function of frequency, the coupling type and strength, and the integration time. In Fig.~\ref{fig:wavepacket} we illustrate a wavepacket described by Eq.~\eqref{eq:phi} for two different distances from source, $R_1$ and $R_2$, corresponding to propagation times, $t_\phi^\prime(R_1)$ and $t_\phi^\prime(R_2)$, respectively. The attenuated amplitude, chirped frequency and enhanced spatial extent of the wave packet are demonstrated.

Scalar fields that couple to various SM fields (electrons, photons, and gluons) will induce variations in fundamental constants such as the electron mass, the fine structure constant, and the strong coupling constant \cite{Arvanitaki:2014faa}. 
Atomic, molecular, and optical (AMO) systems, which are particularly sensitive to variations in fundamental constants, are at the forefront of precision physics, making them ideal platforms for testing such variations \cite{Antypas:2022asj}. 
Different AMO systems are sensitive to different fundamental constants with differing sensitivities, thereby motivating a diverse experimental program incorporating various technologies to search for ultralight scalar dark matter. 
The same experiments that are used for the ambient dark matter  searches can also be sensitive to transient ultralight scalar bursts by being sensitive to temporal variations in fundamental constants.

ALPs that interact with SM fields linearly do not induce variations in fundamental constants due to ALP shift symmetry. 
Instead, ALPs can induce spin precession, photon conversion, and nucleon electric dipole moments (EDMs) \cite{adams2023axion}.
A variety of AMO experiments are currently being employed in order to search for ALP dark matter. 
As with scalars, these same experiments are automatically sensitive to transient ALP bursts.
Here, we do not go into detail of experimental intricacies and take a general approach by taking benchmark sensitivities, consistent with the best projections from planned or proposed experiments.

Here, we discuss representative benchmark scenarios to illustrate the potential sensitivity reach of proposed experiments. Given the diversity of detector types for the various couplings, instead of exhaustive characterization of all possible experimental setups we provide a broad overview of the parameter space that could be probed, leaving detailed, detector-specific analyses for future work. We assume that the integration livetime for experiments searching for the short duration ULB transient bursts is $t_{\rm int,*} = 1$ day throughout, although our analyses can be readily applied to other target values of $t_{\rm int,*}$.

\subsection{Scalars}

As noted above, scalar fields with non-derivative couplings to SM fields—specifically photons, electrons, and gluons—can induce variations in fundamental constants, namely the fine-structure constant $\alpha$, the electron mass $m_e$, and the strong coupling constant $m_q/\Lambda_{\rm QCD}$, respectively.
Recent advancements in quantum science and technology have enabled increasingly precise searches for such variations, particularly in the context of ambient ULB dark matter \cite{Arvanitaki:2014faa,Antypas:2022asj}. A variety of quantum sensors serve as highly-sensitive probes of variations in fundamental constants, including atomic, molecular, and nuclear (i.e. quantum) clocks \cite{Arvanitaki:2014faa}, optical cavities \cite{geraci2019searching}, spectroscopy \cite{VanTilburg:2015oza}, mechanical resonators \cite{Manley:2019vxy}, laser interferometers \cite{Stadnik1}, and atom interferometers \cite{geraci2016sensitivity}. These technologies can also be employed to detect bursts of scalar particles, provided the scalar field couples to SM fields in the same manner.

In order to elucidate the important results specific to the burst searches here, we use a single benchmark quantity $\eta_{\rm{DM}}$ to describe the sensitivity of detectors for a particular coupling. This parameter merges the measurement uncertainty with the detector sensitivity. 
We pick the $\eta_{\rm{DM}}$ for the three cases of interest, i.e. photon, electron, and gluon couplings, $d_e$, $d_{m_e}$, and $d_g$ based on literature values for the highest expected future detector sensitivities in these three cases. 

Quantum clocks and other spectroscopic investigations provide the best sensitivity to scalar couplings in the lowest mass range, as well as the most promising projected reach from proposed experiments. To determine the best prospects for multi-messenger astronomy with scalars, we use the smallest achievable fractional frequency uncertainty, $\eta_{\rm DM}$, from proposed clock and spectroscopy experiments as benchmark values. 
For clock experiments, this value is determined by the ratio of the fractional frequency uncertainty to a sensitivity factor, which depends on the clock frequency's dependence on the fundamental constant that varies due to the $\phi$ coupling. While here we use quantum clocks for reference benchmarks, similar sensitivities can also be achieved with atom interferometers~\cite{geraci2016sensitivity}.

For variations in the fine-structure constant $\alpha$ and the strong coupling constant $\alpha_S$, the best projected experimental reach is from a thorium isomer-based nuclear clock \cite{Arvanitaki:2014faa}, which has seen significant recent advancements \cite{tiedau2024laser,Elwell2024,zhang2024frequency}.
The fractional frequency uncertainty of the nuclear clock is projected to reach $10^{-19}$ at a few days as proposed in \cite{2012PhRvL.108l0802C}. We note that clocks must operate for a certain duration, $t_{\rm int}$, to reach their ultimate sensitivity $\eta_{\rm DM}$. We take $10^6$ seconds integration time as a benchmark in this case. 
The sensitivity of the nuclear clock to variations in $\alpha$ and $\alpha_S$ is further enhanced by factors of $5900$ \cite{beeks2024fine} and $10^5$ \cite{FlaTh06}, respectively. Consequently, we take benchmark sensitivities for probes of $d_e^{(n)}$ and $d_g^{(n)}$ to be $\eta_{\rm DM} = 10^{-19}/5900 \simeq 1.7 \times 10^{-23}$ and $\eta_{\rm DM}=10^{-19}/10^{5} = 10^{-24}$.

The best projection of experimental sensitivities to variations in the electron mass are due to molecular spectroscopy of SrOH, reaching $10^{-17}$ with one day of integration time \cite{SrOH2021} and we adopt $\eta_{\rm DM}=10^{-17}$  as benchmark sensitivity and one day integration time for probing $d_{m_e}^{(n)}$, where $n = 1, 2$. 
 
Although trapped ion nuclear clock promises the $10^{-19}$ uncertainty combined highest sensitivity to variations in $\alpha$, other clock designs, such as lattice clocks, may achieve their proposed operational sensitivities with much shorter integration times. Solid state nuclear clock which uses a macroscopic number of nuclei may enable a superior stability but more studies of the systematics are needed to determine ultimately achievable uncertainty \cite{higgins2024temperature}.
Depending on the signal duration, different systems may exhibit better sensitivity to bursts, even if their sensitivity to ambient DM is lower than that of nuclear clocks. Since both integration time of various detectors and signal duration can vary very significantly, we leave the detailed consideration and comparison of various detector capabilities in a future work. We considered the cases of the integration time of one day and one month and find only small changes in the plots described below.

In order to determine detector sensitivity to bursts, we re-scale the sensitivity of experiments designed to detect signals in the context of ambient DM described above, by adjusting to the duration of the burst which we take to be one day. 

For a given achievable sensitivity to background ultralight scalar DM, $\eta_{\rm DM}$, the rescaled sensitivity for a signal duration $\tilde{t}_*$ with an integration time of $t_{\rm int,*}$ is determined by~\cite{Eby:2021ece} 
\begin{align}
\label{eq:eta_*}
    \eta_* =  \eta_{\rm DM} \frac{t_{\rm int,\, DM}^{1/4}
        {\rm min}\left(\tau_{\rm DM}^{1/4},t_{\rm int,\, DM}^{1/4}\right)}
        {{\rm min}\left(\tilde{t}_*^{1/4},t_{\rm int, *}^{1/4}\right)
        {\rm min}\left(\tau_*^{1/4},t_{\rm int, *}^{1/4}\right)}~,
\end{align}
where $\tau_{\rm DM} = 2\pi/(m_{\rm DM}v^2)$ is the coherence time of non-relativistic, ambient DM with virial velocity $v_{\text{vir}} \simeq 10^{-3}$ \cite{Cooray:2002dia}, and $\tau_*$ is the effective coherence time of the relativistic $\phi$ fields that originate from an astrophysical burst event. In our analysis we compare the sensitivity of conventional DM searches with that of ULB transient searches at a common experimental frequency $f$, where the $\phi$-field energy is $\omega = 2\pi f$. As we scan over frequency we set corresponding non-relativistic DM mass $m_{\rm DM} = \omega$. Hence, $m_{\rm DM}$ is distinct from benchmark ULB mass $m_\phi$. Owing to this distinction, ULB transient searches can be intrinsically more sensitive to lower $m_\phi$ values within any fixed frequency band.

The minimal linear and quadratic couplings detectable for a scalar burst emission event are given by
\begin{align}
\label{eq:direach}
 \phi-{\rm int.}:~~~~~    &d_i^{(1)}  \simeq \frac{\eta_{*}\, \Mpl}{2\sqrt{\pi}\phi_{\rm peak}}, \\
\label{eq:direach2}
 \phi^2-{\rm int.}:~~~~~        &d_i^{(2)}  \simeq \frac{\pi \, \eta_*\, \Mpl^2}{\phi_{\rm peak}^2}~.
\end{align}
Using the expression for $\phi$ with $\rho_{*}$ from Eqs.~\eqref{eq:density}, identifying $\tilde{x}_* = v_g \tilde{t}_*$ with $v_g \simeq 1$, and substituting into Eq.~\eqref{eq:direach}, the sensitivities become  
\begin{align}
\label{eq:d_i(1)}
  \phi-{\rm int.}:~~~~~ &d_i^{(1)}\sim \omega \,\eta_{*}\, \Mpl \sqrt{\dfrac{ R^2 \left(t_* + \dfrac{\delta \omega}{\omega} \dfrac{R}{q^2}\right)}{2 E_{\rm tot}}}, \\
\label{eq:d_i(2)}
 \phi^2-{\rm int.}:~~~~~     &d_i^{(2)} \sim \omega^2 \pi \, \eta_*\, \Mpl^2 \dfrac{2\pi R^2 \left(t_* + \dfrac{\delta \omega}{\omega} \dfrac{R}{q^2}\right)}{E_{\rm tot}}.
\end{align}
Here, we approximate $q = \omega/m_{\rm eff} \simeq \omega/m_{\phi}$. Since $q$ appears in the expression for wave spreading, the relevant effective mass $m_{\rm eff}$ is determined during propagation. Given the low densities of SM matter in interstellar or intergalactic space, the effective mass during propagation is very close to the bare mass.

We use the benchmark dark matter sensitivities and integration times of given detector to determine sensitivities for scalar bursts via Eq.~\eqref{eq:eta_*}. The resulting values are then applied in Eqs.~\eqref{eq:d_i(1)} and \eqref{eq:d_i(2)} to calculate the dilatonic couplings that can be probed for the considered targets. 
We discuss the results in the Sec.~\ref{sec:results}.

\subsection{ALPs}

Unlike scalars, linear ALP interactions with SM fields do not alter fundamental constants. This is a consequence of ALP shift symmetry. ALPs exhibit a range of signatures depending on their couplings to SM constituents. While for scalar interactions, we adopted benchmark sensitivities based on experimental proposals such as thorium-based nuclear clocks for $d_e$ and $d_g$, and SrOH molecular spectroscopy for $d_{m_e}$, our approach for ALPs differs. Given the diverse range of experimental proposals for ALP detection within the relevant frequency range, including both broadband and narrowband searches, we employ a more generalized methodology.

The coupling strength to ambient DM, $g_{\rm DM}$, varies as a function of the signal frequency. Moreover, different experimental proposals assume distinct integration times, further complicating direct comparisons. To establish a benchmark experimental reach, we consider projected sensitivities to $g_{\rm DM}$ across various frequency ranges based on multiple proposed experiments. As a baseline, we assume a constant value for $g_{\rm DM}$ that can be probed for ambient DM over the entire relevant energy range, along with a reference integration time $t_{\rm int,DM}$. We then rescale these values to assess detection prospects for shorter ALP bursts, thereby enabling a broader multi-messenger analysis.

The derivative coupling of ALPs to electrons produces spin-dependent dipole-dipole interactions, which are detectable with torsion pendulums \cite{terrano2019constraints}, magnetometers \cite{chigusa2023light}, and comagnetometers \cite{bloch2020axion}. Future magnetometer experiments with NV centers are projected to probe $g_{\phi e}$ on the order of $2\times10^{-14}~\mathrm{GeV}^{-1}$ \cite{chigusa2023light}, while torsion pendulums may achieve similar sensitivity with an integration time of a year \cite{graham2018spin}. We use $g_{\rm DM}=2\times10^{-14}~\mathrm{GeV}^{-1}$ and $t_{\rm int,DM}=1~\rm year$ as benchmarks for the ALP-electron coupling $g_{\phi e}$.

The ALP-photon coupling $g_{\phi\gamma\gamma}$ induces ALP-photon conversion in magnetic fields \cite{sikivie1983experimental,primakoff1951photo}, which are potentially detectable in various haloscope experiments.
The best projected reach for $g_{\phi\gamma\gamma}$ from proposed experiments in the relevant energy range are from optical ring cavity DANCE \cite{DANCE}, superconducting radio frequency (SRF) cavity \cite{berlin2021heterodyne}, LC lumped element DM-Radio \cite{DMRadio:2022pkf}, as well as other more narrowband searches. 
We use the benchmark $g_{\rm DM}=3\times10^{-16}~\mathrm{GeV}^{-1}$, which is projected to be within the reach of each of these proposals in their respective frequency ranges with integration times of 1 year for DANCE and 5 years for both SRF cavity and DM-Radio. We take our benchmark integration time to be $t_{\rm int,DM}=1\rm~year$.

The ALP-nucleon interactions produce spin-dependent effects that are potentially detectable via various types of haloscope experiments. 
For proton coupling $g_{\phi PP}$, the best projected experimental reach are from proton storage rings \cite{graham2021storage} and NMR-based CASPEr Wind experiment \cite{jackson2020overview}. Together, these experiments are expected to probe a value of $g_{\phi PP} = 10^{-12}~\mathrm{GeV}^{-1}$ with integration times of 1000 seconds for proton storage rings and $10^6$ seconds for CASPEr Wind. We use $t_{\rm int, DM} = 1000~\rm s$ as our benchmark for integration time.
The ALP-neutron coupling $g_{\phi NN}$ is potentially detectable with comagnetometer experiments \cite{bloch2020axion}, electrostatic storage ring \cite{brandenstein2023towards} and CASPEr \cite{jackson2020overview}. We use a benchmark value of $g_{\rm DM} = 10^{-11}~\mathrm{GeV}^{-1}$, which is covered for the majority of the relevant energy range by the cited proposals with integration times of $10^6$ seconds for CASPEr Wind and 5 hours electrostatic storage ring. We take the benchmark $t_{\rm int,DM}=1$ day in our computations.

ALPs can also induce nucleon electric dipole moments (EDMs) via the coupling $g_{\phi P\gamma}$, leading to spin precession, potentially detectable with CASPEr-electric proposal \cite{jackson2020overview}. The projected reach of CASPEr-electric is frequency dependent, however we take the conservative benchmark value of $g_{\rm DM}=10^{-18}~\mathrm{GeV}^{-2}$ with an integration time of $t_{\mathrm{int,DM}}=10^6~\mathrm{s}$.

 The sensitivity of a relativistic ALP burst with non-derivative ALP couplings $g_*$ ($g_{\phi\gamma\gamma}$, $g_{\phi P\gamma}$) scales relative to the cold dark matter with ALP coupling $g_{\mathrm{DM}}$ as~\cite{Eby:2021ece} 
\begin{align}
\label{eq:g_ratio}
    g_* =&~ g_{\rm DM} \sqrt{\frac{\rho_{\rm DM}}{\rho_*}} \nonumber\\
        & \times \frac{t_{\rm int,\,DM}^{1/4}
        {\rm min}\left(\tau_{\rm DM}^{1/4},t_{\rm int,\, DM}^{1/4}\right)}
        {{\rm min}\left(\tilde{t}_*^{1/4},t_{\rm int, *}^{1/4}\right)
        {\rm min}\left(\tau_*^{1/4},t_{\rm int, *}^{1/4}\right)}~.
\end{align}
For derivative couplings ($g_{\phi e}$, $g_{\phi PP}$, $g_{\phi NN}$), an additional velocity ratio $v_{\mathrm{DM}}/v_* \sim 10^{-3}$ modifies the scaling 
\begin{align}
\label{eq:g_ratio_v}
    g_* =&~ g_{\rm DM} \sqrt{\frac{\rho_{\rm DM}}{\rho_*}} \frac{v_{\rm DM}}{ v_* } \nonumber \\
    & \times \frac{t_{\rm int,\,DM}^{1/4}
        {\rm min}\left(\tau_{\rm DM}^{1/4},t_{\rm int,\,DM}^{1/4}\right)}
        {{\rm min}\left(\tilde{t}_*^{1/4},t_{\rm int, *}^{1/4}\right)
        {\rm min}\left(\tau_*^{1/4},t_{\rm int, *}^{1/4}\right)}
\end{align}
For all couplings, we assume $t_{\mathrm{int,*}}=1~\mathrm{day}$ to optimize searches for bursts lasting a day or less, which is what is required for time delays $\delta t$ of less than a day, as per Fig.~\ref{fig:delay_vs_duration}. However, for different signal types and durations, one could choose a different $t_{\rm int,*}$, changing the experimental sensitivities according to Eqs.~(\ref{eq:eta_*}), (\ref{eq:g_ratio}), and (\ref{eq:g_ratio_v}).   

\section{Multimessenger Astronomy}
\label{sec:results}

\begin{figure*}
    \centering
    \includegraphics[width = 1\linewidth]{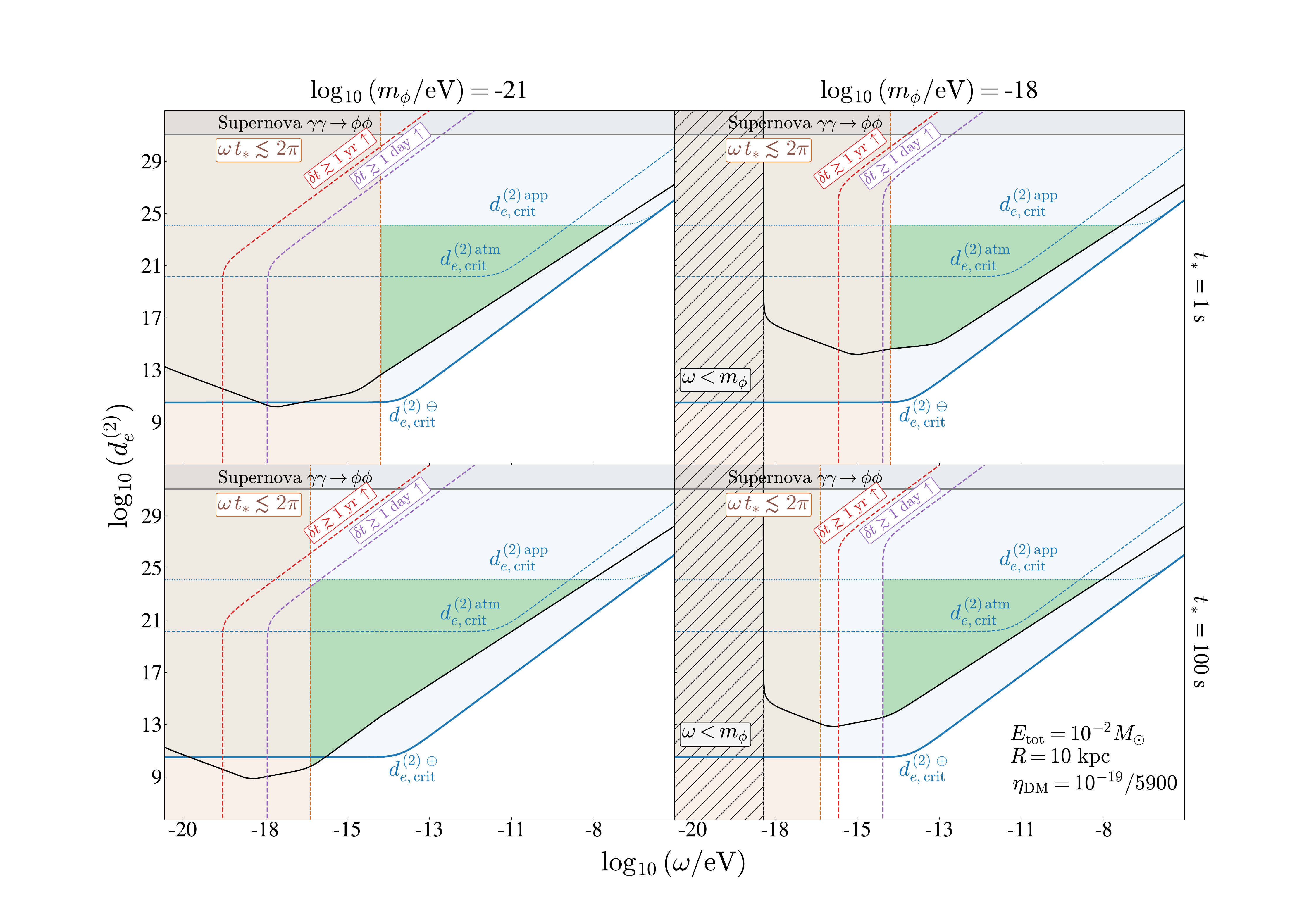}
    \caption{Sensitivity to quadratic dilatonic scalar coupling to photons, $d_{e}^{(2)}$, as a function of the ULB scalar energy $\omega$, considering a relativistic ULB field burst from a source located at $R = 10~\mathrm{kpc}$ from Earth, releasing a total energy of $E_{\rm total} = 10^{-2} M_\odot$ in the form of ULBs. Benchmark ULB masses of $m_{\phi} = 10^{-21}~\mathrm{eV}$ (left column) and $m_{\phi} = 10^{-18}~\mathrm{eV}$ (right column), with intrinsic burst times of $t_* = 1~\mathrm{s}$ (top row) and $t_* = 100~\mathrm{s}$ (bottom row) are displayed.
    The gray region represents parameter space excluded by direct $\phi$ production from supernova photons \cite{olive2008environmental}. The blue solid, dashed, and dotted lines indicate critical screening from the Earth, atmosphere, and experimental apparatus, respectively. The black line shows the projected sensitivity of future experiments with $\eta_\mathrm{DM}=10^{-19}/5900$, based on a Thorium isomer nuclear clock proposal \cite{beeks2024fine}, scaled using Eq.~\eqref{eq:d_i(2)}. The black hatched and brown regions are excluded due to the scalar energy being less than its mass and the uncertainty principle, respectively. Dashed purple and red lines represent detection delays of 1 day and 1 year relative to a light-speed signal. The green region denotes viable parameter space for multimessenger astronomy with scalars, given all constraints and benchmarks.
}
    \label{fig:10kpc_photon_quaddilatonic_coupling}
\end{figure*}

\begin{figure*}
    \centering
    \includegraphics[width = 1\linewidth]{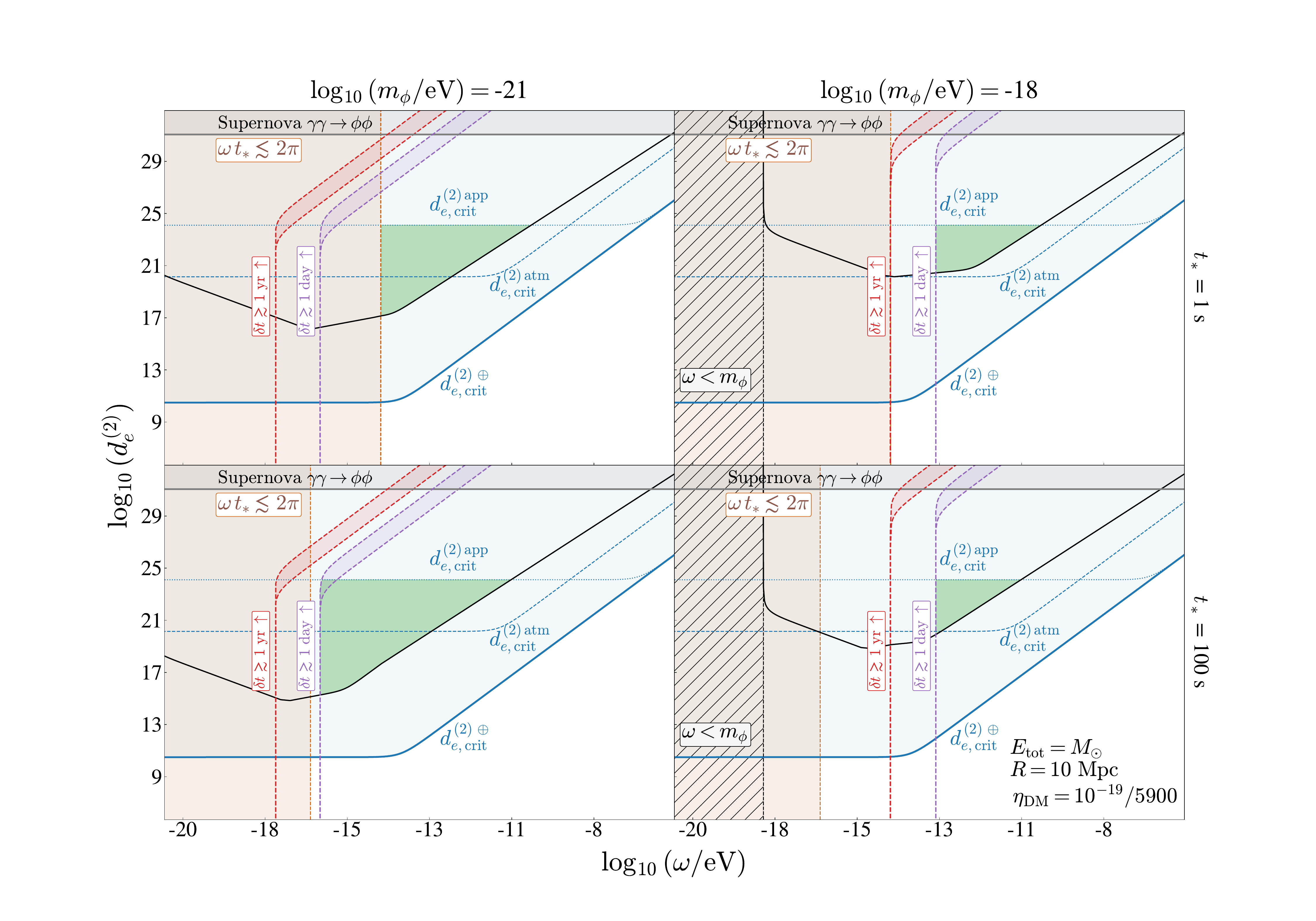}
    \caption{Sensitivity to quadratic dilatonic scalar coupling to photons, $d_{e}^{(2)}$, as a function of the ULB scalar energy $\omega$, considering a relativistic ULB field burst from a source located at $R = 10~\mathrm{Mpc}$ from Earth, releasing a total energy of $E_{\rm total} = M_\odot$ in the form of ULBs. Benchmark ULB masses of $m_{\phi} = 10^{-21}~\mathrm{eV}$ (left column) and $m_{\phi} = 10^{-18}~\mathrm{eV}$ (right column), with intrinsic burst times of $t_* = 1~\mathrm{s}$ (top row) and $t_* = 100~\mathrm{s}$ (bottom row) are displayed.
    The gray region represents parameter space excluded by direct $\phi$ production from supernova photons \cite{olive2008environmental}. Blue solid, dashed, and dotted lines show critical screening from the Earth, atmosphere, and experimental apparatus, respectively. The black line indicates the projected sensitivity of future experiments with $\eta_\mathrm{DM}=10^{-19}/5900$, based on a Thorium isomer nuclear clock proposal \cite{beeks2024fine}, scaled using Eq.~\eqref{eq:d_i(2)}. The black hatched and brown regions are excluded by scalar energy limits and the uncertainty principle. Dashed purple and red lines represent detection delays of 1 day and 1 year relative to a light-speed signal. The green region denotes viable parameter space for multimessenger astronomy with scalars, considering all constraints and benchmarks.
}
    \label{fig:10Mpc_photon_quaddilatonic_coupling}
\end{figure*}

\begin{figure*}
    \centering
    \includegraphics[width = 1\linewidth]{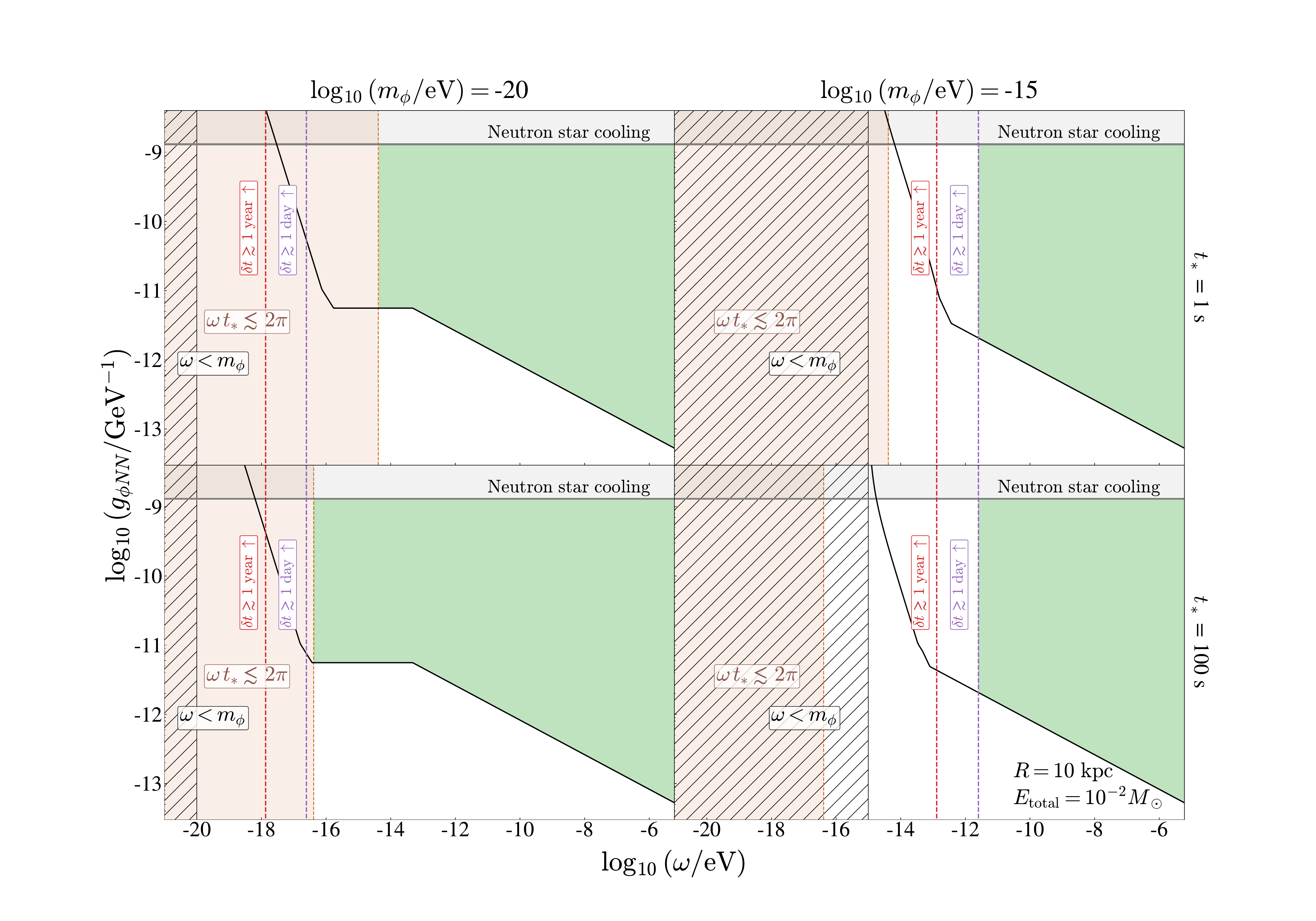}
    \caption{Sensitivity to ALP derivative coupling to neutrons, $g_{\phi NN}$, as a function of the ULB scalar energy $\omega$, considering a relativistic ULB field burst from a source located at $R = 10~\mathrm{kpc}$ from Earth, releasing a total energy of $E_{\rm total} = 10^{-2} M_\odot$ in the form of ULBs. Benchmark ULB masses of $m_{\phi} = 10^{-20}~\mathrm{eV}$ (left column) and $m_{\phi} = 10^{-15}~\mathrm{eV}$ (right column), with intrinsic burst times of $t_* = 1~\mathrm{s}$ (top row) and $t_* = 100~\mathrm{s}$ (bottom row) are displayed. The gray region represents parameter space excluded by neutron star cooling rates considering $\phi$ production~\cite{buschmann2022upper}. The black line shows the projected sensitivity of future experiments with $g_{\rm DM} = 10^{-11}~\mathrm{GeV}^{-1}$ and $t_{\rm int, DM} = 1~\mathrm{day}$, based on proposed comagnetometers \cite{bloch2020axion}, electrostatic storage rings \cite{brandenstein2023towards}, NMR techniques \cite{jackson2020overview}, and superfluid He3 experiments \cite{gao2022axion}, scaled using Eq.~\eqref{eq:g_ratio_v}. The black hatched and brown regions are excluded by ALP energy constraints and the uncertainty principle. Dashed purple and red lines indicate detection delays of 1 day and 1 year relative to a light-speed signal. The green region represents viable parameter space for multimessenger astronomy with ALPs, considering all constraints and benchmarks.
}       \label{fig:10kpc_axion_neutron}
\end{figure*}

\begin{figure*}
    \centering
    \includegraphics[width = 1\linewidth]{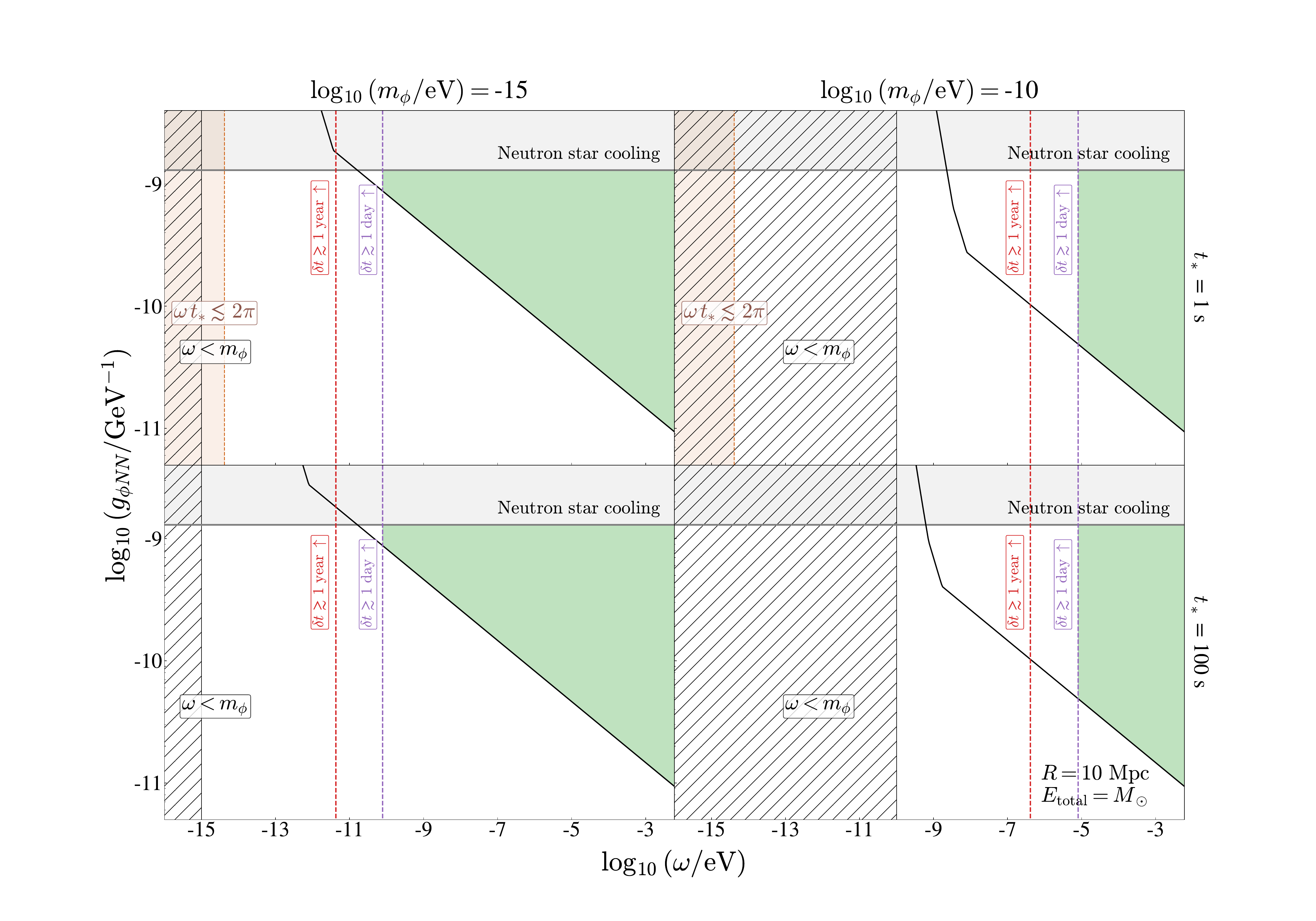}
    \caption{Sensitivity to ALP derivative coupling to neutrons, $g_{\phi NN}$, as a function of the ULB scalar energy $\omega$, considering a relativistic ULB field burst from a source located at $R = 10~\mathrm{Mpc}$ from Earth, releasing a total energy of $E_{\rm total} =  M_\odot$ in the form of ULBs. Benchmark ULB masses of $m_{\phi} = 10^{-15}~\mathrm{eV}$ (left column) and $m_{\phi} = 10^{-10}~\mathrm{eV}$ (right column), with intrinsic burst times of $t_* = 1~\mathrm{s}$ (top row) and $t_* = 100~\mathrm{s}$ (bottom row) are displayed. The gray region represents parameter space excluded by neutron star cooling rates considering $\phi$ production~\cite{buschmann2022upper}. 
    The black line shows the projected sensitivity of future experiments with $g_{\rm DM} = 10^{-11}~\mathrm{GeV}^{-1}$ and $t_{\rm int, DM} = 1~\mathrm{day}$, based on proposed comagnetometers \cite{bloch2020axion}, electrostatic storage rings \cite{brandenstein2023towards}, NMR techniques \cite{jackson2020overview}, and superfluid He3 experiments \cite{gao2022axion}, scaled using Eq.~\eqref{eq:g_ratio_v}. The black hatched and brown regions are excluded by ALP energy constraints and the uncertainty principle. Dashed purple and red lines indicate detection delays of 1 day and 1 year relative to a light-speed signal. The green region denotes viable parameter space for multimessenger astronomy with ALPs, considering all constraints and benchmarks.
}        \label{fig:10Mpc_axion_neutron}
\end{figure*}

\begin{table*}
    \centering
    \begin{tabular}{c|c|c|c|c|c|r}
    \hline \hline
     \multirow{4}{*}{\textbf{Coupling}}   & \multirow{4}{*}{\textbf{\shortstack[c]{ Strongest\\ Constraint}}} & \multicolumn{2}{c|}{\multirow{3}{*}{\textbf{ Critical   Value }}} & \multicolumn{3}{c}{\multirow{2}{*}{\textbf{Sensitivity
     }}} \\
        &  & \multicolumn{2}{c|}{} & \multicolumn{3}{c}{} \\
        &  & \textbf{Earth}   & \textbf{Apparatus}   & \multicolumn{3}{c}{\shortstack[c]{\textit{Galactic source:} $R = 10$ kpc, $E_{\rm tot} = 10^{-2} M_{\odot}$ \\
        \textit{(Extragalactic source:} $R = 10$ Mpc, $E_{\rm tot} = M_{\odot}$\textit{)} }       
        } \\ 
        &  & ($d_{i,\ \rm crit}^{(2)\,\oplus}$) & ($d_{i,\ \rm crit}^{(2)\,\rm app}$) & 
        $m_{\phi} = 10^{-21}$ eV & $m_{\phi} = 10^{-18}$ eV    & Figure
        \\   
        \hline
        $d_e^{(1)}$ & $1 \times 10^{-4}$ ~\cite{Berge:2017ovy}& $\cdots$ & $\cdots$ & $2\times 10^{-6}$ ($3
        \times 10^{-4}$) & $10^{-4}$ ($2\times10^{-1}$) &\ref{fig:10kpc_photon_lindilatonic_coupling} (\ref{fig:10Mpc_photon_lindilatonic_coupling}) \\
        $d_{m_e}^{(1)}$ & $1 \times 10^{-3}$ ~\cite{Berge:2017ovy} & $\cdots$ & $\cdots$ & $1.1$ ($2\times10^{2}$) & $6\times10^{1}$ ($1\times10^{5}$) & \ref{fig:10kpc_electron_lindilatonic_coupling}
        (\ref{fig:10Mpc_electron_lindilatonic_coupling}) \\  
        $d_g^{(1)}$ & $3\times 10^{-6}$ ~\cite{Berge:2017ovy}& $\cdots$ & $\cdots$ & $1\times10^{-7}$ ($2\times10^{-5}$) & $6\times 10^{-6}$ ($1\times10^{-2}$) & \ref{fig:10kpc_gluon_lindilatonic_coupling} (\ref{fig:10Mpc_gluon_lindilatonic_coupling}) \\
        \colorbox{YellowGreen}{$d_e^{(2)}$} & $1 \times10^{31}$ ~\cite{olive2008environmental}& $3\times10^{10}$ & $1\times10^{24}$ & $5\times 10^{9}$ ($8\times 10^{14}$)& $4\times 10^{12}$ ($4\times10^{18}$) & \ref{fig:10kpc_photon_quaddilatonic_coupling} (\ref{fig:10Mpc_photon_quaddilatonic_coupling}) \\
        $d_{m_e}^{(2)}$ & $5 \times 10^{31}$ ~\cite{olive2008environmental}& $2\times10^{11}$ & $1\times10^{25}$ & $3\times 10^{15}$ ($5\times 10^{20}$) & $2\times 10^{18}$ ($2\times 10^{24}$) & \ref{fig:10kpc_electron_quaddilatonic_coupling} (\ref{fig:10Mpc_electron_quaddilatonic_coupling}) \\
        $d_g^{(2)}$ & $5\times 10^{28}$ ~\cite{olive2008environmental}& $6\times10^{7}$& $2\times10^{21}$& $3\times 10^{8}$ ($5\times 10^{13}$) & $2\times 10^{11}$ ($2\times10^{17}$)& \ref{fig:10kpc_gluon_quaddilatonic_coupling} 
        (\ref{fig:10Mpc_gluon_quaddilatonic_coupling}) \\
        \hline
        -- & -- & -- & -- & $m_{\phi} = 10^{-20}$ eV ($10^{-15}$ eV) & $m_{\phi} = 10^{-15}$ eV ($10^{-10}$ eV)& --~~~ \\
        \hline
        $g_{\phi\gamma\gamma}\ \rm(GeV^{-1})$ & $1\times10^{-13}$ ~\cite{sisk2022new}& $\cdots$ & $\cdots$ & $2\times10^{-15}$ ($3\times10^{-13}$)& $2\times10^{-15}$ ($3\times10^{-13}$) & \ref{fig:10kpc_axion_photon}
        (\ref{fig:10Mpc_axion_photon}) \\
        $g_{\phi e}\ \rm(GeV^{-1})$ & $1\times10^{-10}$ ~\cite{capozzi2020axion}& $\cdots$ & $\cdots$ & $5\times10^{-13}$ ($8\times10^{-11}$) & $5\times10^{-13}$ ($8\times10^{-11}$) & \ref{fig:10kpc_axion_electron} (\ref{fig:10Mpc_axion_electron}) \\
        $g_{\phi PP}\ \rm(GeV^{-1})$ & $6\times10^{-10}$ ~\cite{lella2024getting} & $\cdots$ & $\cdots$ & $2\times10^{-15}$ ($3\times10^{-13}$) & $2\times10^{-15}$ ($3\times10^{-13}$) & \ref{fig:10kpc_axion_proton} (\ref{fig:10Mpc_axion_proton}) \\
        \colorbox{YellowGreen}{$g_{\phi NN}\ \rm(GeV^{-1})$} & $1\times10^{-9}$ ~\cite{buschmann2022upper}& $\cdots$ & $\cdots$ & $5\times10^{-14}$  ($9\times10^{-12}$)  & $5\times10^{-14}$ ($9\times10^{-12}$) & \ref{fig:10kpc_axion_neutron} (\ref{fig:10Mpc_axion_neutron}) \\
        $g_{\phi P\gamma}\ \rm(GeV^{-2})$ & $7\times10^{-9}$ ~\cite{caloni2022novel} & $\cdots$ & $\cdots$ & $1\times10^{-17}$ ($2\times10^{-15}$) & $1\times10^{-17}$ ($2\times10^{-15}$) & \ref{fig:10kpc_axion_EDM} (\ref{fig:10Mpc_axion_EDM}) \\
         \hline \hline
    \end{tabular}
    \caption{Sensitivity to SM-$\phi$ couplings from transient astrophysical events from quantum sensors, considering reference Galactic source at $R=10$ kpc with $E_{\rm tot} = 10^{-2} M_{\odot}$ and extragalactic source at $R = 10$ Mpc with $E_{\rm tot} = M_{\odot}$. The sensitivities displayed meet the time delay and emission constraints (defined by the minimum of the green regions in the Figures in the final column). Strongest existing constraints as well as critical coupling values for Earth ($d_{i,\ \rm crit}^{\oplus}$) and apparatus ($d_{i,\ \rm crit}^{\rm app}$) are displayed. For scalars, we consider $m_{\phi} = 10^{-21}$ eV and $m_{\phi} = 10^{-18}$ eV for both sources. For axions, $m_\phi=10^{-20}$ eV and $m_\phi=10^{-15}$ eV, and $m_\phi=10^{-15}$ eV and $m_\phi=10^{-10}$ eV, for the two sources respectively. For all couplings shown, we consider a benchmark experimental integrated livetime of $t_{\rm int,*} = 1$ day. The couplings shaded green are discussed in Sec.~\ref{sec:results}, while the others are discussed in App.~\ref{sec:App_Couplings}. The values in parentheses correspond to the extragalactic source  benchmark. For a given coupling row, when the sensitivity reach is lower than the strongest constraint, there exists viable observation parameter space for our benchmark choices and assumptions. The dots `$\cdots$' indicate that a critical value is not applicable for the given coupling.}
    \label{tab:summarytable}
\end{table*}

We consider two benchmark cases for the study of multimessenger astronomy signals beyond SM with scalars and ALPs: signals originating from sources at Galactic of $\mathcal{O}(10)$ kpc and extragalactic of $\mathcal{O}(10)$ Mpc distance scales.
Here, we present results for scalars quadratically coupled to photons in Fig.~\ref{fig:10kpc_photon_quaddilatonic_coupling} and Fig.~\ref{fig:10Mpc_photon_quaddilatonic_coupling}) as well as ALPs with derivative coupling to neutrons in Fig.~\ref{fig:10kpc_axion_neutron} and Fig.~\ref{fig:10Mpc_axion_neutron}. We present results for other couplings discussed in Sec.~\ref{sec:EFT} in Appendix~\ref{sec:App_Couplings}. 
In all cases, we use the benchmarks and parameters outlined in Secs.~\ref{sec:sources}, \ref{sec:propagation}, and \ref{sec:detection}.

In each figure pair, source distances are either 10 kpc (with total energy $10^{-2}~M_\odot$) or 10 Mpc (with total energy $M_\odot$). Each plot is a $2 \times 2$ coupling constant–ULB energy diagram. Rows correspond to benchmark intrinsic burst durations of $t_* = 1$ second (top) and $t_* = 100$ seconds (bottom), while columns correspond to benchmark mass values: $\rm 10^{-21}~eV$ (left) and $\rm 10^{-18}~eV$ (right) for scalars, and $\rm 10^{-20}~eV$ (left) and $\rm 10^{-15}~eV$ (right) for ALPs. 

The couplings for scalars, ALPs with non-nucleon interactions, and ALPs with nucleon interactions are defined in Eqs.~\eqref{eq:scalar_Lint}, \eqref{eq:ALP_Lint}, and \eqref{eq:ALP_nucl}, respectively. Quadratically coupled scalars can undergo screening due to interactions with dense SM fields, as discussed in Sec.~\ref{sec:screening}. In the plots of quadratic interactions, blue lines denote critical screening from the Earth, atmosphere, or experimental apparatus. Black hatched regions indicate unphysical values where mass exceeds energy, and brown-shaded regions are excluded by the uncertainty principle (see Eq.~\eqref{Eq:uncert}). Purple and red dashed lines represent time delays of 1 day and 1 year, respectively, between ULB and light-speed signals at Earth (see Eq.~\eqref{eq:TimeDelay}). For quadratically coupled scalars, delay times can also depend on screening effects. Details are provided in Sec.~\ref{sec:propagation}.

Projected experimental sensitivities are shown as black lines, rescaled from ultralight experimental benchmarks using Eqs.~\eqref{eq:direach}, \eqref{eq:direach2}, \eqref{eq:g_ratio}, and \eqref{eq:g_ratio_v}. Note that due to the time factors in Eqs.~\eqref{eq:eta_*}, \eqref{eq:g_ratio}, and \eqref{eq:g_ratio_v}, the sensitivity projections have discontinuities in their slopes. This is associated with transitions between different timescale regimes, such as $\tilde{t}_* > t_{\rm int,*}$ or $\tilde{t}_* < t_{\rm int,*}$, as well as $\tau_{\rm DM} > t_{\rm int,DM}$ or $\tau_{\rm DM} < t_{\rm int,DM}$. Gray regions reflect astrophysical or equivalence principle (EP) constraints. Green regions in the plots represent the parameter space where multimessenger astronomy is feasible, taking into account all constraints, physical conditions, and screening effects, based on the chosen benchmark parameters and the best projected sensitivities of future experiments. 

We additionally show a summary of the plots in Table~\ref{tab:summarytable}, where we report the best sensitivity values to the couplings, along with the strongest constraint, for each coupling and benchmark. Additionally,  the critical quadratic couplings are given. The green shaded couplings correspond to those discussed in Sec.~\ref{sec:results}.
 
Quantum sensor networks could significantly enhance sensitivity by correlating signals across multiple detectors, enabling higher precision and better discrimination of signals from noise. Compared to single-sensor setups, networks offer improved localization and timing accuracy for ULBs, allowing more precise source identification and coincidence studies with other messengers such as photons, neutrinos, or GWs\footnote{Here, we do not consider other possible contributions, such as cosmic rays composed of charged particles and nuclei, which are not pristine tracers for distant sources.}. We leave detailed analysis with networks of detectors for future work.
 
Coincidence between scalar or ALP bursts and traditional astrophysical messengers provides an essential diagnostic tool for identifying ULB sources. While photons are subject to dispersion in the intergalactic medium, leading to potential frequency-dependent delays, neutrinos and GWs travel essentially at the speed of light, providing robust reference signals. The observed time lag between a ULB and these messengers can be used to constrain ULB masses and propagation effects, including screening or dispersion, and enable determining the distance to the source.

The association of a scalar or ALP burst with a specific astrophysical event could require dedicated correlation with known astrophysical signals. For example, a ULB burst coinciding with a GW emission event or a supernova neutrino burst would suggests a common origin. Quantum sensor networks, with their enhanced timing and sensitivity, allow triangulation of ULB directions, aiding source identification. Additionally, the observed time delay and burst properties provide insight into coupling constants, propagation effects, and the nature of the source.

\section{Conclusion and Outlook}
\label{sec:conclusion}

The advent of quantum sensors has opened novel and promising avenues for multimessenger astronomy, enabling probing a wide range of unexplored physics beyond SM as predicted by variety of theories. Our analysis demonstrates that multimessenger astronomy incorporating observations by quantum sensors is not only feasible but offers unique opportunities for studying uncharted parameter regimes for ULB scalar and ALP new quantum field interactions. While stringent constraints from EP tests restrict the parameter space for scalar linear couplings, ALP couplings—particularly those involving nucleons—offer significant opportunities. Observations by networks of quantum sensors, combined with traditional astrophysical messengers such as GWs, photons, and neutrinos, can significantly enhance our ability to identify sources of new quantum fields and unravel their origins, providing deeper insights into astrophysical events and fundamental physics.

Extending multimessenger studies to include beyond SM fields could unveil hidden dynamics of astrophysical phenomena that remain insensitive to traditional messengers. The inclusion of scalar or ALP signals provides additional modalities that could reveal a more complete picture of astrophysical events. Beyond the discovery of new fields, this approach holds the potential to uncover deeper insights into their properties and fundamental interactions. Observations of emissions from transient events—such as core-collapse supernovae, binary mergers, or bosenovae—could provide the first evidence of these new quantum fields, offering novel insights into their origin, structure, and role in the cosmos.

Our study addressed several critical challenges for multimessenger astronomy beyond SM, including the impact of matter screening effects and detection prospects. Screening plays a crucial role in assessing the detectability of beyond-SM signals, particularly in the case of quadratic dilatonic couplings. Our detailed analysis of screening and antiscreening effects reveals that screening behaves analogously to potential barriers in quantum mechanics. While positive quadratic dilatonic couplings typically induce screening effects, negative couplings introduce more complex dynamics, encompassing both conventional screening and antiscreening under certain conditions. Importantly, screening is not dictated by specific potential forms but rather by the general behavior of the vacuum as a local minimum.

Analyzing the parameter spaces for linear and quadratic couplings,
we identify significantly larger accessible parameter space for probing quadratic couplings. This is due to weaker astrophysical and laboratory-based constraints compared to linear couplings. For linear scalar couplings, stringent limits arise from EP tests, which search for long-range Yukawa interactions and impose constraints independent of ULB densities in the vicinity of an experiment. In contrast, quadratic couplings lead to weaker, density-independent bounds, as the effective potential falls off as the cube of the distance, thereby evading EP test constraints. This distinction underscores the broader opportunities for probing quadratic couplings in multimessenger studies.

ALP linear couplings further differentiate themselves from pure scalar linear couplings due to the shift symmetry of ALPs, which suppresses direct modifications of fundamental constants. This symmetry ensures that ALP interactions are purely dynamic and oscillatory, avoiding persistent effects such as variations in the fine-structure constant. In contrast, linear scalar couplings directly modify constants and are subject to stringent EP and astrophysical bounds.

Advancing further, a deeper understanding of potential ULB emission sources and their signal event rates is essential. Our general analysis, based on phenomenological parametrizations, underscores the need for future studies that link ULB signal properties to specific source spectra. Multiple burst events—arising from astrophysical phenomena such as supernovae or mergers—offer a promising avenue for investigation, particularly regarding event rates and the interpretation of coincident signals. Integrating such findings into our framework will refine detection strategies and enhance our understanding of the underlying physics.

In conclusion, we have laid the groundwork for exploring ULB signals in multimessenger searches involving astrophysical signals associated with GWs, photons, and neutrinos. Our analysis clarifies the role of screening and antiscreening in detection strategies, differentiates the behavior of linear and quadratic couplings, and underscores the potential of quantum sensor network observations in multimessenger astronomy. Further, our analysis of screening also provides motivation and highlights the  advantages of space-based detector systems, which avoid critical screening by the Earth and its atmosphere. As experimental techniques and detection networks continue to improve, quantum sensors could provide valuable new insights into astrophysical events and the fundamental nature of beyond-SM quantum fields.

\section*{Acknowledgments}
We thank Dmitry Budker and GNOME collaboration for discussion. This work was supported in part by the NSF QLCI Award OMA - 2016244, NSF  PHY-2309254 grant, and the European Research Council (ERC) under the European Union’s Horizon 2020 research and innovation program (Grant Number 856415).
V.T. acknowledges support of the JSPS KAKENHI grant No. 23K13109. This work was supported in part by the World Premier International Research Center Initiative (WPI), MEXT, Japan. 

\appendix

\section{Negative $\beta$ and General Potentials}
\label{app:potential}

We describe how different choices of potentials than Eq.~\ref{eq:phipotscr} affects the results of the screening discussion, in the case that $\beta<0$.

\subsubsection{Theory with $\phi^6$ potential}

In the case of bosenovae, and potentially other production mechanisms, the quartic term is attractive ($\lambda < 0$) and does not stabilize the potential. That is, the potential

\begin{align}
    V(\phi) = \frac{1}{2} m_{\phi}^2 \phi^2 + \frac{\lambda}{4!}\phi^4
\end{align}

\noindent with $\lambda < 0$ does not contain any global or local minima when $m_{\phi}^2 + \beta <0$. To address this, consideration of higher-order terms are required. For the specific case of a $\phi^6$ term, the potential is
\begin{align}
    V = \frac{1}{2}(m_{\phi}^2 + \beta)\phi^2 - \frac{\lambda}{4!} \phi^4 + \frac{\lambda_6}{6!}\phi^6 
\end{align}
and it can be solved exactly. Such potentials can naturally arise in certain axion-like theories (see e.g. \cite{Fan:2016rda}). By revisiting the calculation from the previous section with this specific potential, we obtain additional insights into the dynamics and stability conditions.

The effective mass of $\phi$ becomes

\begin{align}
    m^2_{\rm eff}  
    &= 4|m_{\phi}^2 + \beta| + \delta(\beta)~,
\end{align}
where 
\begin{align}
    \delta(\beta) = \frac{2\lambda(5 \lambda + \sqrt{25\lambda^2 +30|m_{\phi}^2 + \beta |\lambda_6})}{3\,\lambda_6}
\end{align}
results in a $m_{\rm eff}^2 >0$. This also linearly depends on $\beta$, with corrections that scale as $\beta^{1/2}$. 

The index of refraction then takes the form
\begin{align}
     n(\omega) = \frac{k}{\omega} = \sqrt{1 - \frac{4|m_{\phi}^2 + \beta| + \delta(\beta)}{\omega^2}}~.
\end{align}
We can solve for the critical coupling. Under the condition that $\delta(\beta)/(4|m_{\phi}^2 + \beta|) \ll 1$, the critical coupling is  
\begin{align}
    d_i^{(2)} &\gtrsim \Mpl^2\frac{(2\mathcal{R})^{-2} + \omega^2  + 4m_{\phi}^2}{32\pi\, \rho_i}~.
\end{align}

\subsubsection{Theory with $\phi^p$ potential}

In principle, in the potential the first higher-order term with a positive sign need not be $\phi^6$. This could correspond to a more general power $p$. By neglecting the $\phi^4$ term, the potential can be analyzed and solved for a general power $p$.

Let us consider a general potential given by
\begin{align}
    V = \frac{1}{2}(m_{\phi}^2 + \beta)\phi^2 + \frac{\lambda_p}{p!} \phi^{p}~.
\end{align}
Then, the VEV can be found to be
 \begin{align}
      \langle \phi \rangle = \bigg(\frac{(p-1)!|m^2+\beta|}{\lambda_p}\bigg)^{1/(p-2)}~.
  \end{align}

The effective mass of $\phi$ can be then determined to be
\begin{align}
    m_{\phi}^2 
&= -|m^2+\beta| + \frac{\lambda_p}{(p-2)!}\frac{(p-1)!|m^2+\beta|}{\lambda_p} \notag\\
    &= (p-2)|m^2+\beta|~.
\end{align}
This result is consistent with our approximate solution for the $\phi^6$ potential, where the quartic term introduces a small correction. However, it is not guaranteed that the corrections will always remain small for a general potential.

While this represents an idealized scenario, it demonstrates that the overall structure we consider can persist when higher order terms in the potential are considered.
\\
\\

\section{Sensitivity for Additional Couplings}
\label{sec:App_Couplings}

\begin{figure*}
    \centering
    \includegraphics[width = 1\linewidth]{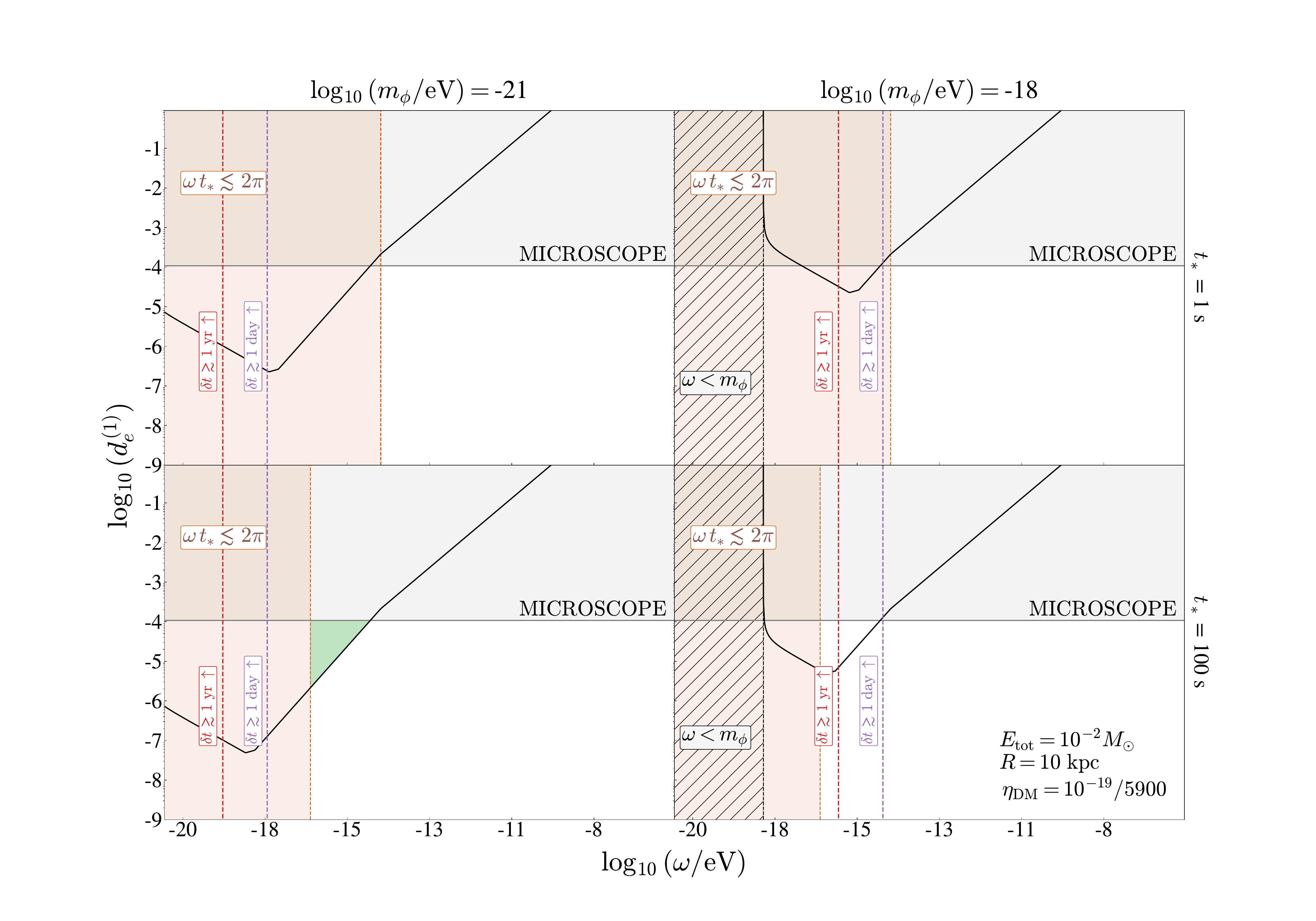}
    \caption{Sensitivity to linear dilatonic scalar coupling to photons, $d_e^{(1)}$, as a function of the ULB scalar energy $\omega$, considering a relativistic ULB field burst from a source located at $R = 10~\mathrm{kpc}$ from Earth, releasing a total energy of $E_{\rm total} = 10^{-2} M_\odot$ in the form of ULBs. Benchmark ULB masses of $m_{\phi} = 10^{-21}~\mathrm{eV}$ (left column) and $m_{\phi} = 10^{-18}~\mathrm{eV}$ (right column), with intrinsic burst times of $t_* = 1~\mathrm{s}$ (top row) and $t_* = 100~\mathrm{s}$ (bottom row) are displayed. 
    The gray line and shaded region indicate the parameter space excluded by MICROSCOPE EP tests~\cite{Berge:2017ovy}. The black line shows the projected sensitivity of future experiments, based on $\eta_\mathrm{DM} = 10^{-19}/5900$, consistent with a Thorium isomer nuclear clock expectation~\cite{beeks2024fine}, scaled for the scalar burst characteristics using Eq.~\eqref{eq:d_i(1)}. The black hatched and brown-shaded regions are excluded due to the scalar energy being less than its mass and the uncertainty principle, respectively. Dashed purple and red lines represent detection delays of 1 day and 1 year, respectively, relative to a signal traveling at the speed of light. The green shaded region denotes the viable parameter space for multimessenger astronomy with scalars, considering all constraints and the given burst benchmarks.
}
    \label{fig:10kpc_photon_lindilatonic_coupling}
\end{figure*}

\begin{figure*}
    \centering
    \includegraphics[width = 1\linewidth]{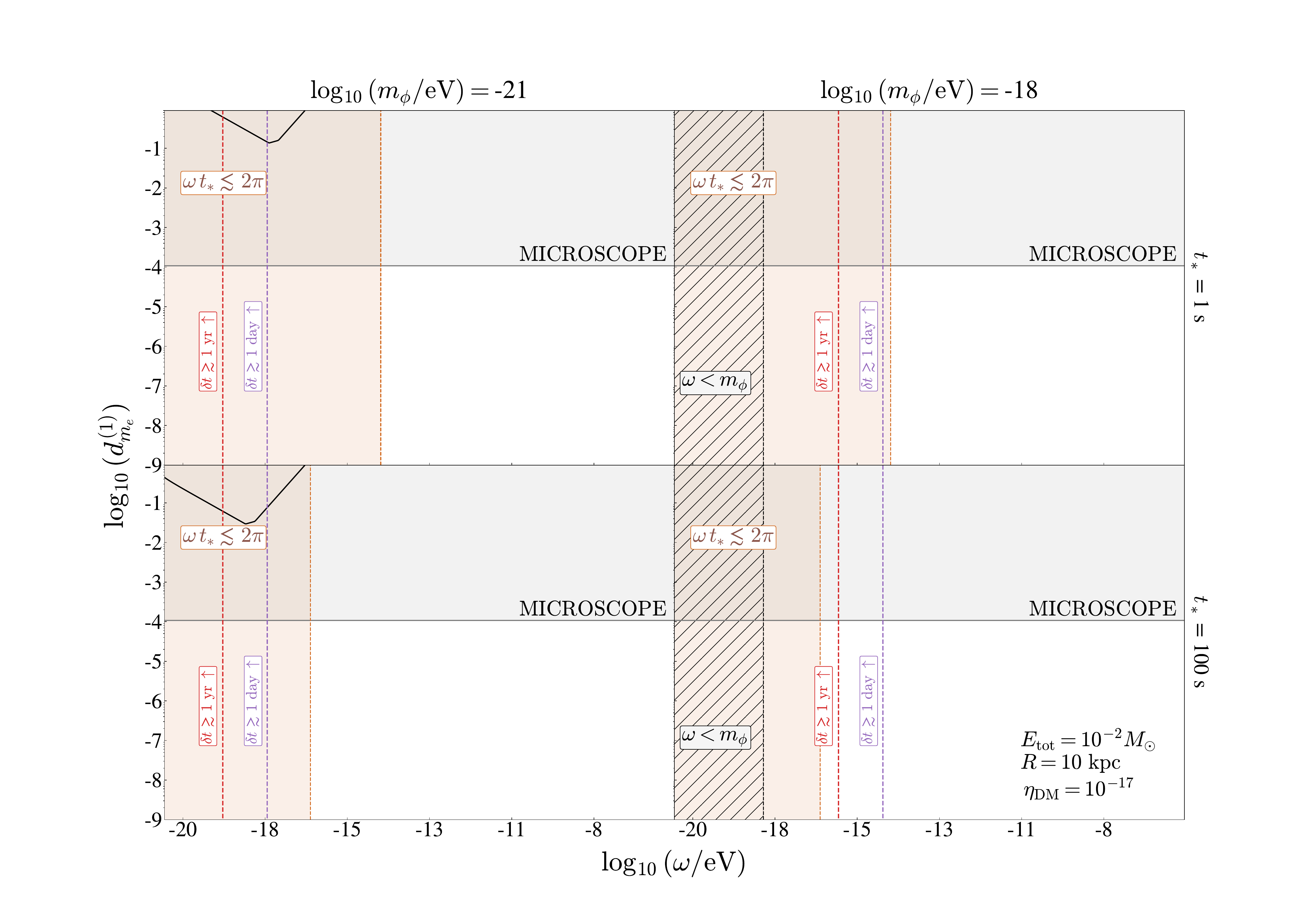}
    \caption{Sensitivity to linear dilatonic scalar coupling to electrons, $d_{m_e}^{(1)}$, as a function of the ULB scalar energy $\omega$, considering a relativistic ULB field burst from a source located at $R = 10~\mathrm{kpc}$ from Earth, releasing a total energy of $E_{\rm total} = 10^{-2} M_\odot$ in the form of ULBs. Benchmark ULB masses of $m_{\phi} = 10^{-21}~\mathrm{eV}$ (left column) and $m_{\phi} = 10^{-18}~\mathrm{eV}$ (right column), with intrinsic burst times of $t_* = 1~\mathrm{s}$ (top row) and $t_* = 100~\mathrm{s}$ (bottom row) are displayed.
    The gray line and shaded region indicate the parameter space excluded by MICROSCOPE EP tests \cite{Berge:2017ovy}. The black line shows the projected sensitivity of future experiments, based on $\eta_\mathrm{DM}=10^{-17}$ from molecular spectra analysis of SrOH \cite{SrOH2021}, scaled for the scalar burst characteristics using Eq.~\eqref{eq:d_i(1)}. The black hatched and   brown-shaded regions are excluded due to the scalar energy being less than its mass and the uncertainty principle, respectively. Dashed purple and red lines represent detection delays of 1 day and 1 year, respectively, relative to a signal traveling at the speed of light. For the given benchmarks, no viable parameter space exists for multimessenger astronomy with scalars.}
    \label{fig:10kpc_electron_lindilatonic_coupling}
\end{figure*}

\begin{figure*}
    \centering
    \includegraphics[width = 1\linewidth]{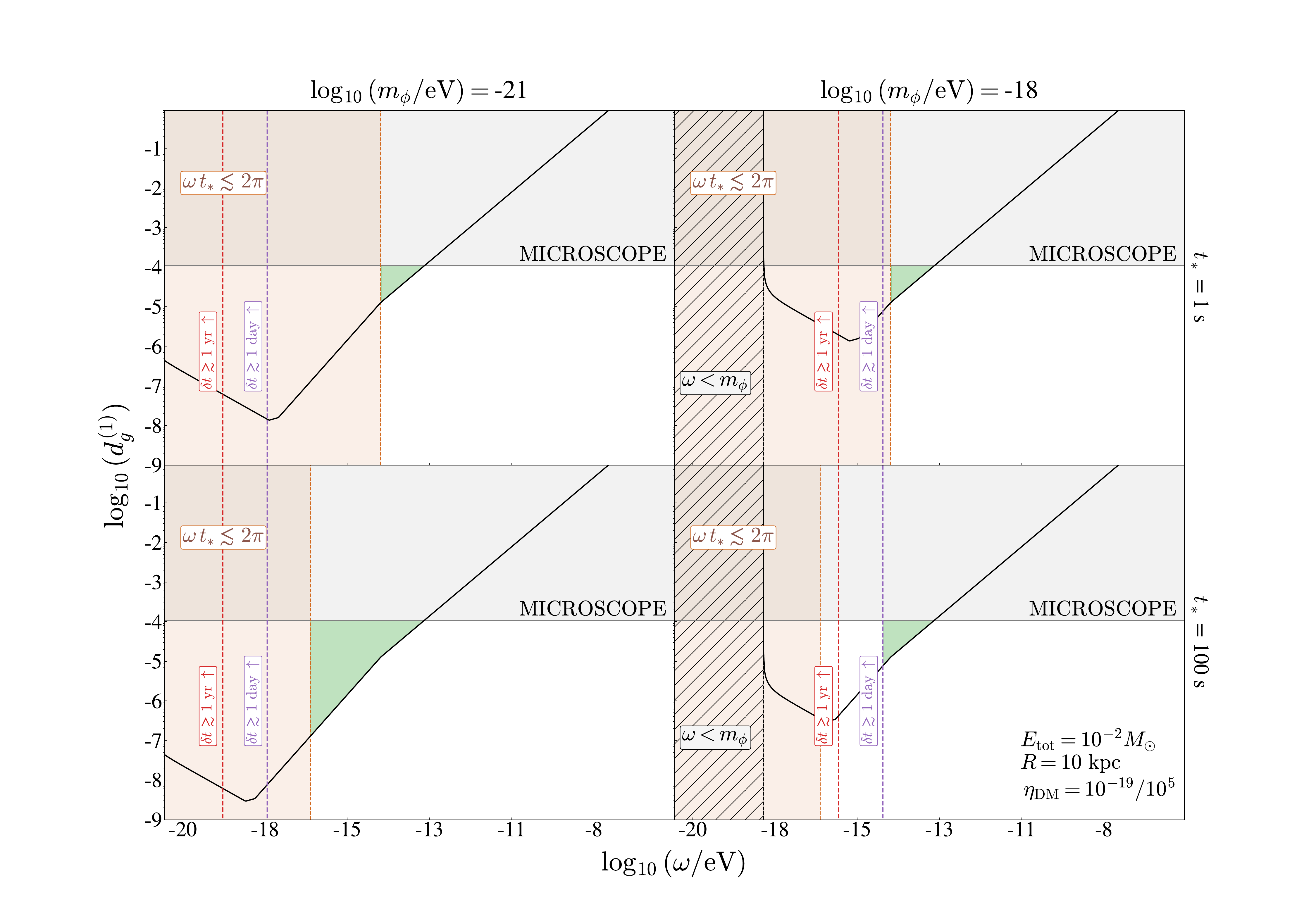}
    \caption{Sensitivity to linear dilatonic scalar coupling to gluons, $d_{g}^{(1)}$, as a function of the ULB scalar energy $\omega$, considering a relativistic ULB field burst from a source located at $R = 10~\mathrm{kpc}$ from Earth, releasing a total energy of $E_{\rm total} = 10^{-2} M_\odot$ in the form of ULBs. Benchmark ULB masses of $m_{\phi} = 10^{-21}~\mathrm{eV}$ (left column) and $m_{\phi} = 10^{-18}~\mathrm{eV}$ (right column), with intrinsic burst times of $t_* = 1~\mathrm{s}$ (top row) and $t_* = 100~\mathrm{s}$ (bottom row) are displayed.
    The gray line and shaded region show the parameter space excluded by MICROSCOPE EP tests \cite{Berge:2017ovy}. The black line represents the projected sensitivity of future experiments, based on $\eta_\mathrm{DM} = 10^{-24}$, consistent with a Thorium isomer nuclear clock proposal \cite{2012PhRvL.108l0802C,FlaTh06}, scaled for the scalar burst characteristics using Eq.~\eqref{eq:d_i(1)}. The black hatched and brown shaded regions are excluded due to the scalar energy being less than its mass and the uncertainty principle, respectively. Dashed purple and red lines indicate detection delays of 1 day and 1 year, respectively, relative to a signal traveling at the speed of light. The green shaded region represents the viable parameter space for multimessenger astronomy with scalars, considering all constraints and the given benchmarks.
}
    \label{fig:10kpc_gluon_lindilatonic_coupling}
\end{figure*}

\begin{figure*}
    \centering
    \includegraphics[width = 1\linewidth]{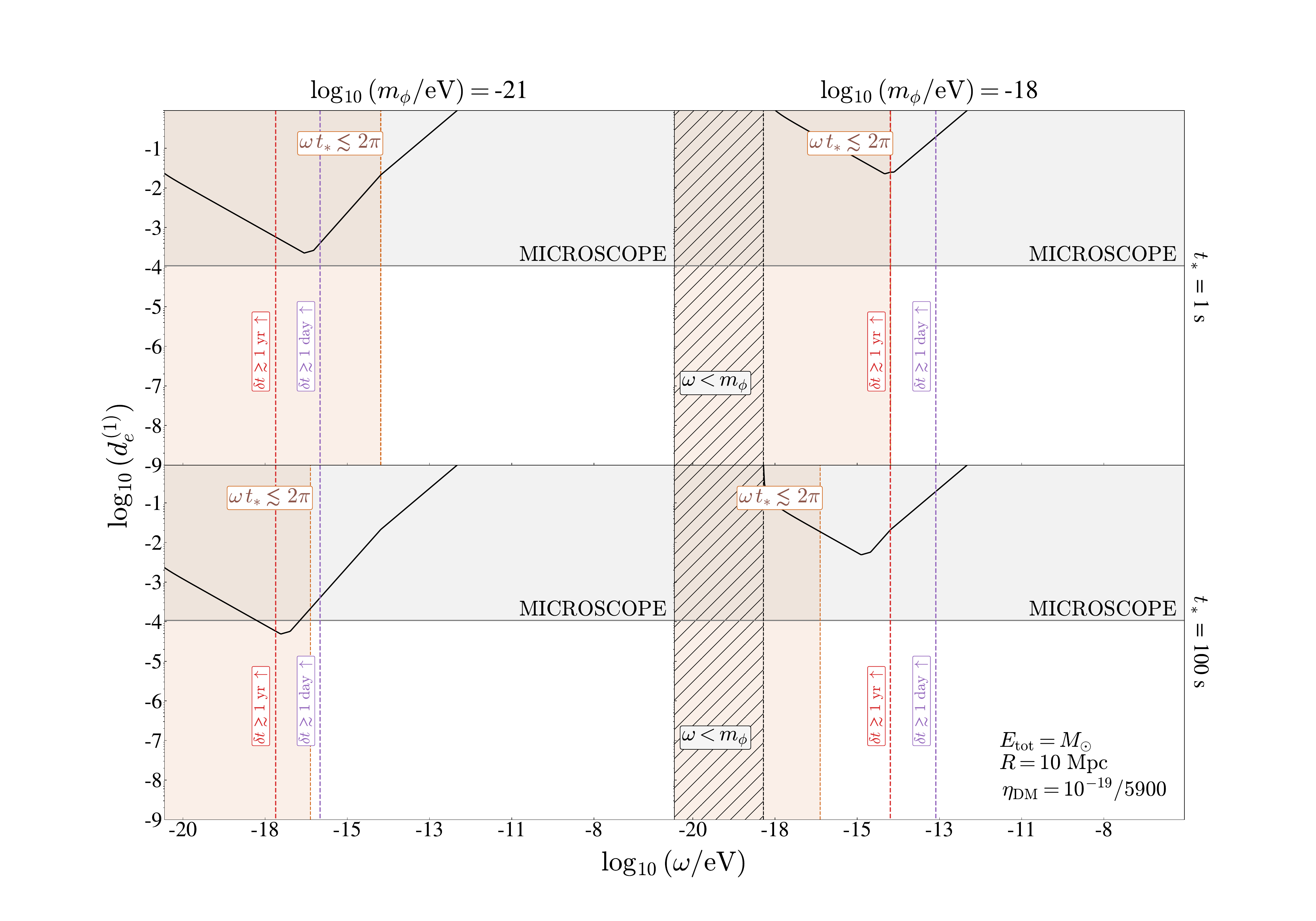}
    \caption{Sensitivity to linear dilatonic scalar coupling to photons, $d_{e}^{(1)}$, as a function of the ULB scalar energy $\omega$, considering a relativistic ULB field burst from a source located at $R = 10~\mathrm{Mpc}$ from Earth, releasing a total energy of $E_{\rm total} = M_\odot$ in the form of ULBs. Benchmark ULB masses of $m_{\phi} = 10^{-21}~\mathrm{eV}$ (left column) and $m_{\phi} = 10^{-18}~\mathrm{eV}$ (right column), with intrinsic burst times of $t_* = 1~\mathrm{s}$ (top row) and $t_* = 100~\mathrm{s}$ (bottom row) are displayed.
The gray region indicates parameter space excluded by MICROSCOPE EP tests \cite{Berge:2017ovy}, and the black line represents the projected sensitivity of future experiments with $\eta_\mathrm{DM}=10^{-19}/5900$, based on a Thorium isomer nuclear clock proposal \cite{beeks2024fine}, scaled using Eq.~\eqref{eq:d_i(1)}. The black hatched and brown regions are excluded due to the scalar energy being less than its mass and the uncertainty principle, respectively. Dashed purple and red lines correspond to detection delays of 1 day and 1 year relative to a light-speed signal. For the given benchmarks, no viable parameter space exists for multimessenger astronomy with scalars.
}
    \label{fig:10Mpc_photon_lindilatonic_coupling}
\end{figure*}

\begin{figure*}
    \centering
    \includegraphics[width = 1\linewidth]{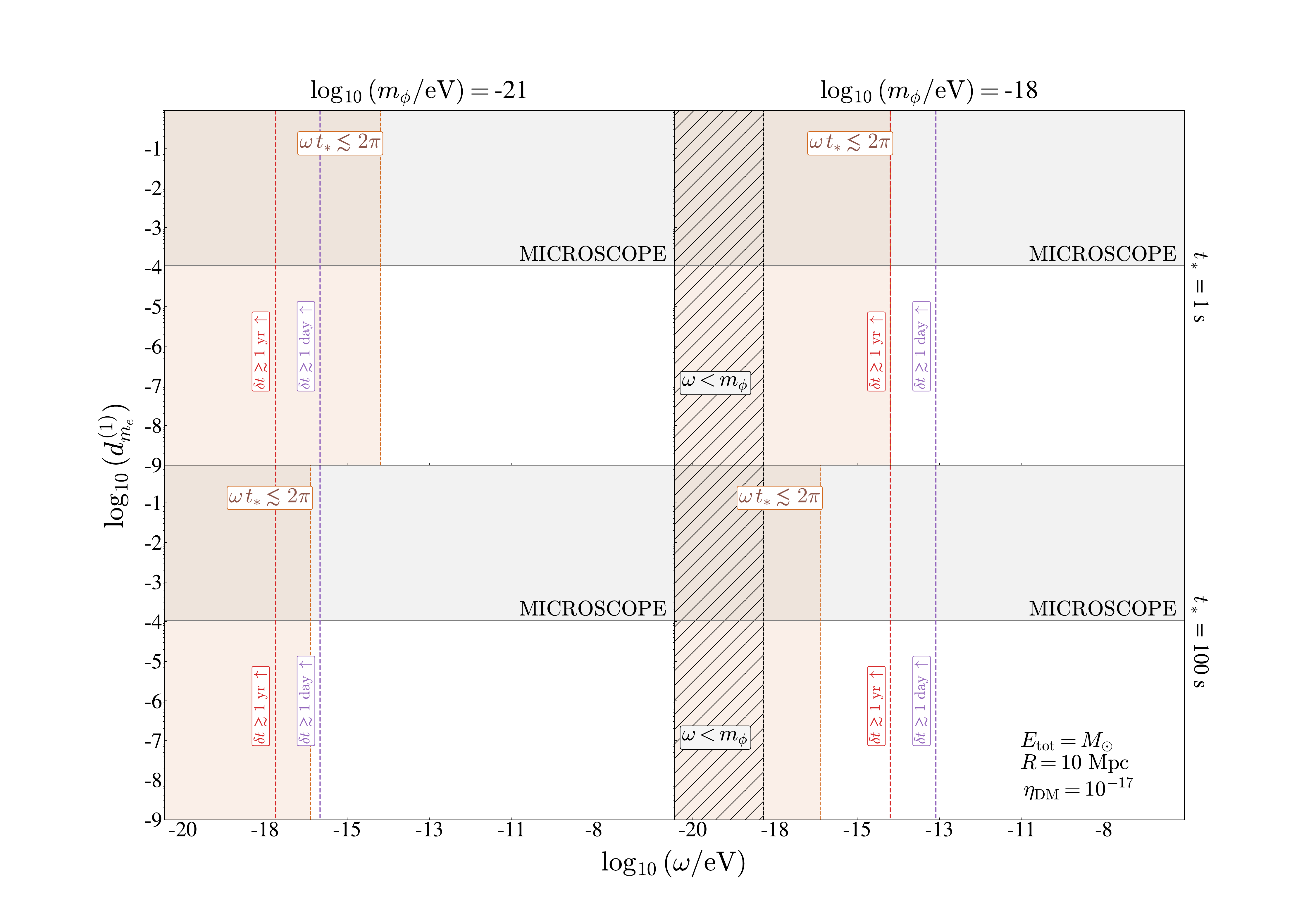}
    \caption{Sensitivity to linear dilatonic scalar coupling to electrons, $d_{m_e}^{(1)}$, as a function of the ULB scalar energy $\omega$, considering a relativistic ULB field burst from a source located at $R = 10~\mathrm{Mpc}$ from Earth, releasing a total energy of $E_{\rm total} = M_\odot$ in the form of ULBs. Benchmark ULB masses of $m_{\phi} = 10^{-21}~\mathrm{eV}$ (left column) and $m_{\phi} = 10^{-18}~\mathrm{eV}$ (right column), with intrinsic burst times of $t_* = 1~\mathrm{s}$ (top row) and $t_* = 100~\mathrm{s}$ (bottom row) are displayed.
    The gray region shows the parameter space excluded by MICROSCOPE EP tests \cite{Berge:2017ovy}, and the black line indicates the projected sensitivity of future experiments with $\eta_\mathrm{DM} = 10^{-17}$, based on molecular spectra analysis of SrOH \cite{SrOH2021}, scaled using Eq.~\eqref{eq:d_i(1)}. The black hatched and brown regions are excluded due to the scalar energy being less than its mass and the uncertainty principle, respectively. Dashed purple and red lines represent detection delays of 1 day and 1 year relative to a light-speed signal. For the given benchmarks, no viable parameter space exists for multimessenger astronomy with scalars.}
    \label{fig:10Mpc_electron_lindilatonic_coupling}
\end{figure*}

\begin{figure*}
    \centering
    \includegraphics[width = 1\linewidth]{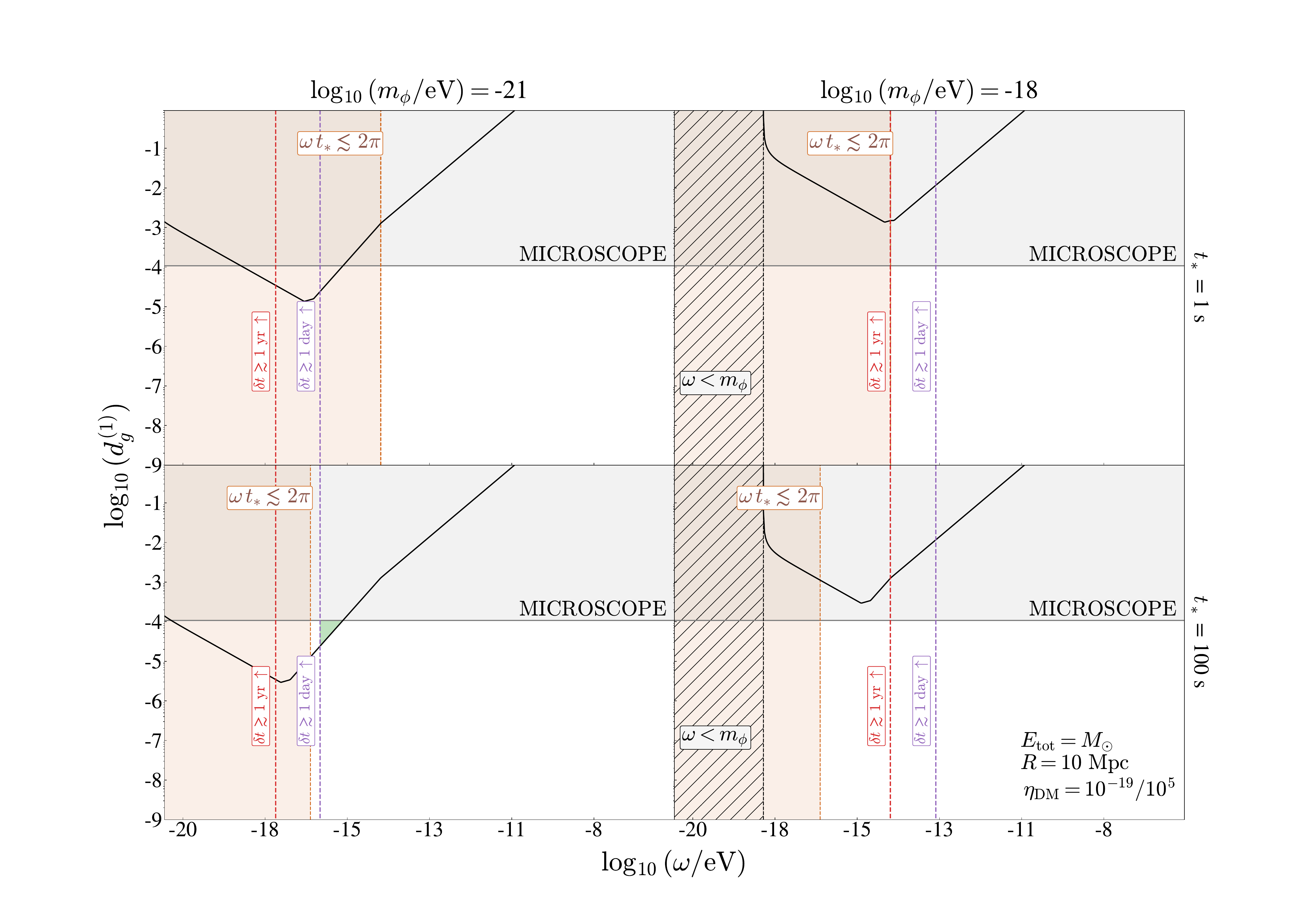}
    \caption{Sensitivity to linear dilatonic scalar coupling to gluons, $d_{g}^{(1)}$, as a function of the ULB scalar energy $\omega$, considering a relativistic ULB field burst from a source located at $R = 10~\mathrm{Mpc}$ from Earth, releasing a total energy of $E_{\rm total} = M_\odot$ in the form of ULBs. Benchmark ULB masses of $m_{\phi} = 10^{-21}~\mathrm{eV}$ (left column) and $m_{\phi} = 10^{-18}~\mathrm{eV}$ (right column), with intrinsic burst times of $t_* = 1~\mathrm{s}$ (top row) and $t_* = 100~\mathrm{s}$ (bottom row) are displayed.
    The gray region represents the parameter space excluded by MICROSCROPE EP tests \cite{Berge:2017ovy}, and the black line indicates the projected sensitivity of future experiments with $\eta_\mathrm{DM}=10^{-24}$, based on a Thorium isomer nuclear clock proposal \cite{2012PhRvL.108l0802C,FlaTh06}, scaled using Eq.~\eqref{eq:d_i(1)}. The black hatched and brown regions are excluded due to the scalar energy being less than its mass and the uncertainty principle, respectively. Dashed purple and red lines correspond to detection delays of 1 day and 1 year relative to a light-speed signal. The green region indicates the viable parameter space for multimessenger astronomy with scalars, given all constraints and benchmarks.
}
    \label{fig:10Mpc_gluon_lindilatonic_coupling}
\end{figure*}

\begin{figure*}
    \centering
    \includegraphics[width = 1\linewidth]{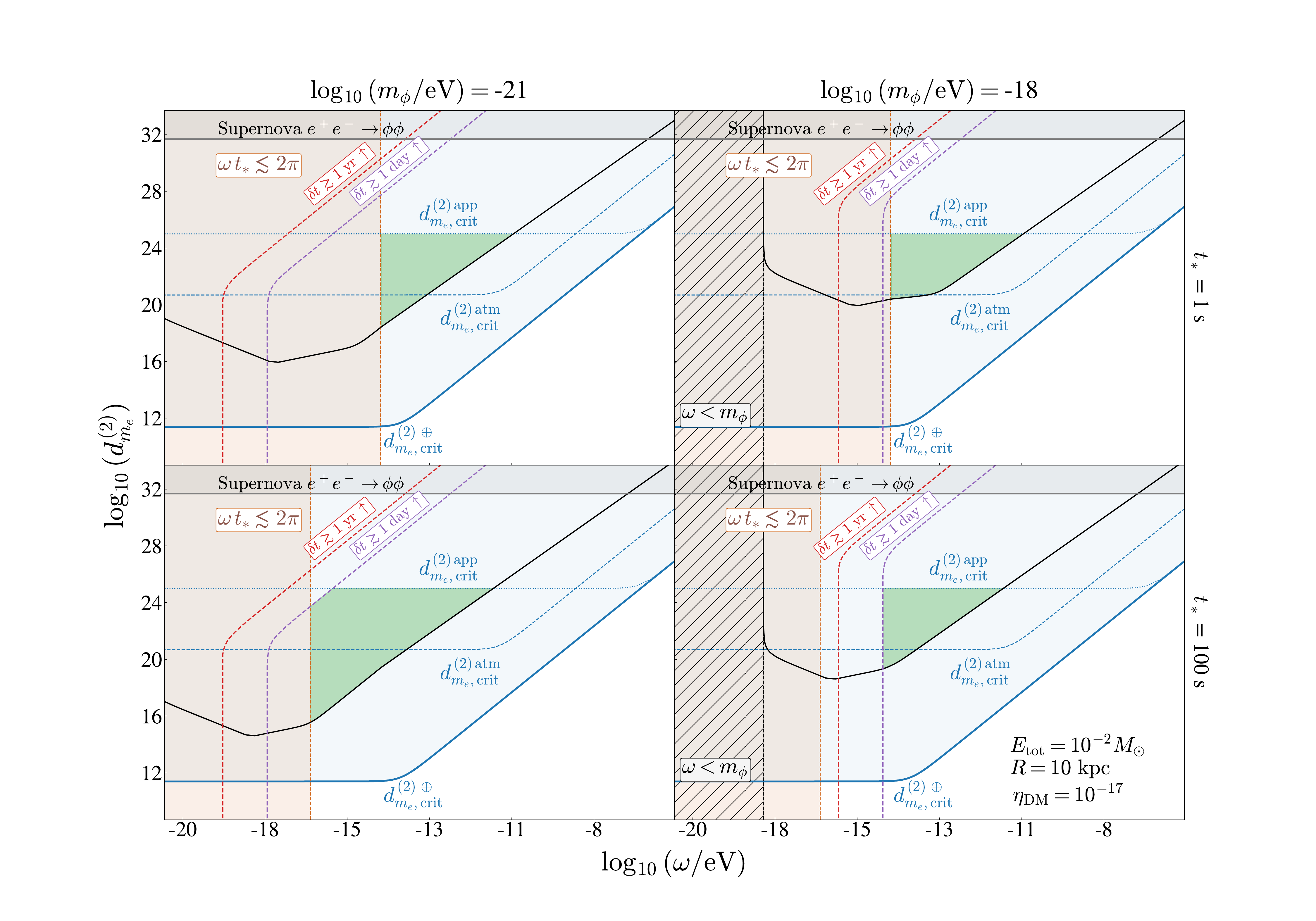}
    \caption{Sensitivity to quadratic dilatonic scalar coupling to electrons, $d_{m_e}^{(2)}$, as a function of the ULB scalar energy $\omega$, considering a relativistic ULB field burst from a source located at $R = 10~\mathrm{kpc}$ from Earth, releasing a total energy of $E_{\rm total} = 10^{-2} M_\odot$ in the form of ULBs. Benchmark ULB masses of $m_{\phi} = 10^{-21}~\mathrm{eV}$ (left column) and $m_{\phi} = 10^{-18}~\mathrm{eV}$ (right column), with intrinsic burst times of $t_* = 1~\mathrm{s}$ (top row) and $t_* = 100~\mathrm{s}$ (bottom row) are displayed.
    The blue solid, dashed, and dotted lines represent critical screening due to the Earth, atmosphere, and experimental apparatus, respectively. The black line shows the projected sensitivity of future experiments with $\eta_\mathrm{DM}=10^{-17}$, based on molecular spectra analysis of SrOH \cite{SrOH2021}, scaled using Eq.~\eqref{eq:d_i(2)}. The black hatched and brown regions are excluded due to the scalar energy being less than its mass and the uncertainty principle, respectively. Dashed purple and red lines indicate detection delays of 1 day and 1 year relative to a light-speed signal. The green region represents the viable parameter space for multimessenger astronomy with scalars, considering all constraints and benchmarks.
}
    \label{fig:10kpc_electron_quaddilatonic_coupling}
\end{figure*}

\begin{figure*}
    \centering
    \includegraphics[width = 1\linewidth]{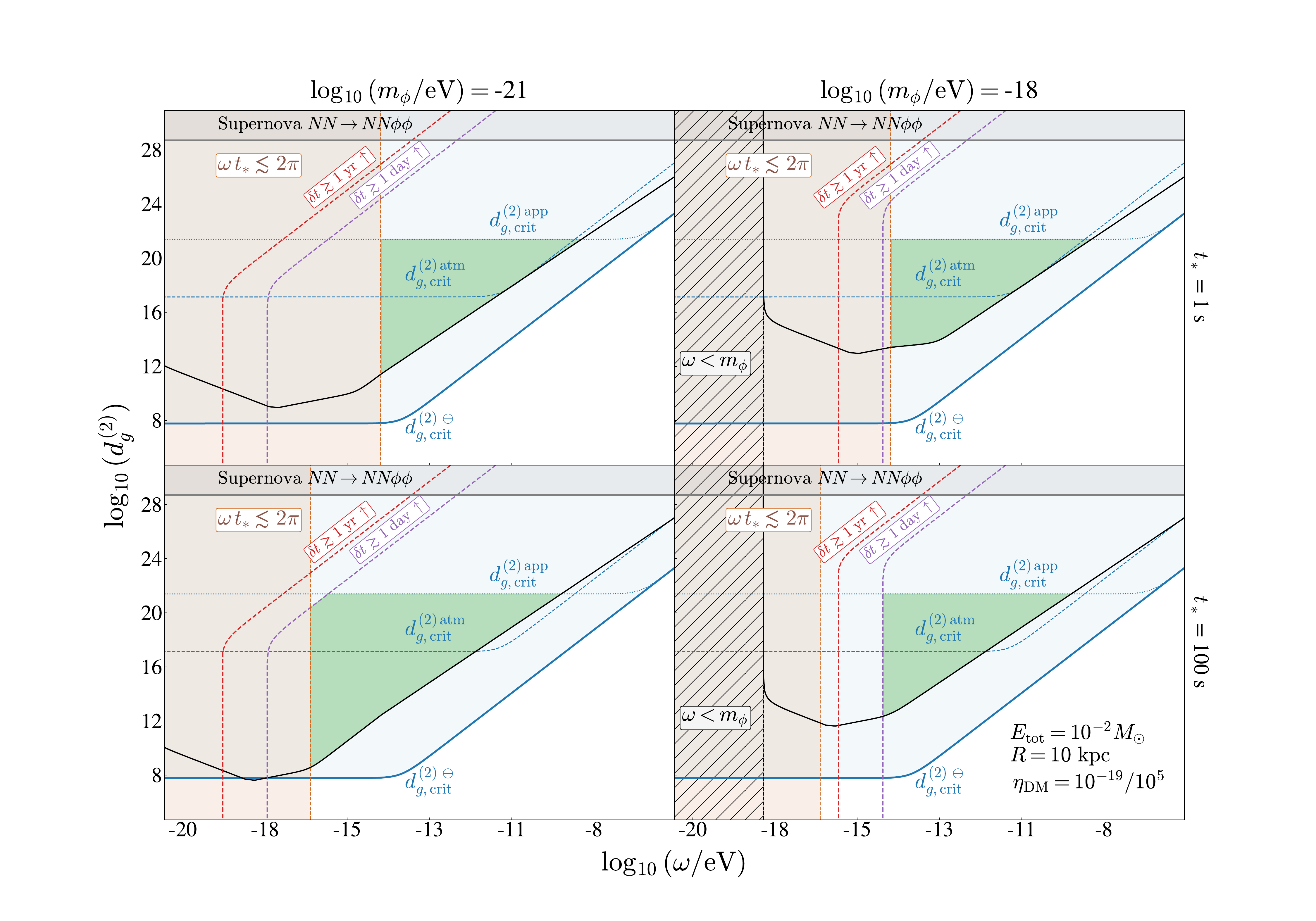}
    \caption{Sensitivity to quadratic dilatonic scalar coupling to gluons, $d_{g}^{(2)}$, as a function of the ULB scalar energy $\omega$, considering a relativistic ULB field burst from a source located at $R = 10~\mathrm{kpc}$ from Earth, releasing a total energy of $E_{\rm total} = 10^{-2} M_\odot$ in the form of ULBs. Benchmark ULB masses of $m_{\phi} = 10^{-21}~\mathrm{eV}$ (left column) and $m_{\phi} = 10^{-18}~\mathrm{eV}$ (right column), with intrinsic burst times of $t_* = 1~\mathrm{s}$ (top row) and $t_* = 100~\mathrm{s}$ (bottom row) are displayed.
The blue solid, dashed, and dotted lines indicate critical screening from the Earth, atmosphere, and experimental apparatus, respectively. The black line represents the projected sensitivity of future experiments with $\eta_\mathrm{DM}=10^{-24}$, based on a Thorium isomer nuclear clock proposal \cite{2012PhRvL.108l0802C,FlaTh06}, scaled using Eq.~\eqref{eq:d_i(2)}. The black hatched and brown regions are excluded by the requirements that scalar energy exceeds its mass and the uncertainty principle. Dashed purple and red lines show detection delays of 1 day and 1 year relative to a light-speed signal. The green region denotes the viable parameter space for multimessenger astronomy with scalars, considering all constraints and benchmarks.
}
    \label{fig:10kpc_gluon_quaddilatonic_coupling}
\end{figure*}

\begin{figure*}
    \centering
    \includegraphics[width = 1\linewidth]{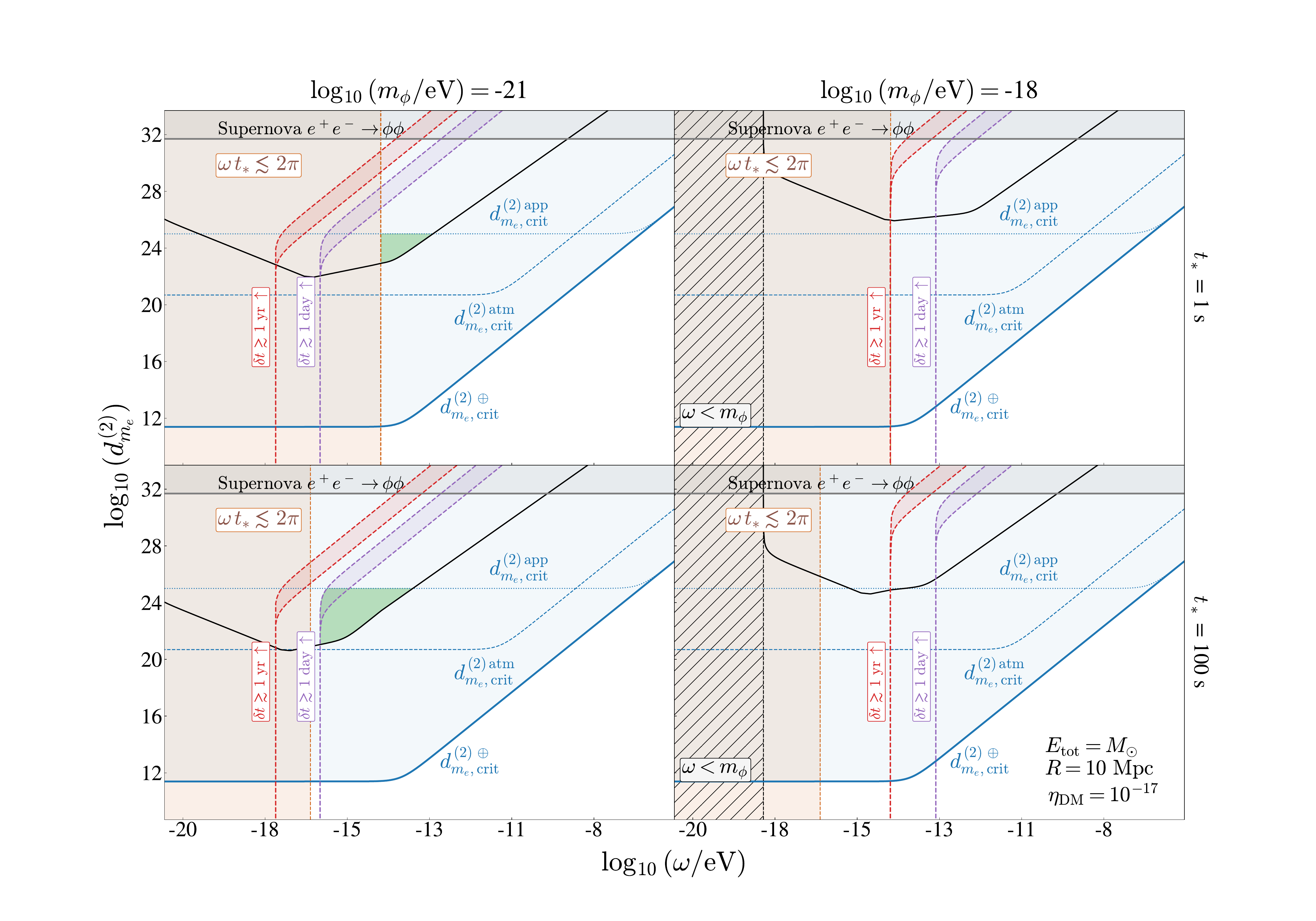}
    \caption{Sensitivity to quadratic dilatonic scalar coupling to electrons, $d_{m_e}^{(2)}$, as a function of the ULB scalar energy $\omega$, considering a relativistic ULB field burst from a source located at $R = 10~\mathrm{Mpc}$ from Earth, releasing a total energy of $E_{\rm total} = M_\odot$ in the form of ULBs. Benchmark ULB masses of $m_{\phi} = 10^{-21}~\mathrm{eV}$ (left column) and $m_{\phi} = 10^{-18}~\mathrm{eV}$ (right column), with intrinsic burst times of $t_* = 1~\mathrm{s}$ (top row) and $t_* = 100~\mathrm{s}$ (bottom row) are displayed.
    The blue solid, dashed, and dotted lines represent critical screening from the Earth, atmosphere, and experimental apparatus, respectively. The black line indicates the projected sensitivity of future experiments with $\eta_\mathrm{DM}=10^{-17}$, based on the SrOH molecular spectra analysis \cite{SrOH2021}, scaled using Eq.~\eqref{eq:d_i(2)}. The black hatched and brown regions are excluded by scalar energy limits and the uncertainty principle. Dashed purple and red lines show detection delays of 1 day and 1 year relative to a light-speed signal. The green region represents the viable parameter space for multimessenger astronomy with scalars, considering all constraints and benchmarks.
}
    \label{fig:10Mpc_electron_quaddilatonic_coupling}
\end{figure*}

\begin{figure*}
    \centering
    \includegraphics[width = 1\linewidth]{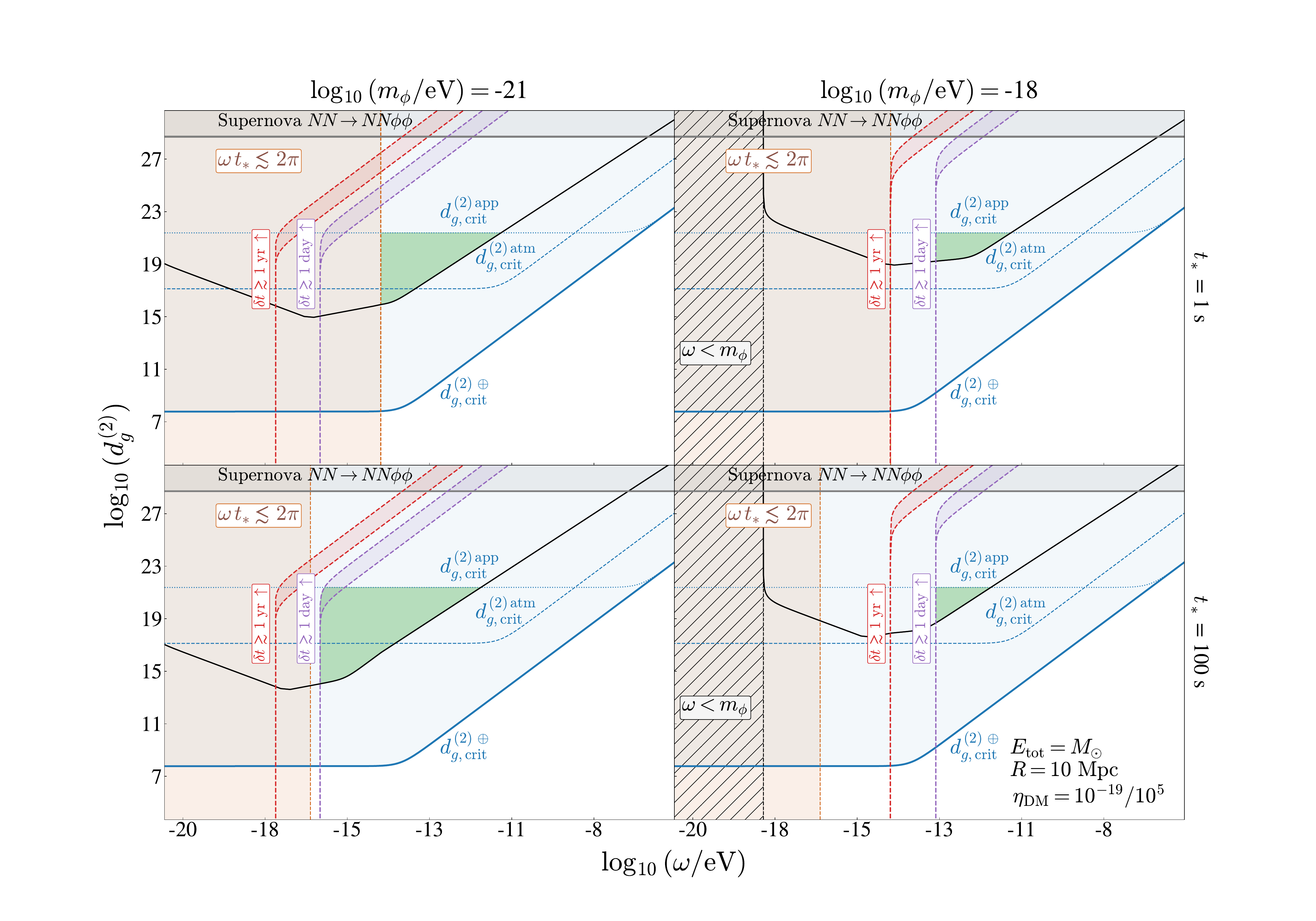}
    \caption{Sensitivity to quadratic dilatonic scalar coupling to gluons, $d_{g}^{(2)}$, as a function of the ULB scalar energy $\omega$, considering a relativistic ULB field burst from a source located at $R = 10~\mathrm{Mpc}$ from Earth, releasing a total energy of $E_{\rm total} = M_\odot$ in the form of ULBs. Benchmark ULB masses of $m_{\phi} = 10^{-21}~\mathrm{eV}$ (left column) and $m_{\phi} = 10^{-18}~\mathrm{eV}$ (right column), with intrinsic burst times of $t_* = 1~\mathrm{s}$ (top row) and $t_* = 100~\mathrm{s}$ (bottom row) are displayed.
The blue solid, dashed, and dotted lines indicate critical screening from the Earth, atmosphere, and experimental apparatus, respectively. The black line shows the projected sensitivity of future experiments with $\eta_\mathrm{DM}=10^{-24}$, based on a Thorium isomer nuclear clock proposal \cite{2012PhRvL.108l0802C,FlaTh06}, scaled using Eq.~\eqref{eq:d_i(2)}. The black hatched and brown regions are excluded by scalar energy constraints and the uncertainty principle. Dashed purple and red lines represent detection delays of 1 day and 1 year relative to a light-speed signal. The green region denotes viable parameter space for multimessenger astronomy with scalars, considering all constraints and benchmarks.
}
    \label{fig:10Mpc_gluon_quaddilatonic_coupling}
\end{figure*}

\begin{figure*}
    \centering
    \includegraphics[width = 1\linewidth]{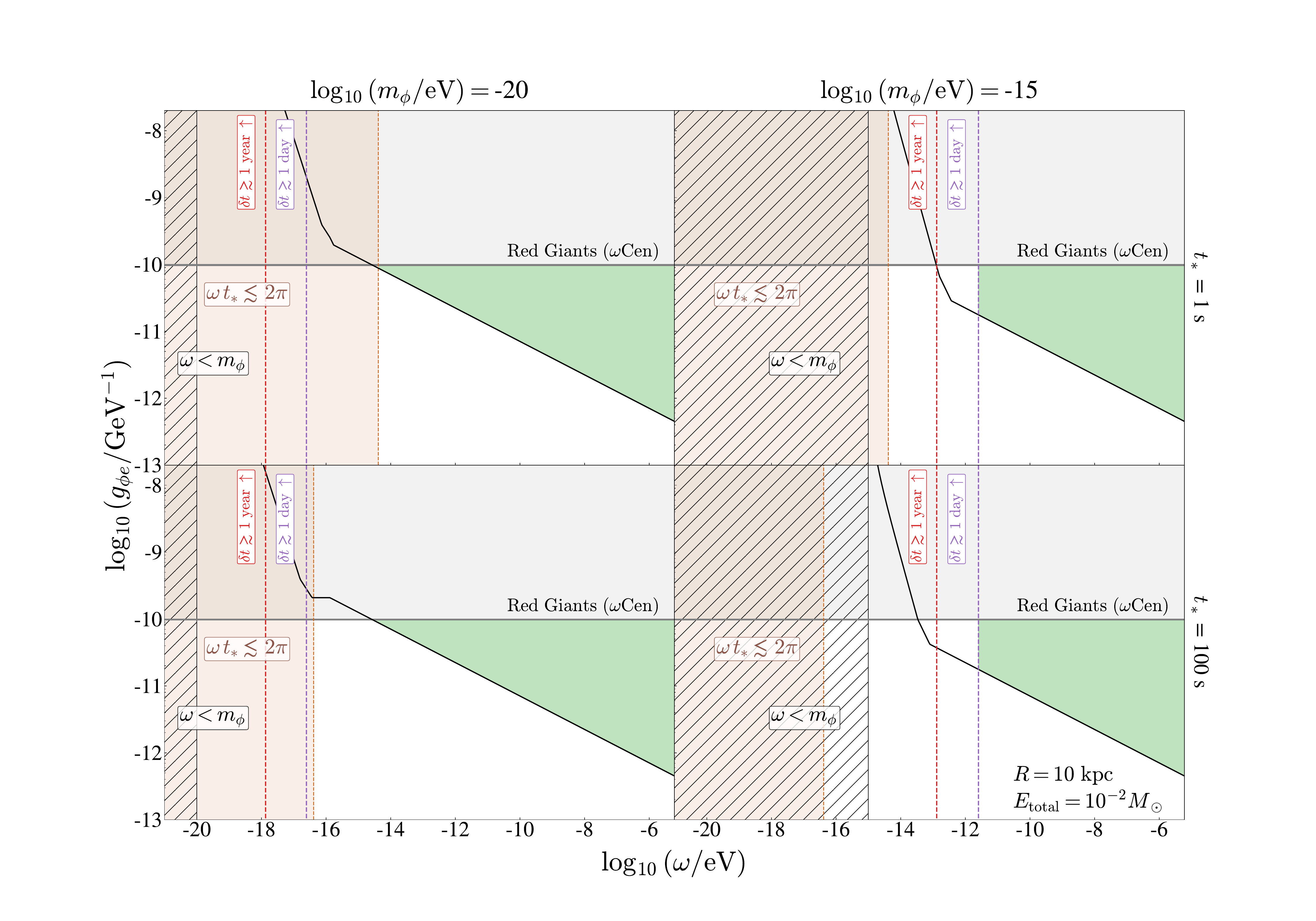}
    \caption{Sensitivity to ALP derivative coupling to electrons, $g_{\phi e}$, as a function of the ULB scalar energy $\omega$, considering a relativistic ULB field burst from a source located at $R = 10~\mathrm{kpc}$ from Earth, releasing a total energy of $E_{\rm total} = 10^{-2} M_\odot$ in the form of ULBs. Benchmark ULB masses of $m_{\phi} = 10^{-20}~\mathrm{eV}$ (left column) and $m_{\phi} = 10^{-15}~\mathrm{eV}$ (right column), with intrinsic burst times of $t_* = 1~\mathrm{s}$ (top row) and $t_* = 100~\mathrm{s}$ (bottom row) are displayed.
    The gray region represents parameter space excluded by cooling in tip of the red-giant branch in $\omega$ Centauri from $\phi$ emission~\cite{capozzi2020axion}. 
    The black line shows the projected sensitivity of future experiments with $g_{\rm DM} = 2\times10^{-11}~\mathrm{GeV}^{-1}$ and $t_{\rm int, DM} = 1~\mathrm{year}$, based on proposed magnetometers \cite{chigusa2023light} and torsion pendulums \cite{graham2018spin}, scaled using Eq.~\eqref{eq:g_ratio_v}. The black hatched and brown regions are excluded due to ALP energy constraints and the uncertainty principle. Dashed purple and red lines indicate detection delays of 1 day and 1 year relative to a light-speed signal. The green region denotes viable parameter space for multimessenger astronomy with scalars, considering all constraints and benchmarks.
}
    \label{fig:10kpc_axion_electron}
\end{figure*}

\begin{figure*}
    \centering
    \includegraphics[width = 1\linewidth]{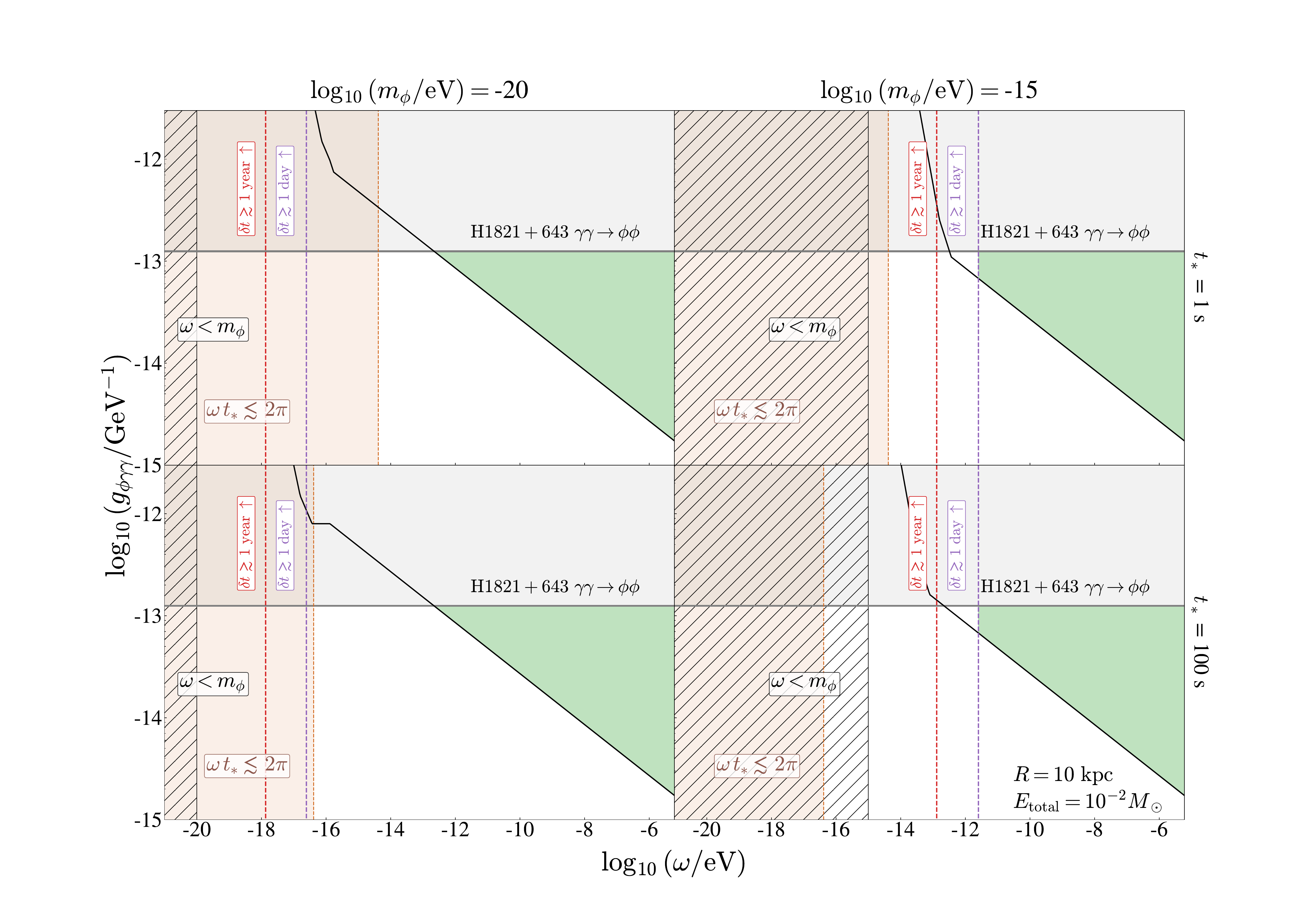}
    \caption{Sensitivity to ALP linear coupling to photons, $g_{\phi \gamma\gamma}$, as a function of the ULB scalar energy $\omega$, considering a relativistic ULB field burst from a source located at $R = 10~\mathrm{kpc}$ from Earth, releasing a total energy of $E_{\rm total} = 10^{-2} M_\odot$ in the form of ULBs. Benchmark ULB masses of $m_{\phi} = 10^{-20}~\mathrm{eV}$ (left column) and $m_{\phi} = 10^{-15}~\mathrm{eV}$ (right column), with intrinsic burst times of $t_* = 1~\mathrm{s}$ (top row) and $t_* = 100~\mathrm{s}$ (bottom row) are displayed.
   The gray region represents parameter space excluded by Chandra observations of spectral distortions from $\gamma$-$\phi$ oscillations in luminous quasar H1821+643~\cite{sisk2022new}.  The black line shows the projected sensitivity of future experiments with $g_{\rm DM} = 3\times10^{-16}~\mathrm{GeV}^{-1}$ and $t_{\rm int, DM} = 1~\mathrm{year}$, based on proposed superconducting cavities \cite{berlin2021heterodyne} and lumped-element detectors \cite{DMRadio:2022pkf}, scaled using Eq.~\eqref{eq:g_ratio}. The black hatched and brown regions are excluded by ALP energy constraints and the uncertainty principle. Dashed purple and red lines indicate detection delays of 1 day and 1 year relative to a light-speed signal. The green region denotes viable parameter space for multimessenger astronomy with scalars, considering all constraints and benchmarks.
}
    \label{fig:10kpc_axion_photon}
\end{figure*}

\begin{figure*}
    \centering
    \includegraphics[width = 1\linewidth]{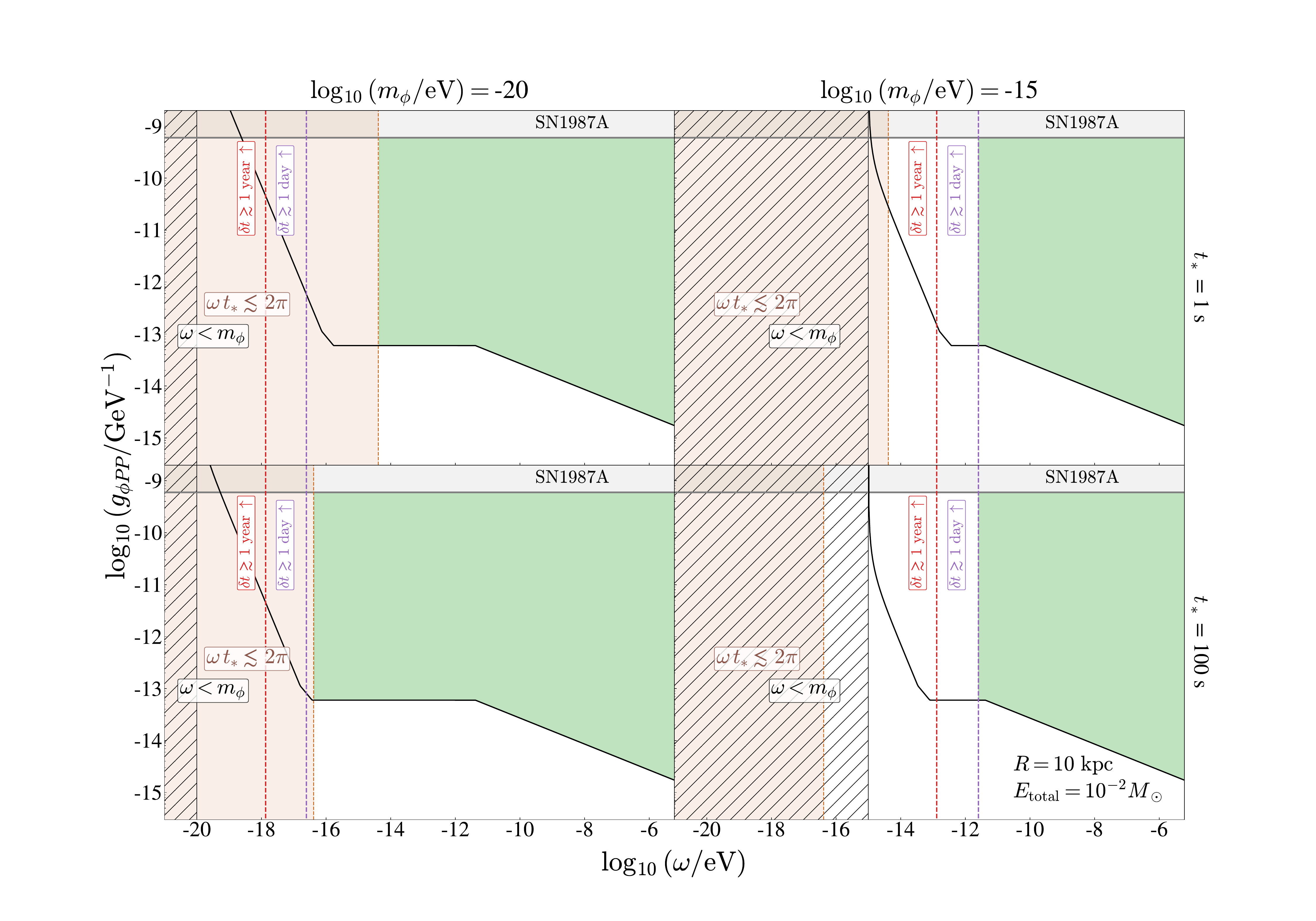}
    \caption{Sensitivity to ALP derivative coupling to protons, $g_{\phi PP}$, as a function of the ULB scalar energy $\omega$, considering a relativistic ULB field burst from a source located at $R = 10~\mathrm{kpc}$ from Earth, releasing a total energy of $E_{\rm total} = 10^{-2} M_\odot$ in the form of ULBs. Benchmark ULB masses of $m_{\phi} = 10^{-20}~\mathrm{eV}$ (left column) and $m_{\phi} = 10^{-15}~\mathrm{eV}$ (right column), with intrinsic burst times of $t_* = 1~\mathrm{s}$ (top row) and $t_* = 100~\mathrm{s}$ (bottom row) are displayed.
    The gray region represents parameter space excluded by ALP signals in the Kamiokande-II neutrino experiment associated with supernova SN1987A~\cite{lella2024getting}.  
    The black line shows the projected sensitivity of future experiments with $g_{\rm DM} = 10^{-12}~\mathrm{GeV}^{-1}$ and $t_{\rm int, DM} = 1000~\mathrm{s}$, based on proposed NMR techniques \cite{jackson2020overview} and proton storage rings \cite{graham2021storage}, scaled using Eq.~\eqref{eq:g_ratio_v}. The black hatched and brown regions are excluded by ALP energy constraints and the uncertainty principle. Dashed purple and red lines indicate detection delays of 1 day and 1 year relative to a light-speed signal. The green region represents viable parameter space for multimessenger astronomy with ALPs, considering all constraints and benchmarks.
}
    \label{fig:10kpc_axion_proton}
\end{figure*}

\begin{figure*}
    \centering
    \includegraphics[width = 1\linewidth]{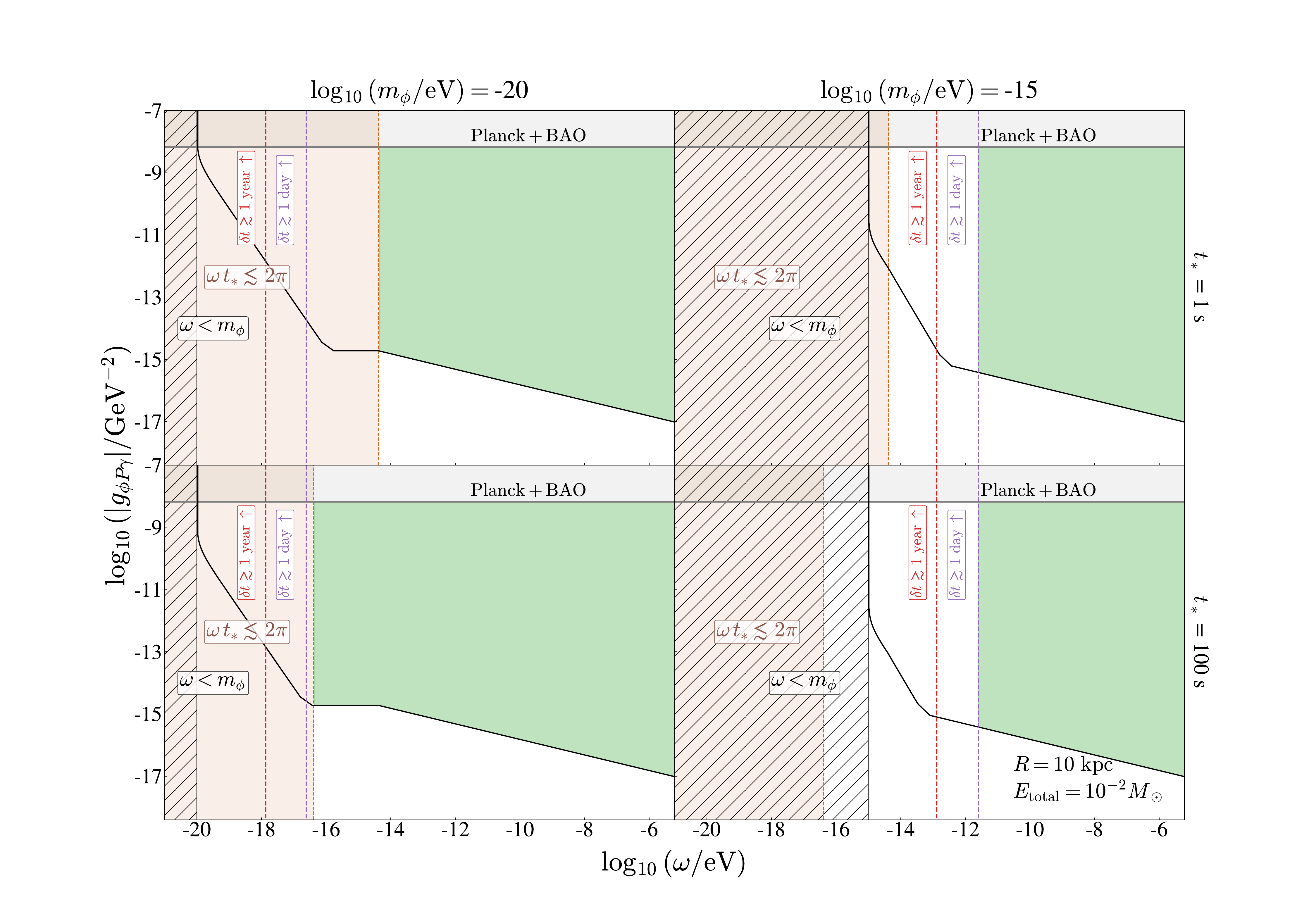}
    \caption{Sensitivity to ALP EDM coupling to protons, $|g_{\phi P \gamma}|$, as a function of the ULB scalar energy $\omega$, considering a relativistic ULB field burst from a source located at $R = 10~\mathrm{kpc}$ from Earth, releasing a total energy of $E_{\rm total} = 10^{-2} M_\odot$ in the form of ULBs. Benchmark ULB masses of $m_{\phi} = 10^{-20}~\mathrm{eV}$ (left column) and $m_{\phi} = 10^{-15}~\mathrm{eV}$ (right column), with intrinsic burst times of $t_* = 1~\mathrm{s}$ (top row) and $t_* = 100~\mathrm{s}$ (bottom row) are displayed. The gray region represents parameter space excluded by thermally-produced ALP signals in cosmic microwave background as observed by Planck2018 and baryon acoustic oscillations (BAO) data \cite{caloni2022novel}. The black line shows the projected sensitivity of future experiments with $g_{\rm DM} = 10^{-18}~\mathrm{GeV}^{-2}$ and $t_{\rm int, DM} = 10^6~\mathrm{s}$, based on proposed NMR techniques \cite{jackson2020overview}, scaled using Eq.~\eqref{eq:g_ratio}. The black hatched and brown regions are excluded by ALP energy constraints and the uncertainty principle. Dashed purple and red lines indicate detection delays of 1 day and 1 year relative to a light-speed signal. The green region denotes viable parameter space for multimessenger astronomy with ALPs, considering all constraints and benchmarks.
}       \label{fig:10kpc_axion_EDM}
\end{figure*}

\begin{figure*}
    \centering
    \includegraphics[width = 1\linewidth]{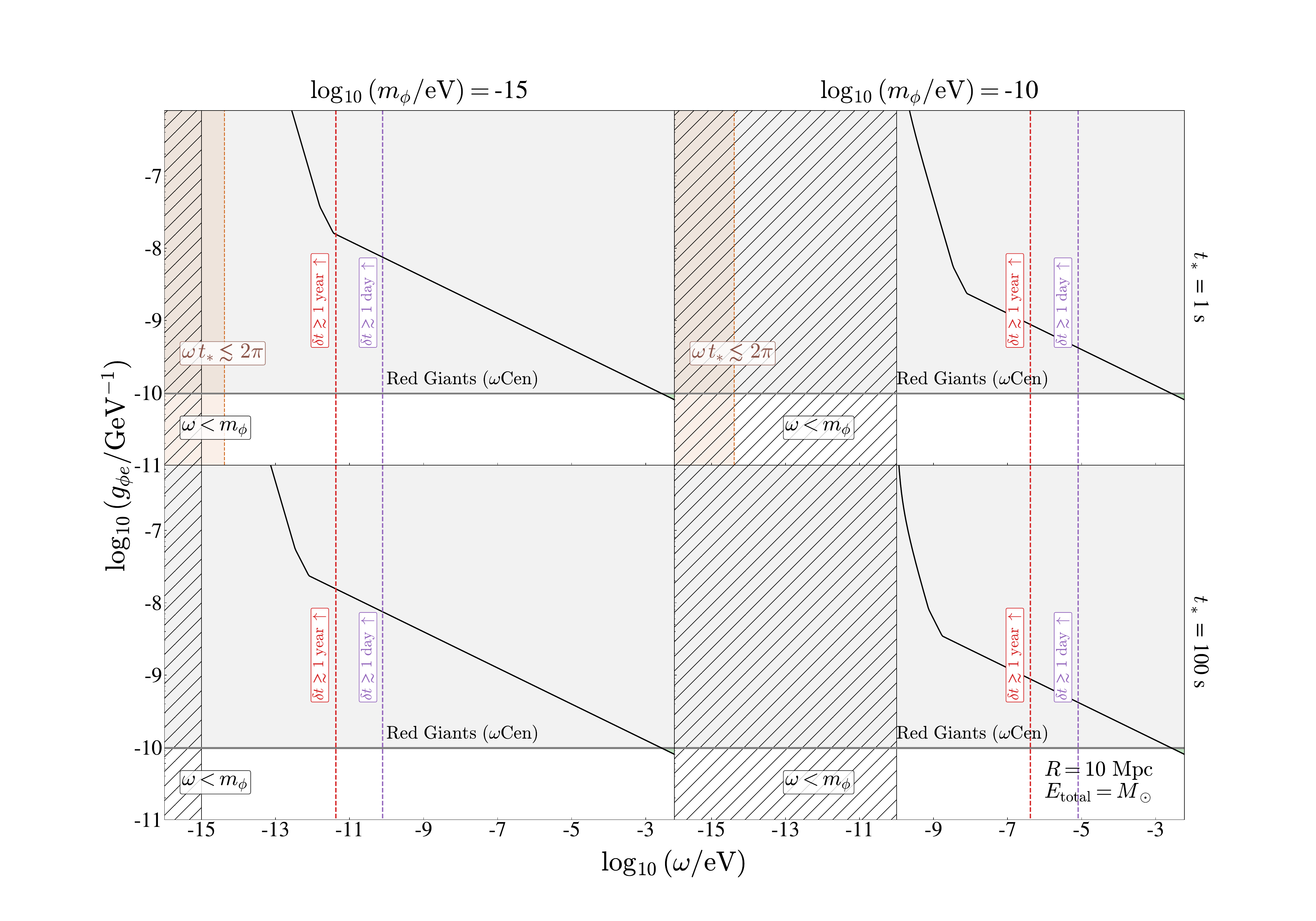}
    \caption{Sensitivity to ALP derivative coupling to electrons, $g_{\phi e}$, as a function of the ULB scalar energy $\omega$, considering a relativistic ULB field burst from a source located at $R = 10~\mathrm{Mpc}$ from Earth, releasing a total energy of $E_{\rm total} = M_\odot$ in the form of ULBs. Benchmark ULB masses of $m_{\phi} = 10^{-15}~\mathrm{eV}$ (left column) and $m_{\phi} = 10^{-10}~\mathrm{eV}$ (right column), with intrinsic burst times of $t_* = 1~\mathrm{s}$ (top row) and $t_* = 100~\mathrm{s}$ (bottom row) are displayed. The gray region represents parameter space excluded by cooling in tip of the red-giant branch in $\omega$ Centauri from $\phi$ emission~\cite{capozzi2020axion}.
    The black line shows the projected sensitivity of future experiments with $g_{\rm DM} = 2\times10^{-11}~\mathrm{GeV}^{-1}$ and $t_{\rm int, DM} = 1~\mathrm{year}$, based on proposed magnetometers \cite{chigusa2023light} and torsion pendulums \cite{graham2018spin}, scaled using Eq.~\eqref{eq:g_ratio_v}. The black hatched and brown regions are excluded by ALP energy constraints and the uncertainty principle. Dashed purple and red lines indicate detection delays of 1 day and 1 year relative to a light-speed signal. For these benchmarks, no viable parameter space exists for multimessenger astronomy with ALPs.
}       \label{fig:10Mpc_axion_electron}
\end{figure*}

\begin{figure*}
    \centering
    \includegraphics[width = 1\linewidth]{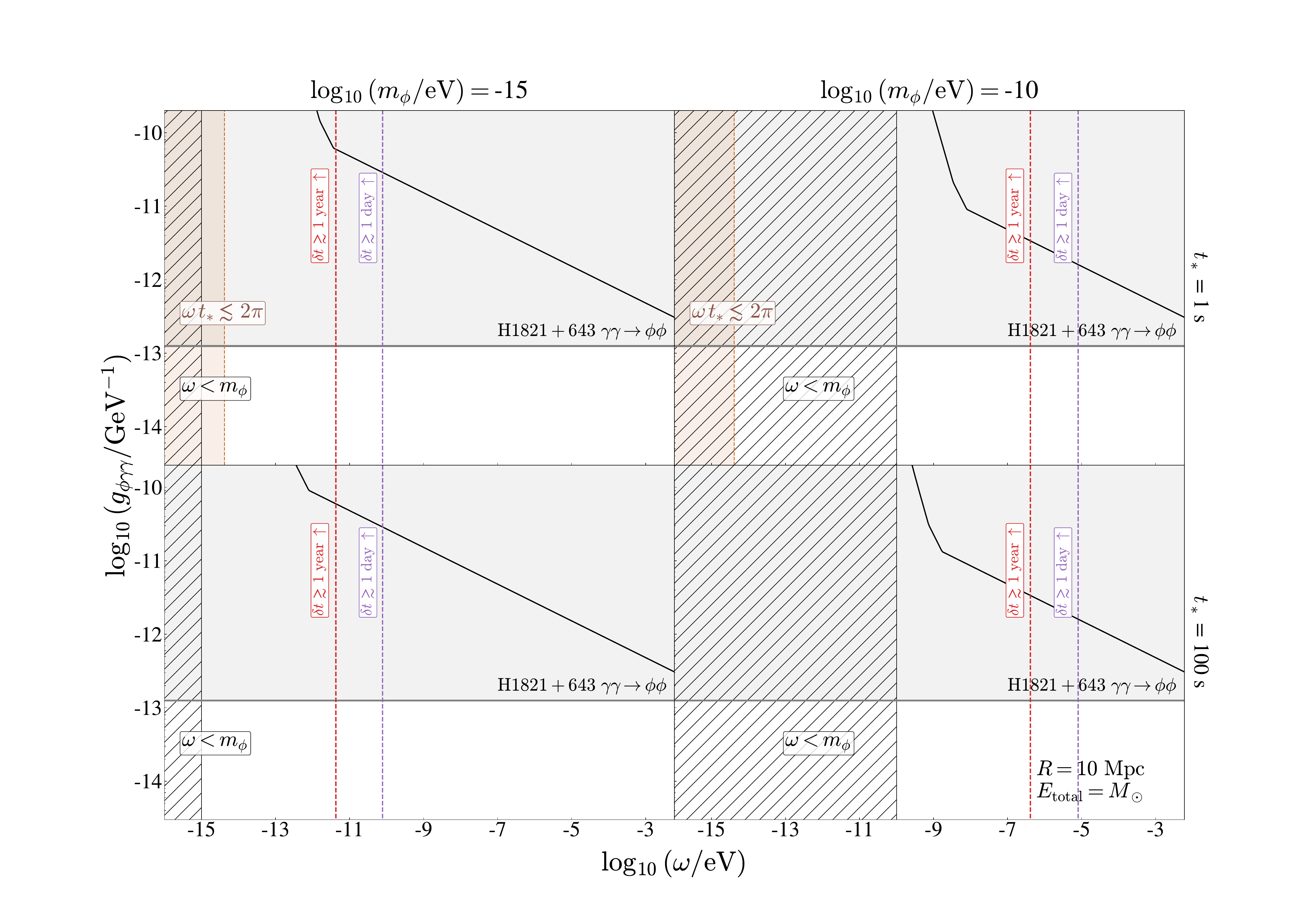}
    \caption{Sensitivity to ALP linear coupling to photons, $g_{\phi \gamma\gamma}$, as a function of the ULB scalar energy $\omega$, considering a relativistic ULB field burst from a source located at $R = 10~\mathrm{Mpc}$ from Earth, releasing a total energy of $E_{\rm total} =  M_\odot$ in the form of ULBs. Benchmark ULB masses of $m_{\phi} = 10^{-15}~\mathrm{eV}$ (left column) and $m_{\phi} = 10^{-10}~\mathrm{eV}$ (right column), with intrinsic burst times of $t_* = 1~\mathrm{s}$ (top row) and $t_* = 100~\mathrm{s}$ (bottom row) are displayed.
   The gray region represents parameter space excluded by Chandra observations of spectral distortions from $\gamma$-$\phi$ oscillations in luminous quasar H1821+643~\cite{sisk2022new}.
    The black line shows the projected sensitivity of future experiments with $g_{\rm DM} = 3\times10^{-16}~\mathrm{GeV}^{-1}$ and $t_{\rm int, DM} = 1~\mathrm{year}$, based on proposed superconducting cavities \cite{berlin2021heterodyne} and lumped-element detectors \cite{DMRadio:2022pkf}, scaled using Eq.~\eqref{eq:g_ratio}. The black hatched and brown regions are excluded by ALP energy constraints and the uncertainty principle. Dashed purple and red lines indicate detection delays of 1 day and 1 year relative to a light-speed signal. For these benchmarks, no viable parameter space exists for multimessenger astronomy with ALPs.
}        \label{fig:10Mpc_axion_photon}
\end{figure*}

\begin{figure*}
    \centering
    \includegraphics[width = 1\linewidth]{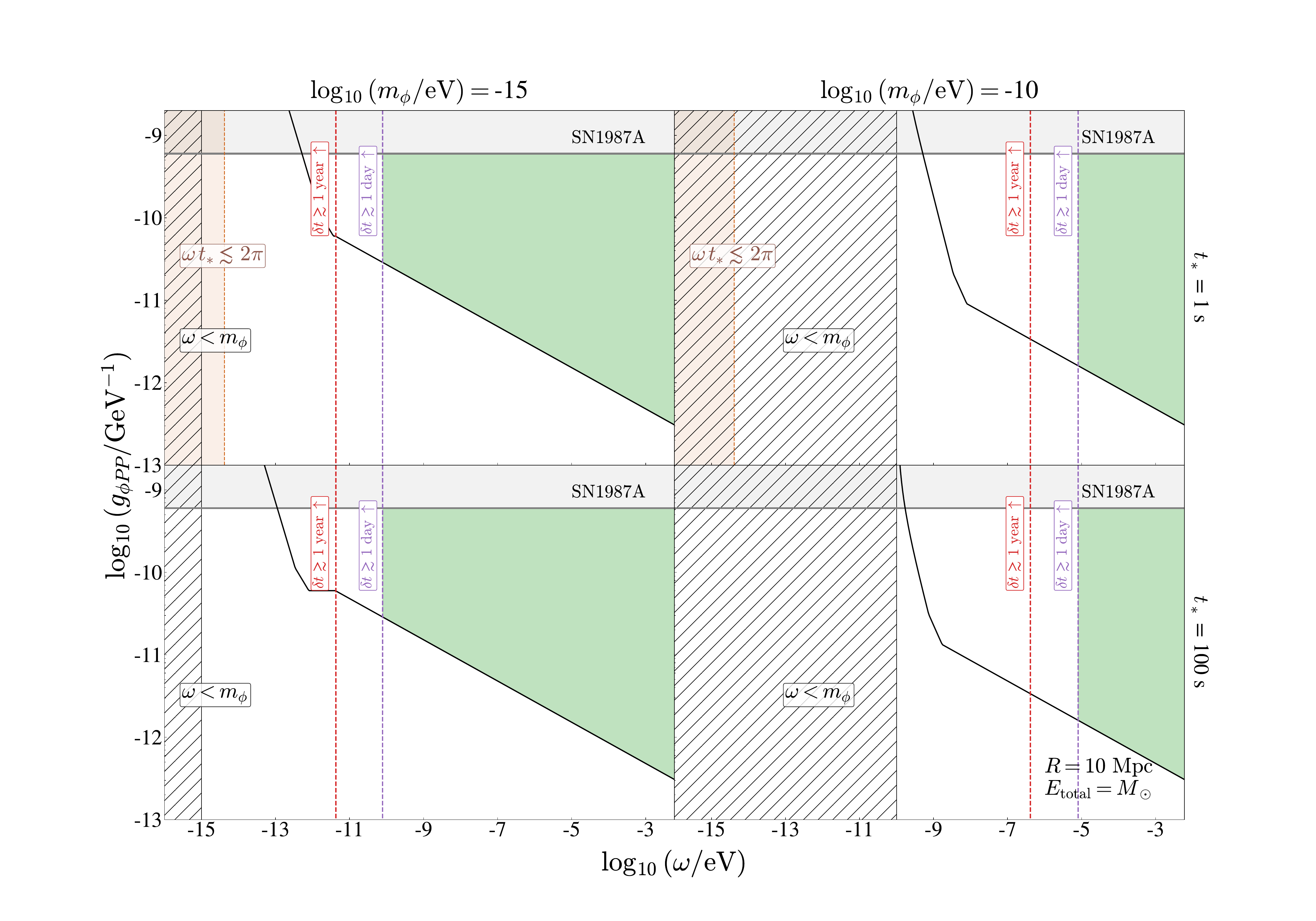}
    \caption{Sensitivity to ALP derivative coupling to protons, $g_{\phi PP}$, as a function of the ULB scalar energy $\omega$, considering a relativistic ULB field burst from a source located at $R = 10~\mathrm{Mpc}$ from Earth, releasing a total energy of $E_{\rm total} =  M_\odot$ in the form of ULBs. Benchmark ULB masses of $m_{\phi} = 10^{-15}~\mathrm{eV}$ (left column) and $m_{\phi} = 10^{-10}~\mathrm{eV}$ (right column), with intrinsic burst times of $t_* = 1~\mathrm{s}$ (top row) and $t_* = 100~\mathrm{s}$ (bottom row) are displayed.
    The gray region represents parameter space excluded by ALP signals in the Kamiokande-II neutrino experiment associated with supernova SN1987A~\cite{lella2024getting}. 
    The black line shows the projected sensitivity of future experiments with $g_{\rm DM} = 10^{-12}~\mathrm{GeV}^{-1}$ and $t_{\rm int, DM} = 1000~\mathrm{s}$, based on proposed NMR techniques \cite{jackson2020overview} and proton storage rings \cite{graham2021storage}, scaled using Eq.~\eqref{eq:g_ratio_v}. The black hatched and brown regions are excluded by ALP energy constraints and the uncertainty principle. Dashed purple and red lines indicate detection delays of 1 day and 1 year relative to a light-speed signal. The green region represents viable parameter space for multimessenger astronomy with ALPs, considering all constraints and benchmarks.
}        
    \label{fig:10Mpc_axion_proton}
\end{figure*}

\begin{figure*}
    \centering
    \includegraphics[width = 1\linewidth]{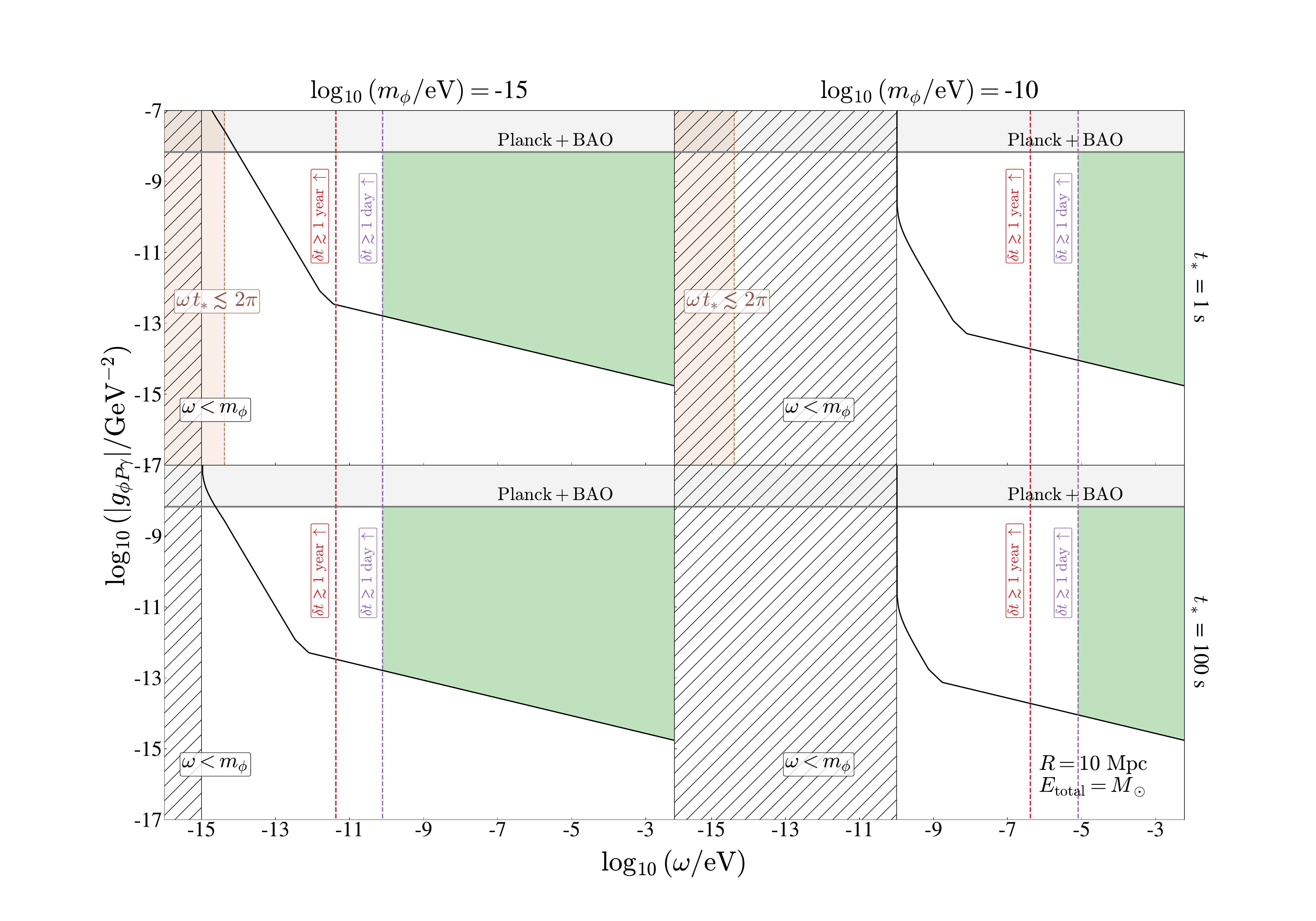}
    \caption{Sensitivity to ALP EDM coupling to protons, $|g_{\phi P \gamma}|$, as a function of the ULB scalar energy $\omega$, considering a relativistic ULB field burst from a source located at $R = 10~\mathrm{Mpc}$ from Earth, releasing a total energy of $E_{\rm total} =  M_\odot$ in the form of ULBs. Benchmark ULB masses of $m_{\phi} = 10^{-15}~\mathrm{eV}$ (left column) and $m_{\phi} = 10^{-10}~\mathrm{eV}$ (right column), with intrinsic burst times of $t_* = 1~\mathrm{s}$ (top row) and $t_* = 100~\mathrm{s}$ (bottom row) are displayed. The gray region represents parameter space excluded by thermally-produced ALP signals in cosmic microwave background as observed by Planck2018 and baryon acoustic oscillations (BAO) data \cite{caloni2022novel}.
    The black line shows the projected sensitivity of future experiments with $g_{\rm DM} = 10^{-18}~\mathrm{GeV}^{-2}$ and $t_{\rm int, DM} = 10^6~\mathrm{s}$, based on proposed NMR techniques \cite{jackson2020overview}, scaled using Eq.~\eqref{eq:g_ratio}. The black hatched and brown regions are excluded by ALP energy constraints and the uncertainty principle. Dashed purple and red lines indicate detection delays of 1 day and 1 year relative to a light-speed signal. The green region denotes viable parameter space for multimessenger astronomy with ALPs, considering all constraints and benchmarks.
}        \label{fig:10Mpc_axion_EDM}
\end{figure*}

For completeness, we present the sensitivity reach for all the couplings discussed in Sec.~\ref{sec:EFT}. 
These include linear scalar couplings to photons (Figs.~\ref{fig:10kpc_photon_lindilatonic_coupling}, \ref{fig:10Mpc_photon_lindilatonic_coupling}), electrons (Figs.~\ref{fig:10kpc_electron_lindilatonic_coupling}, \ref{fig:10Mpc_electron_lindilatonic_coupling}), and gluons (Figs.~\ref{fig:10kpc_gluon_lindilatonic_coupling}, \ref{fig:10Mpc_gluon_lindilatonic_coupling}), as well as quadratic couplings to electrons (Figs.~\ref{fig:10kpc_electron_quaddilatonic_coupling}, \ref{fig:10Mpc_electron_quaddilatonic_coupling}), and gluons (Figs.~\ref{fig:10kpc_gluon_quaddilatonic_coupling}, \ref{fig:10Mpc_gluon_quaddilatonic_coupling}). For ALPs, we consider derivative couplings to electrons (Figs.~\ref{fig:10kpc_axion_electron}, \ref{fig:10Mpc_axion_electron}), linear couplings to photons (Figs.~\ref{fig:10kpc_axion_photon}, \ref{fig:10Mpc_axion_photon}), derivative couplings to protons (Figs.~\ref{fig:10kpc_axion_proton}, \ref{fig:10Mpc_axion_proton}), and EDM couplings to nucleons (Figs.~\ref{fig:10kpc_axion_EDM}, \ref{fig:10Mpc_axion_EDM}). 

For some couplings (Figs.~\ref{fig:10kpc_electron_lindilatonic_coupling}, \ref{fig:10Mpc_photon_lindilatonic_coupling}, \ref{fig:10Mpc_electron_lindilatonic_coupling}, \ref{fig:10Mpc_axion_electron}, and \ref{fig:10Mpc_axion_photon}), the best projections from proposed experiments are not sufficiently strong to overcome astrophysical constraints after rescaling the sensitivities for ULB bursts. This is especially relevant for sources at greater distances (e.g., $\mathcal{O}(10)$ Mpc) or for scalar linear couplings, where stringent constraints from EP tests \cite{Berge:2017ovy} significantly limit the allowed parameter space.
Multimessenger astronomy prospects with ultralight scalars are viable for linear couplings to photons and gluons for a Galactic source, and for quadratic couplings to photons, electrons and gluons. For ALPs, multimessenger astronomy prospects remain viable for the three nuclear couplings (to protons, neutrons, and nucleons via EDM), while the parameter space for electron and photon couplings is excluded due to tighter experimental and astrophysical constraints. 

\bibliographystyle{JHEP}
\bibliography{ref}

\end{document}